\documentclass[10pt,journal,compsoc]{IEEEtran}
\usepackage{dblfloatfix}
\usepackage{setspace}
\usepackage{mathrsfs}
\usepackage[boxed,ruled,vlined,linesnumbered]{algorithm2e}
\usepackage{multirow}
\usepackage{times}
\usepackage{hhline}
\usepackage{url}
\usepackage{framed}
\usepackage{longtable}
\usepackage{mdframed}

\newcommand\encircle[1]{%
\tikz[baseline=(X.base)]
   \node (X) [draw, shape=circle, inner sep=-1.5pt, fill=black, text=white] {\strut #1};}

\newcommand*\wrapletters[1]{\wr@pletters#1\@nil}
\def\wr@pletters#1#2\@nil{#1\allowbreak\if&#2&\else\wr@pletters#2\@nil\fi}
\usepackage{tikz}
\usetikzlibrary{automata,matrix,shapes,arrows,positioning,chains,calc}
\usetikzlibrary{snakes}
\usetikzlibrary{arrows,scopes}
\usetikzlibrary{positioning,chains,fit,shapes,calc}
\usepackage{caption}
\usepackage{subcaption}
\usepackage{amsmath}
\SetKwHangingKw{Local}{Local}
\SetKwHangingKw{Global}{Global}
\SetKwHangingKw{Constant}{Constant}
\SetKwHangingKw{Input}{Input}
\SetKwHangingKw{Alias}{Alias}
\SetKwHangingKw{Output}{Output}
\SetKwHangingKw{SideEffect}{Side effects}
\SetKwHangingKw{InternalEvent}{Internal event}
\SetKwHangingKw{External}{External}
\SetKwHangingKw{ExternalEvent}{External event}
\SetKwHangingKw{Precondition}{Precondition :}
\SetKwHangingKw{Postcondition}{Postcondition :}
\SetKwHangingKw{Upon}{Upon}
\SetKwHangingKw{DoForever}{Do Forever}
\SetKwHangingKw{PVab}{Persistent variables:}

\usepackage{theorem}
\usepackage{wrapfig}
\usepackage{amssymb}
\usepackage{xspace}

\newtheorem{theorem}{Theorem}

\newcommand{\qed}{\hfill $\blacksquare$}
\newcommand{\qedClaim}{\hfill \ensuremath{\Box}}
\newenvironment{proof}{\noindent {\bf Proof.}\ }{\qed\par\vskip 0mm\par}

\newcommand{\B}{\vspace*{-\smallskipamount}}
\newcommand{\BB}{\vspace*{-\medskipamount}}
\newcommand{\BBB}{\vspace*{-\bigskipamount}}

\newcommand{\remove}[1]{}

\usepackage{csquotes}
\usepackage{enumitem}
\usepackage{tablefootnote}
\usepackage{fnpct}
\newcommand{\BigO}[0]{{\cal O}\xspace}

\ifCLASSOPTIONcompsoc
  \usepackage[nocompress]{cite}
\else
  \usepackage{cite}
\fi
\ifCLASSINFOpdf
\else
\fi

\IEEEoverridecommandlockouts

\begin{document}
\title{Privacy-Preserving Secret Shared Computations using MapReduce\thanks{\textbf{Manuscript received 29 June 2018; accepted 01 Aug. 2019. DOI: 10.1109/TDSC.2019.2933844. \copyright 2019 IEEE. Personal use of this material is permitted. Permission from IEEE must be obtained for all other uses, including reprinting/republishing this material for advertising or promotional purposes, collecting new collected works for resale or redistribution to servers or lists, or reuse of any copyrighted component of this work in other works.}
\protect\\
\textbf{The final published version of this paper may differ from this accepted version.}
\protect\\
\textbf{Corresponding and first author:} S. Sharma (e-mail: \texttt{shantanu.sharma@uci.edu}).\protect\\
{An extended abstract of this work was accepted in the Annual IFIP WG 11.3 Working Conference on Data and Applications Security and Privacy (DBSec), 2016~\cite{DBLP:conf/dbsec/DolevL016}.} \protect\\
Shlomi Dolev is with Ben-Gurion University of the Negev, Beer-Sheva, Israel. 
\protect\\
Yin Li is with Xinyang Normal University and Henan Key Lab of Analysis and Application of Educating Big Data, China. 
\protect\\
Peeyush Gupta, Sharad Mehrotra, and Shantanu Sharma are with University of California, Irvine, USA. \protect\\
\textnormal{\textbf{S. Dolev's} research was partially supported by the Lynne and William Frankel Center for Computer Science, the Rita Altura Trust Chair in Computer Science and also supported by a grant from the Israeli Ministry of Science, Technology and Space, Infrastructure
Research in the Field of Advanced Computing and Cyber Security, the
Israel National Cyber Bureau, the Israel \& the Japan Science and Technology Agency (JST), and the German Research Funding
Organization (DFG, Grant\#8767581199).}\protect\\
\textnormal{\textbf{Y. Li's} work is supported by National Natural Science Foundation of China (Grant no. 61402393, 61601396).}\protect\\
\textnormal{\textbf{S. Mehrotra's} work is based on research sponsored by DARPA under agreement number FA8750-16-2-0021 and partially supported by NSF grants 1527536 and 1545071. The U.S. Government is authorized to reproduce and distribute reprints for Governmental purposes notwithstanding any copyright notation thereon. The views and conclusions contained herein are those of the authors and should not be interpreted as necessarily representing the official policies or endorsements, either expressed or implied, of DARPA or the U.S. Government.}}}


\author{Shlomi Dolev, Peeyush Gupta, Yin Li, Sharad Mehrotra, Shantanu Sharma}

\markboth{IEEE Transactions on Dependable and Secure Computing, Accepted 01 Aug. 2019.}%
{Dolev \MakeLowercase{\textit{et al.}}: Privacy-Preserving Secret Shared Computations using MapReduce}
\IEEEtitleabstractindextext{
\begin{abstract}
Data outsourcing allows data owners to keep their data at \emph{untrusted} clouds that do not ensure the privacy of data and/or computations. One useful framework for fault-tolerant data processing in a distributed fashion is MapReduce, which was developed for \emph{trusted} private clouds. This paper presents algorithms for data outsourcing based on Shamir's secret-sharing scheme and for executing privacy-preserving SQL queries such as count, selection including range selection, projection, and join while using MapReduce as an underlying programming model. Our proposed algorithms prevent an adversary from knowing the database or the query while also preventing output-size and access-pattern attacks. Interestingly, our algorithms do not involve the database owner, which only creates and distributes secret-shares once, in answering any query, and hence, the database owner also cannot learn the query. Logically and experimentally, we evaluate the efficiency of the algorithms on the following parameters: (\textit{i}) the number of communication rounds (between a user and a server), (\textit{ii}) the total amount of bit flow (between a user and a server), and (\textit{iii}) the computational load at the user and the server.\B
\end{abstract}

\begin{IEEEkeywords}
Computation and data privacy, data and computation outsourcing, distributed computing, MapReduce, Shamir's secret-sharing.\BBB
\end{IEEEkeywords}}
\maketitle

\IEEEdisplaynontitleabstractindextext
\IEEEpeerreviewmaketitle
\ifCLASSOPTIONcompsoc
\IEEEraisesectionheading{\section{Introduction}\label{sec:introduction}}
\else

\BBB\BB
\section{Introduction}
\label{sec:introduction}
\B
\fi
The past few years have witnessed a huge amount of sensitive data generation due to several applications, \textit{e}.\textit{g}., location tracking sensors, web crawling, social networks, and body-area networks. Such real-time data assists users in several ways such as suggesting new restaurants, music, videos, alarms for health checkups based on the user's history; hence, it carries a potential threat to the user's privacy. MapReduce~\cite{DBLP:conf/osdi/DeanG04} was introduced by Google in 2004 and has emerged as a programming model for fast, parallel, and fault-tolerant processing of large-scale data at a \emph{trusted} private cloud. The huge amount of data creates hurdles to process it at a private cloud due to limited resources. Therefore, data and computation outsourcing, which move databases and computations from a trusted private cloud to an untrusted centralized (public) cloud, become a prominent solution. However, outsourcing jeopardizes the security and privacy of the data and computations. At present, due to constraints such as limited network bandwidth and network latency, uploading data to far sided clouds is not a trivial task, and edge or fog computing overcomes such a problem to some extent~\cite{DBLP:conf/sigcomm/BonomiMZA12}. However, edge or fog computing also suffers from the same security/privacy issues in data processing.

Recently, some works based on encryption~\cite{DBLP:conf/pet/BlassPMO12,DBLP:conf/fc/MayberryBC13,epic,DBLP:conf/oopsla/TetaliLMM13,crypsis} and trusted hardware~\cite{DBLP:conf/sp/SchusterCFGPMR15,DBLP:conf/uss/DinhSCOZ15} have been proposed to execute MapReduce computations in a secure and privacy-preserving manner at the cloud. Encryption-based secure MapReduce techniques~\cite{DBLP:conf/pet/BlassPMO12,DBLP:conf/fc/MayberryBC13,epic,DBLP:conf/oopsla/TetaliLMM13,crypsis} provide computationally secure frameworks but are limited to only count and selection queries on a \emph{non-skewed} dataset. These techniques also inherit disadvantages of encryption techniques. For example, MrCrypt~\cite{DBLP:conf/oopsla/TetaliLMM13}, which is based on homomorphic encryption, may prevent information leakage when mixed with ORAM~\cite{DBLP:conf/stoc/Goldreich87}, but it incurs a significant time delay~\cite{DBLP:conf/ccs/OhrimenkoCFGKS15,DBLP:conf/icml/Gilad-BachrachD16}. Prism~\cite{DBLP:conf/fc/MayberryBC13}, which is based on searchable encryption, also leaks the keywords and index traversal. Trusted-hardware-based MapReduce solutions, \textit{e}.\textit{g}., M2R~\cite{DBLP:conf/uss/DinhSCOZ15}, VC3~\cite{DBLP:conf/sp/SchusterCFGPMR15}, which are based on Intel Software Guard Extensions (SGX)~\cite{sgx}, also reveal access-patterns due to side-channel attacks (such as cache-line, branch shadowing, and page-fault attacks~\cite{DBLP:conf/ccs/WangCPZWBTG17,DBLP:conf/eurosec/GotzfriedESM17}) on SGX. Hence, these solutions are not secure. Also, as stated in~\cite{DBLP:conf/ccs/OhrimenkoCFGKS15}, these solutions suffer from information leakage during intermediate data traversal from mappers to reducers. Thus, the existing solutions fail to provide completely secure computations or common SQL (\textit{e}.\textit{g}., count, selection, join, and range) queries.

The following example shows that hiding access-patterns and preventing attacks based on output-size is important to achieve the data/query security, even if the query is hidden and the data is non-deterministically\footnote{Non-deterministic encryption achieves ciphertext indistinguishably, so that the adversary cannot know which two ciphertexts contain an identical cleartext, by just observing the encrypted data.} encrypted. Consider an employee relation denoted by $E(\mathit{SSN}, \mathit{Name})$ and a cancer hospital relation denoted by $H(\mathit{SSN}, \mathit{Name}, \mathit{Treatment})$. Also, consider that both relations are stored at a single (public) server\footnote{Hereafter, the word ``server'' is used to indicate public/edge servers, since our proposed algorithms are independent of the type of computing resources.} that may be an honest-but-curious adversary, which executes the query correctly but tries to learn more about encrypted data. Assume a query that finds the employees' details suffering from cancer, \textit{i}.\textit{e}., \texttt{SELECT E.SSN, E.Name FROM E INNER JOIN H ON E.SSN=H.SSN}. Before executing this query, an adversary is not aware of the number of employees suffering from cancer. The use of an access-pattern-hiding technique for joining these two relations can hide the fact which two tuples of the relation join, \textit{i}.\textit{e}., have a common $\mathit{SSN}$. However, access-pattern-hiding techniques cannot prevent output-size-based attacks. Thus, after executing such a query, the adversary can know the number of employees suffering from cancer based on output-size; thereby, gains more information than before the query execution.

A completely secure solution must achieve (\textit{i}) data privacy (\textit{i}.\textit{e}., the data or metadata should not be revealed to the adversary before/during/after a computation), and (\textit{ii}) query privacy (\textit{i}.\textit{e}., the user query should be hidden from the database (DB) owner and the servers).

\BBB
\subsection*{Contribution}
\B
In this paper, our goal is to support privacy-preserving SQL queries, \textit{e}.\textit{g}., select, project, and join, while guaranteeing information-theoretic data and query privacy. Here, the algorithms are developed for MapReduce; however, they can be generalized for the standard database systems and any other frameworks. Our contribution can be summarized, as follows:

\begin{itemize}[leftmargin=0.1in]
  \item \textit{Information-theoretically secure outsourcing.} We provide Shamir's secret-sharing (SSS)~\cite{DBLP:journals/cacm/Shamir79} based information-theoretically secure data and computation outsourcing technique that prevents an adversary from knowing the database or the query. The proposed SSS-based mechanism creates different shares of multiple occurrences of the same cleartext value, preventing the frequency-count attack (\textit{i}.\textit{e}., finding the number of tuples containing an identical value). The techniques are designed in such a way that after outsourcing the DB to servers, the DB owner does not need to be online during the query execution.

  \item \textit{Privacy-preserving query execution by third parties.} We provide privacy-preserving algorithms for queries: count (\S\ref{subsec:Count query execution}), selection (\S\ref{subsec:Search query execution}), project (\S\ref{subsubsec:Multiple occurrences of the joining value}), join (\S\ref{subsec:Join Query Execution}), and range (\S\ref{subsec:Range Query Execution}). The queries are executed in an oblivious manner so that the adversary cannot distinguish any two queries and data satisfying (based on access-patterns) to the queries. Moreover, our proposed approaches do not reveal the output or the query based on output-sizes. Our proposed approaches use an existing string-matching technique on secret-shared, known as accumulating-automata (AA)~\cite{DBLP:conf/ccs/DolevGL15} that originally supported only count queries.

  \item \textit{Minimum leakage and complex queries.} During a MapReduce computation, an adversary having some background knowledge can learn some information by observing output-sizes, as shown in the previous example. Our proposed algorithms minimize information leakage. We find a relation among the workload at the user, the number of communication rounds, and the data privacy. We will see in \S\ref{subsec:Search query execution} that in the case of skewed data, as the user is willing to perform fewer interpolation tasks on the secret-shared outputs of the servers, information leakage at the servers increases that may breach the data privacy.

\item \textit{Analysis of the algorithms.} We mathematically and experimentally analyze our algorithms on the following three parameters: (\textit{i}) the total amount of dataflow between the user and a server, (\textit{ii}) the number of interaction rounds between the user and a server, and (\textit{iii}) the computational workload at the user and the servers. \S\ref{sec:Experimental Evaluation} experimentally evaluates communication cost and computational cost at the server and user. \S\ref{sec:Complexity Analysis} provides theorems and proofs to show the above three parameters for each algorithm.

\end{itemize}

\BBB
\subsection*{Related Work}
\B
\noindent\textbf{An overview of Shamir's secret-sharing (SSS)}. SSS~\cite{DBLP:journals/cacm/Shamir79} is a cryptographic algorithm developed by Adi Shamir in 1979. The main idea of SSS is to divide a secret value, say $S$, into $c$ non-communicating servers such that no one can know the secret $S$ until they can collect $c^{\prime}< c$ shares, where $c^{\prime}$ is the threshold of SSS. Particularly, the secret owner randomly selects a polynomial of degree $c^{\prime}-1$ with $c^{\prime}-1$ random coefficients: $f(x)= a_0+a_1x+a_2x^2+\ldots+a_{c^{\prime}-1}x^{c^{\prime}-1} \in \mathbb{F}_P[x]$, where $a_0=S$, $P$ is a prime number, and $\mathbb{F}_P$ is a finite field of order $P$. The secret owner distributes the secret $S$ to $c$ servers by placing $x=1,2,\ldots, c$ in $f(x)$. The secret $S$ is constructed by performing Lagrange interpolation~\cite{corless2013graduate} on any $c^{\prime}$ shares. Note that in this paper we assume $c^{\prime}=c$, since we are not dealing with fault tolerance or a malicious adversary; however, it will not affect the algorithm's design.

\bgroup
\def\arraystretch{1.3}
\begin{table*}[!t]
\centering
\caption{Comparison of different algorithms with our algorithms.}
\label{table:Summary of privacy}
\BB
\centering
\scriptsize
\begin{tabular}{|p{3.8cm}|p{2.5cm}|p{2.6cm}|p{1.5cm}|p{1.5cm}|l|l|}

\hline Algorithms & Communication cost & \multicolumn{2}{ c| }{Computational cost} & $\#$ rounds & Matching & Based on \\

\hhline{~~|-|-|~|} & {~} & User  & Server &  & & \\ \hline

\multicolumn{7}{c}{\textbf{Count query}} \\\hline

EPiC~\cite{epic}  &$\BigO(1)$ & $\BigO(1)$ & $\BigO(n)$ & 1 & Online& E \\ \hline

\textbf{Our solution} \S\ref{subsec:Count query execution} &$\BigO(1)$ & $\BigO(1)$ & $nw$ & 1 & Online & SSS\\ \hline

\multicolumn{7}{c}{\textbf{Selection query when one value has one tuple}} \\\hline

Chor et al.~\cite{DBLP:journals/iacr/ChorGN98}  &$\BigO(nmw)$ &$\BigO(1)$ & $\BigO(nmw)$ & $\log_2 n$ & Online& SSS \\ \hline

PRISM~\cite{DBLP:conf/pet/BlassPMO12}&  $\BigO((nm)^{\frac{1}{2}}w)$ & $\BigO((nm)^{\frac{1}{2}}w)$  &  $\BigO(nmw)$ & $q$ & & E \\ \hline

\textbf{Our solution} \S\ref{subsec:Unary Occurrence of a Pattern}  & $\BigO(mw)$ & $\BigO(mw)$ & $\BigO(nmw)$ & 1 & Online & SSS \\ \hline

\multicolumn{7}{c}{\textbf{Selection query when one value has multiple tuples}} \\\hline

rPIR~\cite{DBLP:journals/iacr/LiMD14}  & $\BigO(nm)$ & $\BigO(1)$ & $\BigO(nmw)$ & 1 & No & SSS \\ \hline

PIRMAP~\cite{DBLP:conf/fc/MayberryBC13}  &  $\BigO(nmw)$ & $\BigO(mw)$ & $\BigO(nmw)$ & 1 & No & E \\ \hline

Goldberg~\cite{DBLP:conf/fc/LueksG15}  & $\BigO(n+m)$ & $\BigO(m)$ & $\BigO(nm)$ & 2 & Offline & SSS \\ \hline

Emekci et al.~\cite{DBLP:journals/isci/EmekciMAA14}  &$\BigO(\ell m)$ & $\BigO(\ell m)$ & $\BigO(n)$ & 2 & Offline & vSS \\ \hline

\textbf{Our one-round solution}: knowing addresses \S\ref{subsubsec:Multiple Occurrences of a Pattern} &
$\BigO(n)$ &
$\BigO(n)$ &
$\BigO(nw)$ &
1 &
Online & SSS \\ \hline

\textbf{Our tree-based solution}: knowing addresses \S\ref{subsubsec:Multiple Occurrences of a Pattern} & $\BigO\big((\log_{\ell}n+\log_2{\ell})\ell\big)$ & $\BigO\big((\log_{\ell}n+\log_2{\ell})\ell\big)$ & $\BigO\big((\log_{\ell}n+\log_2{\ell})\ell nw\big)$ & $\BigO(\lfloor \log_{\ell}n\rfloor +\lfloor \log_2{\ell} \rfloor)$ & Online & SSS \\ \hline

\textbf{Our solution}: fetching tuples \S\ref{subsubsec:Multiple Occurrences of a Pattern} &
$\BigO(\ell mw)$ &
$\BigO(\ell mw)$ &
$\BigO(\ell nmw)$ &
1 &
Online & SSS \\ \hline

\multicolumn{7}{c}{\textbf{Join queries}} \\ \hline

\textbf{Our solution}: Non-PK/FK Join \S\ref{subsec:Join Query Execution} &  $\BigO(2nwk+2k\ell^2mw)$ & $\BigO(2nw+2k\ell^2mw)$ & $\BigO(2\ell^2kmw)$ &  $\BigO(2k)$ & Online & SSS \\ \hline

\multicolumn{7}{c}{\textbf{Range queries}} \\ \hline

\textbf{Range-based count query}: \S\ref{subsec:Range Query Execution} &$\BigO(1)$ & $\BigO(1)$ & $\BigO(nw)$ & 1 & Online & SSS\\ \hline

\textbf{Range-based selection query}: \S\ref{subsec:Range Query Execution}&
$\BigO(n+\ell mw)$ &
$\BigO(n+\ell mw)$ &
$\BigO(\ell nmw)$ &
2 &
Online & SSS \\ \hline\hline

\multicolumn{7}{|p{15.5cm}|}{\textbf{Notations:} Online: perform query execution at the server. Offline: perform query execution at the user. E: encryption-decryption based. SSS: Shamir's Secret-sharing. vSS: a variant of SSS. $n$: $\#$ tuples, $m$: $\#$ attributes, $\ell$: $\#$ occurrences of a predicate ($\ell \leq n$), $w$: bit-length of a word.} \\\hline
\end{tabular}
\BBB
\end{table*}
\egroup

\smallskip
\noindent\textbf{An overview of security techniques.} PRISM~\cite{DBLP:conf/pet/BlassPMO12}, PIRMAP~\cite{DBLP:conf/fc/MayberryBC13}, EPiC~\cite{epic}, MrCrypt~\cite{DBLP:conf/oopsla/TetaliLMM13}, and Crypsis~\cite{crypsis} provide privacy-preserving MapReduce computations in the cloud on encrypted data. However, all these protocols/systems support only count and selection queries on encrypted data while either incurring significant overhead in terms of time or do not provide data/query security (as a tradeoff between preserving data privacy and utility). Details of security and privacy concerns in MapReduce may be found in~\cite{security-review}. There are some other encryption-based systems~\cite{DBLP:journals/cacm/PopaRZB12,DBLP:conf/cidr/ArasuBEKKRV13}
and trusted hardware-based systems~\cite{DBLP:journals/iacr/WangDDB06,DBLP:conf/fpl/ArasuEKKRV13,DBLP:conf/sp/SchusterCFGPMR15,DBLP:conf/uss/DinhSCOZ15} to execute SQL queries without using MapReduce as a programming model. These systems also result in information leakage due to deterministic or order-preserving encryption~\cite{DBLP:conf/ccs/NaveedKW15,DBLP:conf/ccs/KellarisKNO16} or limited operations. Searchable techniques are designed for keyword searches on encrypted or cleartext data. For example, searchable encryption~\cite{DBLP:conf/sp/SongWP00,DBLP:journals/tifs/DuWHW18} allows searching on encrypted data, while function secret-sharing (FSS)~\cite{DBLP:conf/eurocrypt/BoyleGI15} allows searching on cleartext data. Searchable encryption mixed with oblivious-RAM (ORAM)~\cite{DBLP:conf/stoc/Goldreich87} hides access-patterns, while PIR and FSS hide access-patterns by default. However, these searchable techniques are not secure against output-size attacks and/or not information-theoretically secure. Other work (\textit{e}.\textit{g}.,~\cite{DBLP:journals/tkde/WangDCCZCH18}) proposed to operate arithmetic operations over encrypted data, based on specialized cryptographic techniques. There are work on verifying SQL queries on outsourced cleartext data, such as IntegriDB~\cite{DBLP:conf/ccs/ZhangKP15} and vSQL~\cite{DBLP:conf/sp/ZhangGKPP17} (or~\cite{DBLP:conf/sp/ZhangGKPP18}). We do not describe these systems in detail, since the paper focuses on MapReduce-based computations.

There is also work on secret-sharing-based query execution, \textit{e}.\textit{g}.,~\cite{DBLP:conf/icde/EmekciAAG06,DBLP:journals/iacr/LiMD14,DBLP:journals/isci/EmekciMAA14,DBLP:journals/isci/XiangLCGY16}. \cite{DBLP:conf/icde/EmekciAAG06} supports privacy-preserving join using secret-sharing and requires that two different DB owners share some information for constructing \textit{an identical share for identical values} in their relations. However, sharing information among DB owners is not trivial when they are governed by different organizations and policies, and moreover, by following this approach, a malicious DB owner may be able to obtain another relation.~\cite{DBLP:journals/isci/EmekciMAA14} provides a technique for data outsourcing using a variation of SSS. However,~\cite{DBLP:journals/isci/EmekciMAA14} suffers from two major limitations: (\textit{i}) in order to answer a query, the DB owner has to work on all the shares, hence, the DB owner (not the servers) incurs the overhead of secure computing; and (\textit{ii}) a third-party without involving the DB owner cannot directly issue any query on secret-shares. Also,~\cite{DBLP:journals/isci/EmekciMAA14} provides a way to construct polynomials that can maintain the orders of the secrets. However, these polynomials are based on an integer ring (no modular reduction) rather than a finite field; thus, it has a potential security risk. There is also work~\cite{DBLP:conf/fc/LueksG15,DBLP:journals/iacr/LiMD14,DBLP:journals/iacr/ChorGN98} that provide searching operations on secret-shares. In~\cite{DBLP:conf/fc/LueksG15}, a data owner builds a Merkle hash tree~\cite{DBLP:conf/crypto/Merkle87} according to a query. In~\cite{DBLP:journals/iacr/LiMD14}, a user knows the addresses of the desired tuples, so they can fetch all those tuples obliviously from the servers without searching at the server. Similar ideas can also be found in~\cite{DBLP:journals/iacr/ChorGN98}.

Our proposed algorithms overcome the disadvantages of the existing secret-sharing-based data outsourcing techniques~\cite{DBLP:conf/icde/EmekciAAG06,DBLP:journals/isci/EmekciMAA14,DBLP:conf/fc/LueksG15,DBLP:journals/iacr/LiMD14,DBLP:journals/iacr/ChorGN98}. In our approach, (\textit{i}) unlike~\cite{DBLP:conf/icde/EmekciAAG06}, the DB owner do not need and share information, (\textit{ii}) unlike~\cite{DBLP:journals/isci/EmekciMAA14}, have an identical share for multiple occurrences of a value, and (\textit{iii}) unlike~\cite{DBLP:journals/isci/EmekciMAA14,DBLP:journals/isci/XiangLCGY16}, a third party can directly execute queries in the servers without revealing queries to the servers/DB owner, and without involving the DB owner in answering a query.

Table~\ref{table:Summary of privacy} summarizes all the results of this paper and a comparison with the existing algorithms, based on the five criteria: (\textit{i}) communication cost, (\textit{ii}) computational cost at the user and the server, (\textit{iii}) number of rounds, (\textit{iv}) matching of a keyword in an online or offline manner, and (\textit{v}) dependence of secret-sharing. Note that in offline string-matching operations, the user needs to download the whole database and then searches for the query predicate. Thus, the number of rounds is decreased, while the communication cost and computational cost at the user increase. In contrast, online operations refer to a strategy where the server performs the desired operation without sending the whole database to the user. From Table~\ref{table:Summary of privacy}, it is also clear that our count query algorithm requires an identical amount of communication as encryption-based Epic~\cite{epic}. However, we perform some more work at the server due to our secret-sharing-based data encoding technique. In case of a selection query, where one value has only one tuple, we perform better than existing SSS-based selection query algorithms in terms of communication and computational cost. However, when a value has multiple tuples, our algorithms work as good as existing algorithms; nonetheless, we provide more secure ways to prevent information leakage. To the best of our knowledge, there is no prior work on secure join queries using secret-sharing.

\BB
\section{The Model}
\label{section:System and Adversary Settings}
\B
We assume the following three entities in our model; see Figure~\ref{fig:system_settings}:
\begin{figure}[h]
\BBB
\centering
\includegraphics[scale=0.37]{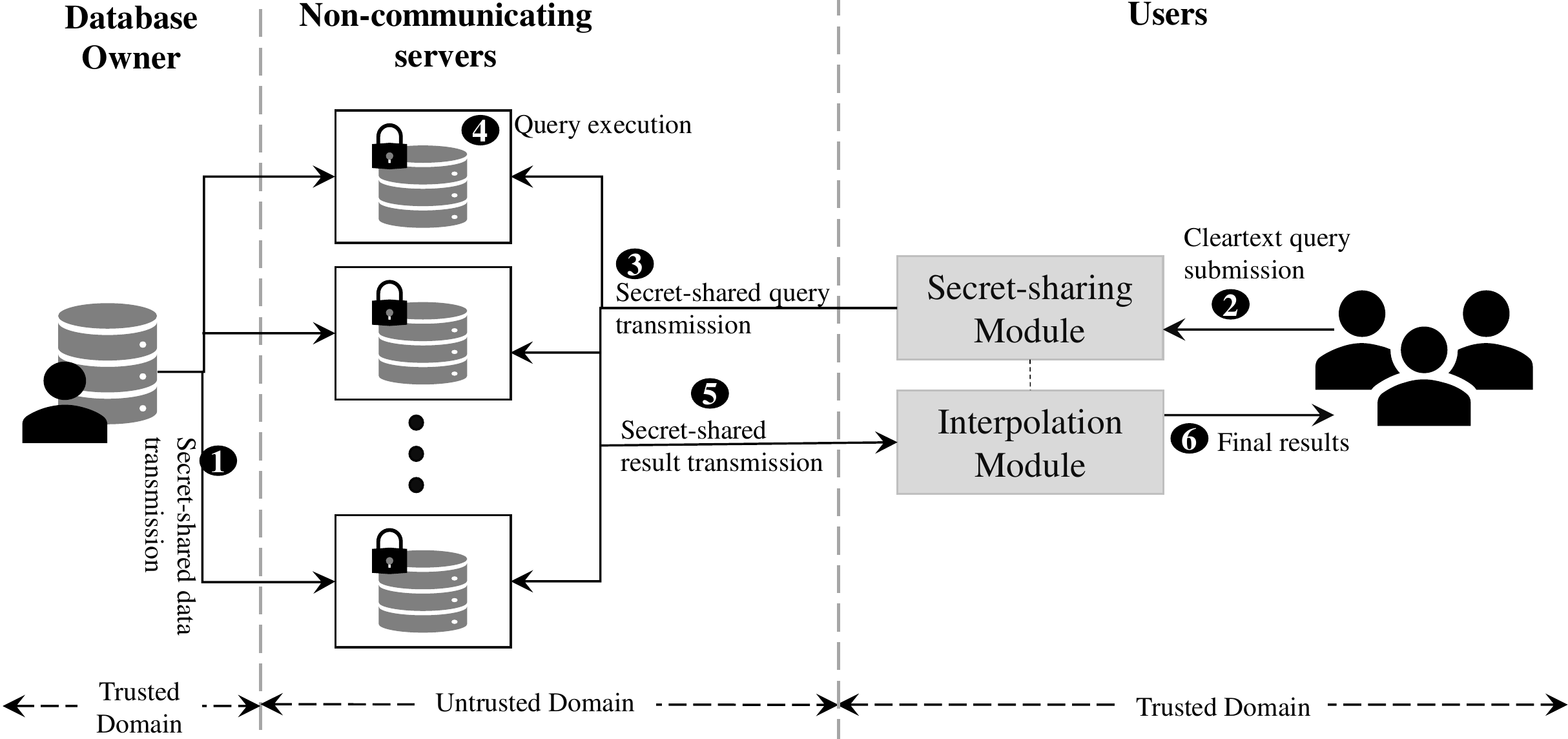}
\BBB
\caption{The system architecture.}
\BB
\label{fig:system_settings}
\B
\end{figure}
\begin{enumerate}[leftmargin=0.1in]
  \item The \emph{trusted} database (DB) owner, who owns a dataset, creates secret-shares of the dataset, and outsources the $i^{\mathit{th}}$ share to the $i^{\mathit{th}}$ server (\encircle{1} in Figure~\ref{fig:system_settings}). An algorithm for creating secret-shares is described in \S\ref{subsec:Data Model Creation and Distribution of Secret-Shares of a Relation}.

  \item $c>2$ \emph{untrusted} and \emph{non-communicating} servers, which store the secret-shared database and execute users' queries (\encircle{4}). The non-communicating servers do not exchange data with each other before/during/after the computation, and these servers only exchange data with the user (or the DB owner). We use $c$ servers to provide privacy-preserving computations using SSS. Note that a single non-trustworthy server cannot provide privacy-preserving computations using secret-shares. Since we are using MapReduce as our programming model, each server deploys a \textit{master process} that executes the computation by allocating the \textit{map tasks} and the \textit{reduce tasks}. After the computation, the servers provide secret-shared outputs to the user (\encircle{5}).

  \item An (authenticated and authorized) user, who wishes to execute queries on the secret-shared data at the servers. The users submit the query to a secret-sharing-module at their ends and that converts the query predicate to $c$ secret-shared query predicates (\encircle{2}). These secret-shared query predicates are submitted to (the master process at) $c$ servers (\encircle{3}).
      Also, the user has an interpolation module that receives the secret-shared outputs from the servers (\encircle{5})and executes Lagrange interpolation~\cite{corless2013graduate} over them to produce the final output (\encircle{6}).

      It will be clear soon that the user performs the minimum work in the case of count queries (\S\ref{subsec:Count query execution}), single tuple fetch queries (\S\ref{subsubsec:A unique occurrence of the joining value}), and primary-foreign keys-based join queries (\S\ref{subsubsec:A unique occurrence of the joining value}). However, the user performs some more work in case of fetching multiple tuples (\S\ref{subsubsec:Multiple Occurrences of a Pattern}) and non-primary-foreign keys-based join queries (\S\ref{subsubsec:Multiple occurrences of the joining value}). Nevertheless, the work performed by the user is very less as compared to the servers, which will be experimentally validated in \S\ref{sec:Experimental Evaluation}.
\end{enumerate}
Table~\ref{table:Notations used in the paper} shows the notations used in this paper.

\bgroup
\def\arraystretch{1.1}
\begin{table}[!t]
\centering
\scriptsize
\begin{tabular}{|l|p{6.7cm}|}
\hline Notations & Meaning \\ \hline
$c$  & \# non-communicating servers holding secret-shares \\\hline
$c^{\prime}$ & Threshold of Shamir's secret-sharing \\\hline
$R$ & A relation (or table) in cleartext \\\hline
$n$ & Number of tuples (or rows) in $R$ \\\hline
$m$ & Number of attributes (or columns) in $R$ \\\hline
$A_i$ & The $i^{\mathit{th}}$ attribute\\\hline
$p$ & A counting/searching predicate \\\hline
$\ell$ & \# occurrences of $p$ \\\hline
$[R]^i$ & The $i^{\mathit{th}}$ secret-shared relation $[R]$ \\\hline
$\mathit{RID}$ & Row/Tuple-id attribute \\\hline
$[v_i]_j^k$ & The $j^{\mathit{th}}$ alphabet of the $i^{\mathit{th}}$ secret-shared value $[v]$ at the server $k$\\\hline
$w$ & Maximum bit-length \\\hline
$\mathit{Len}(x)$ & Length of a sting $x$ \\\hline
\end{tabular}
\BB
\caption{Notations used in the paper.}
\label{table:Notations used in the paper}
\BBB
\end{table}
\egroup

\BB
\subsection{Data Model: Creation and Distribution of Secret-Shares of a Relation}
\label{subsec:Data Model Creation and Distribution of Secret-Shares of a Relation}
We first provide a simplified and insecure algorithm, and then, a secure algorithm~\cite{DBLP:conf/ccs/DolevGL15} for creating secret-shares based on SSS. We explain the creation of secret-shares with the help of an example of a relation, $\mathit{Employee}$; see Table~\ref{fig:database1}.

\parskip 0pt
\setlength{\parindent}{15pt}

\smallskip\noindent\textbf{Non-secret-shared encoded data.} Assume that a database contains only English words. Since the English alphabet consists of 26 letters, each letter can be represented by a unary vector with 26 bits. Hence, the letter `A' is represented as $(1_1,0_2,0_3,\ldots,0_{26})$, where the subscript represents the position of the letter; since `A' is the first letter, the first value in the vector is one and others are zero. Similarly, `B' is $(0_1,1_2,0_3,\ldots,0_{26})$, `J' is $(0_1,\ldots,0_9,1_{10},0_{11},\ldots,0_{26})$, and so on. Now, if the database owner sends these vectors to a server, then the server can easily deduce words.

The reason for using unary representation here is that it is very easy to verify two identical letters. The expression $\mathit{output}=\sum^{r}_{i=0}u_i\times v_i$, compares two letters, where $(u_0,u_1,\ldots, u_r)$ and $(v_0,v_1,\ldots, v_r)$ are two unary representations. It is clear that whenever any two letters are identical, $\mathit{output}$ is equal to one; otherwise, $\mathit{output}$ is equal to zero.

\smallskip\noindent\textbf{Secret-shared encoded data.} When outsourcing a vector to the servers, we use SSS and make secret-shares of every bit by selecting different polynomials of an identical degree; see Algorithm~\ref{alg:Algorithm for creating secret-shares} in Appendix~\ref{app_sec:Pseudocodes}.\footnote{Appendix~\ref{app_sec:Pseudocodes} provides algorithms' pseudocodes and their descriptions.} For example, the DB owner creates secret-shares of each bit of the vector of `A' ($(1_1,0_2,0_3,\ldots,0_{26})$) by using 26 polynomials of an identical degree, since the length of the vector is 26. Following that, the DB owner can create secret-shares of all the other letters and distribute them to different servers.

Since we use SSS, a single server cannot infer a secret. Moreover, it is important to emphasize that we use \textit{different} polynomials for creating secret-shares of each occurrence of each letter; thereby multiple occurrences of a word in a database have different secret-shares. Therefore, a server is also unable to know the number of occurrences of a word in the whole dataset. Following that, the two occurrences of the word \texttt{John} in our example (see Table~\ref{fig:database1}) have two different secret-shares.

\begin{table}[!t]
\scriptsize
  \centering
  \begin{tabular}{|l|l|l|l|l|l|l|l|}
    \hline
    EId & FirstName & LastName & DateofBirth & Salary & Dept & $\mathit{RID}$  \\ \hline\hline
    E101 & Adam & Smith      & 12/07/1975 & 1000 & Sale & 1 \\ \hline
    E102 & John & Taylor       & 10/30/1985 & 2000 & Design&2   \\ \hline
    E103 & Eve  & Smith      & 05/07/1985 & 500  & Sale & 3 \\ \hline
    E104 & John & Williams   & 04/04/1990 & 5000 & Sale & 4  \\ \hline
  \end{tabular}
  \BB
  \caption{A relation: \textit{Employee}.}
  \label{fig:database1}
\BBB
\end{table}

\smallskip\noindent\textit{Secret-shares of numeral values}. We follow a similar approach for creating secret-shares of numeral values as used for alphabets. In particular, the DB owner creates a unary vector of length 10 and places all the values zero except only one according to the position of a number. For example, `1' becomes $(1_1,0_2,\ldots,0_{10})$, `0' becomes $(0_1,0_2,\ldots,1_{10})$, and so on. After that, the DB owner uses SSS to create secret-shares of every bit of a vector by selecting different polynomials of an identical degree and sends them to multiple servers. Note that by following the same procedure we can also create secret-shares of special letters or symbols.

\smallskip\noindent\textbf{Data outsourcing.} Let $R$ be a relation of $n$ tuples and $m$ attributes, denoted by $A_1,A_2, \ldots, A_m$. Let $t_{\mathit{ij}}$ be a value at the $i^{\mathit{th}}$ tuple and $j^{\mathit{th}}$ attribute. The DB owner adds one more attribute, entitled $\mathit{RID}$, where the $i^{\mathit{th}}$ ($1\leq i\leq n$) value of the attribute $\mathit{RID}$ contains the number $i$. The DB owner creates $c$ shares of each value $t_{\mathit{ij}}$ of $A_i$ ($1\leq i\leq m$) and $\mathit{RID}$ attributes using the above-mentioned approach. This step results in $c$ relations: $[R]^i(A_1,A_2,\ldots,A_m,\mathit{RID})$, where $[R]^i$ ($1\leq i\leq c$) denotes $i^{\mathit{th}}$ secret-shared relation. The secret-shared relation $[R]^i$ is outsourced to the $i^{\mathit{th}}$ server. Table~\ref{fig:database1} shows an Employee relation with one new additional attribute, namely $\mathit{RID}$.

\smallskip
\noindent\textbf{Reducing the size of secret-shared data.} The above-mentioned approach for creating secret-shares of non-numerical values results in a dataset of significantly large size and a large number of servers, which depends on the maximum length of a value in cleartext. For example, if there are two names, for example, Jacqueline and Amy, then we may need 260 zeros or ones to represent each of the names using unary representation. However, we can reduce the size of unary representation by appropriately mapping the words to a smaller domain. For example, we may map the word Jacqueline to 1 and Amy to 2, and thereby need only at most 10 zeros or ones to represent the newly obtained values, which reduce the size of secret-shared data significantly.

\noindent Hashing can map large strings/text to smaller length strings; however, hashing can result in a collision (two different values are mapped to an identical hashed value). In case of collisions, results received from servers can be erroneous. The user can overcome errors in results for selection and join queries by filtering un-desired tuples; however, in the case of count queries, there is no way to know collisions and fix errors in query results that occurred due to hash collisions. Nevertheless, the probability of collision is extremely low. For a perfect hash function that generates $k$-bit hashes, the probability of collision for $n$ distinct elements is $P_{\mathit{collision}} = n^2/2^{k+1}$. For example, for a 256-bit hash function, the probability of collision in mapping all possible 32-bit integers is $ 2^{64}/2^{256+1} =  1/2^{193}$, which is extremely low. Note that cryptographic hash functions, for example, SHA2 (\S8.6 of~\cite{dan}) can provide a higher level of collision resistance, which is quite comparable to a perfect hash function.

\noindent\textbf{Note.} The DB owner may use binary representation for representing secret-shares, due to its compact representation as compared to unary representation. However, during string-matching operations over binary secret-shared data, polynomial degree increases significantly and that requires more number of shares during interpolation as compared to the number of shares required in the case of unary secret-shared data. For example, consider a decimal number, say $n$ ($= 300$), having $l_d$ ($= 3$) digits in decimal, and takes $l_b$ ($= 9$) digits in binary ($100101100$). Note that unary and binary representation take $30$ and $9$ numbers to represent $300$, respectively. However, when the user wishes to know the result of the string-matching operation in one communication round, we need at least $2l_d+1$ and $2l_b+1$ servers for string-matching, when using unary and binary representation, respectively.

\BB
\subsection{Query Model}
\label{subsec:query model}
\B
Let $q$ be a query on a relation, say $R(A_1,A_2,\ldots,A_m)$. Let $[R]^1(A_1,A_2,\ldots,A_m,\mathit{RID})$, $[R]^2(A_1,A_2,\ldots,A_m,\mathit{RID})$, $\ldots$, $[R]^c(A_1,A_2,\ldots,A_m,\mathit{RID})$ be secret-shared relations of the relation $R$ at $c$ servers, one relation at each server. The query $q$ is transformed to $c$ queries, say $q_1, q_2,\ldots, q_c$, that are transmitted to servers such that the query $q_i$ is executed at the $i^{\mathit{th}}$ server. The $c$ servers execute the query and provide the secret-shared outputs. The interpolation module at the user computes the final output by executing a query, say $q_{\mathit{interpolate}}$, that interpolates the received partial results.

\noindent\textit{Correctness}. Consider an ideal execution of the query $q$, where the DB owner executes the query and provides the answer to the user. The query execution of the secret-shared data is correct when the real execution of the query and the ideal execution of the query provide an identical answer, \textit{i}.\textit{e}.,\footnote{We do not provide any result verification approach in this paper, since we are assuming an honest-but-curious adversary.}
\B
$$q(R)\equiv q_{\mathit{interpolate}} (q_1([R]^1), q_2([R]^2), \ldots, q_{c}([R]^{c}))$$

\BB
\subsection{Adversarial Settings}\label{subsec:Adversarial Settings}
\B
We consider an honest-but-curious adversary, which is widely considered in the standard security settings for the public cloud~\cite{DBLP:conf/infocom/YuWRL10,DBLP:journals/tpds/CaoWLRL14,DBLP:conf/pet/OlumofinG10}. The honest-but-curious adversary stores the database, performs assigned computations correctly, and returns answers, but tries to breach the privacy of data or (MapReduce) computations, by analyzing data, computations, dataflow, or output-sizes. However, such an adversary does not modify or delete information/computation.

In our setting, the adversary cannot launch an attack against the DB owner, who is trustworthy. Hence, the adversary cannot access the secret-sharing algorithm at the DB owner. We follow the restriction of the standard SSS that the adversary cannot collude with all (or possibly the majority of) the servers, or the communication channels between the DB-owner/user and the majority of the servers. This can be achieved by either encrypting the traffic between user and servers, or by using anonymous routing~\cite{DBLP:journals/cacm/GoldschlagRS99}. Note that if the adversary can collude with the communication channels between the majority of servers and user, then the secret-sharing technique is not applicable. We assume that the adversary knows the following publicly available information: (\textit{i}) auxiliary information about the relations, \textit{e}.\textit{g}., number of attributes, number of tuples, organizations to which the relations belong; (\textit{ii}) the size of outputs transferred from mappers to reducers and from the server to the user; and (\textit{iii}) any background knowledge. Having above-mentioned information, the adversarial goal is to learn the database/query and the tuple satisfying the query.

\noindent\textbf{Aside.} The overhead in considering Byzantine faults is high, so systems will use other means to detect and replace Byzantine participants (which is orthogonal to our scope). It should also be noted that standard techniques based on Berlekamp-Welch algorithm~\cite{welch1986error}, where additional secret-shares are used to encode the data can be directly applied here, enabling us to cope with a malicious adversary. For instance, the method given in~\cite{DBLP:conf/algocloud/DolevL15} may be used. However, any verification algorithm is error prone, as it is based on some redundancy used for error-detection/error-correction. Verifiability of the received shares may be required, since noise in the transmission channel can corrupt some bits. In this case, a suitable cryptographic hash function can be applied to the secret-shared answers and the corresponding hash values can be published on a bulletin board by each server. On receiving the secret-shared answers from the servers, the user can compute the hash value on the received shares and compare it against the corresponding hash value published on the bulletin board. If they match, then the received shares are correct in the sense that transmitted shares by the servers are  accurately received at the user. Otherwise, the user discards the received shares and may request the server to resend the secret-shared answers.

\BB
\subsection{Security Goals and Analysis}
\label{subsec:Security Goals}
\B
The security goals under the above-mentioned system and adversarial settings are twofold:

\smallskip
\noindent(\textit{i}) \textbf{Data privacy} requires that the stored input data, intermediate data during computations, and output data are not revealed to the servers, and the secret value can only be reconstructed by the DB owner or an authorized user. In addition, the two or more occurrences of a value in the relation must be different at the servers to prevent frequency-count analysis, while data is at the rest. Data privacy also must prevent any statistical inference on the data after the query execution, so that the adversary cannot find how many numbers of tuples satisfy a query.

\bgroup
\def\arraystretch{.9}
\begin{table*}[!t]
  \centering
  \scriptsize
\begin{tabular}{|l|l|l|l|l|l|l|l|}\hline
  \multicolumn{6}{|c|}{Computation on}\\\hline

  Operations & Server 1             & Server 2           & Server 3    & Server 4 & Server 5      \\\hline
  Multiplication& $2\times  3=  6$    & $3\times 5 = 15$ &  $4\times 7 =28$
  &  $5\times 9 =45$ &  $6\times 11 =66$ \\\hline

  Multiplication& $5\times 4= 20$   & $10\times 8 = 80$ &$15\times 12 =180$
  &  $20\times 16 =320$ &  $25\times 20 =500$ \\\hline

  Add ($N_1$ value) & 26 & 95 & 208 & 365 & 566\\\hline\hline

  Multiplication& $2\times 3 =  6$    & $4\times  6 = 24$ & $6\times9=54$
  &  $8\times 12 =96$ &  $10\times 15 =150$ \\\hline

  Multiplication & $9\times 8=72$ & $17\times  15 =255$ &  $25\times22=550$
  &  $33\times 29 =957$ &  $41\times 36 =1626$\\\hline

  Add ($N_2$ value) & 78 & 279 & 604 & 1053 & 1366 \\\hline\hline

  Multiplication ($N_3$ value)& 2028 & 26505 & 125632 & 384345 & 920316\\\hline

\end{tabular}
\BB
\caption{Multiplication of shares and addition of final shares by the servers.}
\label{tab:cloud multiply}
\BBB
\end{table*}
\egroup

\smallskip
\noindent(\textit{ii}) \textbf{Query privacy} requires that the user query must be hidden from the servers, as well as, the DB owner. Additionally, the servers cannot distinguish two or more queries of an \textit{identical type} based on the output-sizes. Note that all the count queries are of the same type based on output-sizes, since they return an identical number of bits as compared to selection or join queries. Following that the servers cannot distinguish two or more selection or join queries based on the output-size. In addition, query privacy requires that all the filtering operations (\textit{e}.\textit{g}., selection) are pushed down before the join operation while not revealing selection predicates and qualified tuples. Thus, in short, to achieve the query privacy, we need to prevent access-patterns and output-size attacks. Note that since we are using MapReduce, we need to prevent access-patterns and output-size attacks at the map and reduce phases, as well.

\BB
\section{An Example of String-Matching using Accumulating Automata}
\label{sec:Preliminarily Example}
\B
Before going into details of algorithms, we provide an example to understand the working of string-matching algorithm over secret-shares, since our algorithms for count, selection (or tuple fetch), and join queries are based on string-matching over secret-shares, which is executed using accumulating automata (AA)~\cite{DBLP:conf/ccs/DolevGL15}.

\noindent\textbf{Notations.} Let $s$ be a string in cleartext, and $s_j$ be the $j^{\mathit{th}}$ alphabet of the string. Let $[s]^i$ be the $i^{\mathit{th}}$ share of the string. Let $\mathit{Len(s)}$ be the length of the cleartext string $s$, and $\mathit{Len([s])}$ be the length of the secret-shared string $[s]$. Let $p$ be a cleartext query predicate, and $[p]^i$ be the $i^{\mathit{th}}$ secret-shared query predicate.

\medskip The string-matching algorithm consists of the following steps:

\smallskip
\noindent\textit{DB owner: Secret-shared database outsourcing.} Assume that there are only two alphabets: J and O; thus, we can represent J as $\langle 1,0\rangle$ and O as $\langle 0,1\rangle$. If the DB owner outsources a word `JO' by creating a vector $\langle 1,0,0,1\rangle$, then the adversary will know the string. Hence, the DB owner creates secret-shares of each value of the vector using polynomials of an identical degree, as shown in Table~\ref{tab:Secret-shares by DB owner}. Note that by using this type of unary representation, in this example, the length of secret-shared string $\mathit{Len([s])}$ will twice the length of the original string. Thus, in this example, the $j^{th}$ alphabet of the cleartext string $s$ is mapped to two secret-shared numbers. Further, note that we create five shares. It will be clear soon that to execute string-matching algorithm over a cleartext string, $s$, of length, $\mathit{Len(s)}$, we need at least $2\mathit{Len(s)}+1$ shares.

\bgroup
\def\arraystretch{.86}
\begin{table}[h]
\B
\centering
  \scriptsize
  \begin{tabular}{|p{.8cm}|l|p{.7cm}|p{.8cm}|p{.8cm}|p{.8cm}|p{.8cm}|}\hline

  Vector values & Polynomials & First shares & Second shares & Third shares & Fourth shares & Fifth shares\\\hline
  1 & $1+x$  & 2 & 3  & 4  & 5 & 6  \\\hline
  0 & $0+5x$ & 5 & 10 & 15 & 20& 25 \\\hline
  0 & $0+2x$ & 2 &  4 & 6  & 8 & 10 \\\hline
  1 & $1+8x$ & 9 & 17 & 25 & 33& 41 \\\hline
\end{tabular}
\BB
\caption{Secret-shares of a vector $\langle 1,0,0,1\rangle$, created by the DB owner.}
\label{tab:Secret-shares by DB owner}
\BBB
\end{table}
\egroup

\smallskip
\noindent\textit{User: Secret-shared query generation.} A user submits its query to the secret-sharing module that creates unary vectors for each letter of the query predicate, and then, creates secret-shares of each value of the unary vectors using any polynomial of the same degree as used by the DB owner,\footnote{We do not need to create an identical share of a searching pattern that is identical to the outsourced data. Note that the secure cryptographic techniques (\textit{e}.\textit{g}., searchable encryption~\cite{DBLP:conf/sp/SongWP00}, oblivious keyword search~\cite{DBLP:journals/jc/OgataK04}, or function-secret-sharing~\cite{DBLP:conf/eurocrypt/BoyleGI15}) do not create a searching pattern that is identical to the outsourced data.} before sending them to the servers. For example, if the user wants to search for JO, then the secret-sharing module creates a unary vector as $\langle 1,0,0,1\rangle$, creates secret-shares of this vector (see Table~\ref{tab:Secret-shares by user}), and sends to servers.

\bgroup
\def\arraystretch{.86}
\begin{table}[h]
\B
  \centering
  \scriptsize
  \begin{tabular}{|p{.8cm}|l|p{.7cm}|p{.8cm}|p{.8cm}|p{.8cm}|p{.8cm}|}\hline
  Vector values & Polynomials & First shares & Second shares & Third shares & Fourth shares & Fifth shares\\\hline

  1 & $1+2x$ & 3 & 5 & 7 & 9 & 11 \\\hline
  0 & $0+4x$ & 4 & 8 & 12& 16& 20 \\\hline
  0 & $0+3x$ & 3 & 6 & 9 & 12& 15 \\\hline
  1 & $1+7x$ & 8 & 15& 22& 29& 36 \\\hline
 \end{tabular}
 \BB
\caption{Secret-shares of a vector $\langle 1,0,0,1\rangle$, created by the user.}
\label{tab:Secret-shares by user}
\BB
\end{table}
\egroup

\smallskip
\noindent\textit{Servers: String-matching operation.} Each server holds: (\textit{i}) a secret-shared string $[s]$ (for example, `JO' of secret-shared form), and (\textit{ii}) the secret-shared searching predicate $[p]$ (for example, `JO' of secret-shared form). For executing the string-matching operation, the server creates an automaton with $\mathit{Len(s)}+1$ nodes, of which the first $\mathit{Len(s)}$ nodes perform bit-wise multiplication between string and predicate vectors, and the last node keeps the final result by multiplying outputs of all the first $\mathit{Len(s)}+1$ nodes. Table~\ref{tab:steps1} shows all the steps taken by the automaton. Further, we explain the steps given in Table~\ref{tab:steps1} to show how servers search a pattern `JO' over the string `JO'; see Table~\ref{tab:cloud multiply}.

\begin{table}[!h]
\begin{center}
\scriptsize
\centering
\begin{tabular}{|p{7.5cm}|}
\hline
\textsc{Step} $1$: $N_1=[s]_1 \odot [p]_1$\\
\textsc{Step} $2$: $N_2=[s]_2 \odot [p]_2$\\

\indent \vdots

\textsc{Step} ${\mathit{Len(s)}}$: $N_{\mathit{Len(s)}}=[s]_{\mathit{Len(s)}} \odot [p]_{\mathit{Len(s)}}$\\

\textsc{Step} $\mathit{Len(s)}+1$: $N_{\mathit{Len(s)}+1}=N_1\times N_2\times \ldots \times N_{\mathit{Len(s)}}$\\\hline

\textbf{Notations}: $\odot$: bit-wise multiplication and addition of all values in a vector.\\\hline

\end{tabular}
\BB
\caption{The accumulating automation steps executed by the server for performing string-matching operation.}
\label{tab:steps1}
\BBB
\end{center}
\end{table}

\smallskip
\noindent\textit{User: Result reconstruction.} Once the user receives the results from all the servers, the interpolation module performs Lagrange interpolation to obtain the final answer, as follows:
\centerline{\scriptsize
$
\frac{(x-2)(x-3)(x-4)(x-5)}{(1-2)(1-3)(1-4)(1-5)}\times 2028 +
\frac{(x-1)(x-3)(x-4)(x-5)}{(2-1)(2-3)(2-4)(2-5)}\times 26505 +$}
\centerline{\scriptsize
$\frac{(x-1)(x-2)(x-4)(x-5)}{(3-1)(3-2)(3-4)(3-5)}\times 125632 +
\frac{(x-1)(x-2)(x-3)(x-5)}{(4-1)(4-2)(4-3)(4-5)}\times 384345 +$}
\centerline{\scriptsize
$\frac{(x-1)(x-2)(x-3)(x-4)}{(5-1)(5-2)(5-3)(5-4)}\times 920316 = 1
$}
\noindent Now, note that the final answer is 1, which shows that the secret-shared string at the server matches the user query.

\BB
\section{Privacy-Preserving Query Processing on Secret-Shares using MapReduce}
\label{sec:Query Processing on Secret-Shares using MapReduce in the Clouds}
\B
This section presents privacy-preserving algorithms for performing three fundamental operations, \textit{i}.\textit{e}., count, selection, and join queries, on secret-shared relations.

\BB
\subsection{Count Query}
\label{subsec:Count query execution}
\B
This section presents an oblivious counting algorithm that finds the number of occurrences of a query keyword/pattern, $p$, in an attribute, $A_i$, of a relation, $R$. Note that count query algorithm for MapReduce is an extension of string-matching algorithm AA. Algorithm~\ref{alg:count algorithm} in Appendix~\ref{app_sec:Pseudocodes} presents pseudocode of counting algorithm. The algorithm proceeds as follows:

\smallskip
\noindent\textit{User: secret-shared query generation.} The user sends the count query having a predicate $p$ to the secret-sharing module. The secret-sharing module transforms the query predicate $p$ into $c$ secret-shared predicates, denoted by $[p]^j$ (where $1\leq j\leq c$), as suggested in \S\ref{subsec:Data Model Creation and Distribution of Secret-Shares of a Relation}. The secret-shared count query predicates are sent to $c$ servers.

\smallskip
\noindent\textit{Servers: count query execution}. Now, a server holds the following: (\textit{i}) a relation of secret-shared form, (\textit{ii}) the secret-shared counting predicate $[p]$, and (\textit{iii}) the code of mappers. A mapper at each server, say $k\in [1,c]$, performs the string-matching algorithm (as explained in \S\ref{sec:Preliminarily Example}) that compares each value $[v_j]^k$ ($1\leq j\leq n$) of the attribute $A_i$ of the relation $[R]^k$ with the query predicate $[p]^k$. If the value $[v_j]^k$ and the predicate $[p]^k$ match, it will result in one of secret-share form; otherwise, the result will be zero of secret-share form. While executing the string-matching operation on the next value $[v_{j+1}]^k$, the output of the previous matching operation is added, leading to provide the final output after executing the string-matching operation on all values of the attribute $A_i$. Since the sum of the values (one or zero) is of secret-shared form, the servers cannot learn the exact answer, and the mapper sends the final value to the user. The count query execution steps are given in Table~\ref{tab:steps_count}. Note that all the steps are identical to the steps given in Table~\ref{tab:steps1}, except the last step for counting the occurrences.

\begin{table}[h!]
\BB
\begin{center}
\centering
\scriptsize
\begin{tabular}{|p{8.5cm}|}
\hline
\textsc{Step} $1^{\mathrm{r}}$: $N_1^{\mathrm{r}}=[v_{\mathrm{r}}]_1 \odot [p]_1$\\
\textsc{Step} $2^{\mathrm{r}}$: $N_2^{\mathrm{r}}=[v_{\mathrm{r}}]_2 \odot [p]_2$\\

\indent \vdots

\textsc{Step} ${\mathit{Len(v_r)}}^{\mathrm{r}}$: $N_{\mathit{Len(v_r)}}^{\mathrm{r}}=[v_{\mathrm{r}}]_{\mathit{Len(v_r)}} \odot [p]_{\mathit{Len(v_r)}}$\\

\textsc{Step} $(\mathit{Len(v_r)}+1)^{\mathrm{r}}$: $\mathit{count}= \mathit{count}+ N_1^{\mathrm{r}} \times N_2^{\mathrm{r}} \times \ldots \times N_{\mathit{Len(v_r)}}^{\mathrm{r}}$\\\hline

\textbf{Notations.} $1\leq {\mathrm{r}} \leq n$: the number of the tuples in the relation. $\odot$: bit-wise multiplication and addition of all values in a vector.\\\hline

\end{tabular}
\BB
\caption{The accumulating automation steps executed by the server for count query.}
\label{tab:steps_count}
\BBB
\end{center}
\end{table}

\noindent\textit{User: result reconstruction.} The user performs an interpolation operation on the outputs obtained from servers, and it will produce the final answer to the count query.

\noindent\textbf{Example.} Suppose the count query on the employee relation (see Table~\ref{fig:database1}) is: \texttt{SELECT
COUNT(*) FROM Employee WHERE
FirstName = `John'}. Table~\ref{tab:example_count_query} shows how do the servers execute this query. Note that for explanation purpose, we show cleartext values; however, servers execute operations on secret-shares.

\begin{table}[!h]
\BB
\scriptsize
  \centering
  \begin{tabular}{|l|l|l|l|l|l|l|}
    \hline
    FirstName & String-matching resultant & $\mathit{Count}$ value\\ \hline
    Adam & 0 & 0\\\hline
    John & 1 & 1\\\hline
    Eve  & 0 & 0\\\hline
    John & 1 & 2\\\hline
  \end{tabular}
  \BB
  \caption{Count query execution example, when counting \texttt{FirstName = `John'}.}
  \label{tab:example_count_query}
\BBB
\end{table}

\smallskip
\noindent\textbf{Remark: Use of multiple mappers and a reducer.} In MapReduce, one can use many mappers to process the query. Thus, each mapper processes a different split of the relation and produces output of the form of a $\langle \mathit{key}, \mathit{value}\rangle$ pair, where the $\mathit{key}$ is an identity of an input split on which the operation was performed, and the corresponding $\mathit{value}$ is the final summation of secret-shared values obtained as the string-matching resultant on the attribute $A_i$. Here, a single reducer at a server processes all $\langle \mathit{key}, \mathit{value}\rangle$ pairs and adds $\mathit{value}$ of each split to provide a single secret-shared value to the user.

\noindent\textbf{Aside.} If a user searches \texttt{John} in a database containing names like `John' and `Johnson,' then our algorithm will show two occurrences of \texttt{John}. However, this is a problem associated with string-matching. In order to search a predicate precisely, we may use the terminating symbol for indicating the end of the predicate. In the above example, we can use ``\texttt{John }'', which is the searching predicate ending with a whitespace, to obtain the correct answer.

\BB\subsection{Selection Queries}
\label{subsec:Search query execution}
\B
This section presents oblivious algorithms for selection queries. The proposed algorithms first execute count query Algorithm~\ref{alg:count algorithm} for finding the number of tuples containing a selection predicate, say $p$, and then, after obtaining addresses of tuples\footnote{The address of a tuple indicates an index of the tuple, \textit{e}.\textit{g}., the fifth tuple.} containing $p$, fetches such tuples. Specifically, we provide 2-phased selection query algorithms, where:
\parskip 0pt
\setlength{\parindent}{15pt}

\noindent \textsc{Phase 0}: Count the occurrence of $p$.

\noindent \textsc{Phase 1}: Finding addresses of tuples containing $p$.

\noindent \textsc{Phase 2}: Fetching all the tuples containing $p$.

\smallskip We classify the selection query based on the number of tuples with a value, as follows:

\noindent\emph{One value holds only one tuple}. For example, the \texttt{Salary} attribute contains four values, and each value appears in one tuple; see Table~\ref{fig:database1}. In this case, there is \emph{no need} to know the address of the tuple containing $p$, \textit{i}.\textit{e}., implementing \textsc{Phase 1}. \S\ref{subsec:Unary Occurrence of a Pattern} presents an algorithm for retrieving tuples in the case of one value with one tuple.

\noindent\emph{Multiple values hold multiple tuples}.
For example, the \texttt{FirstName} attribute contains four values, and two employees have the same first name as \texttt{John}; see Table~\ref{fig:database1}. In this case, we need to know the address of the tuple containing $p$ (\textsc{Phase 1}) before retrieving the desired tuples. \S\ref{subsubsec:Multiple Occurrences of a Pattern} presents two algorithms for retrieving tuples when a value can appear in multiple tuples.

\B
\subsubsection{\textbf{One Value --- One Tuple}}
\label{subsec:Unary Occurrence of a Pattern}
In the case of one tuple per value, Algorithm~\ref{alg:single word fetch} (pseudocode is given in Appendix~\ref{app_sec:Pseudocodes}) fetches the entire tuple in a privacy-preserving manner without revealing the selection predicate $p$ and the tuple satisfying $p$, as follows:

\parskip 0pt
\setlength{\parindent}{15pt}

\smallskip
\noindent\textit{User: secret-shared query generation.} Consider the following query: \texttt{SELECT * FROM R WHERE $A_j$ = }$p$. The secret-sharing module at the user creates $c$ secret-shares of $p$ and sends them to $c$ servers, as did in count query~\ref{subsec:Count query execution}.

\smallskip
\noindent\textit{Servers: selection query execution.} The server executes a map function that matches the secret-shared predicate $[p]$ with each secret-shared value $[v_i]$ ($i=1,2,\ldots,n$) of the attribute $A_j$. Consequently, the map function results in either 0 or 1 of secret-shared form. If $[p]$ matches $[v_i]$ value of the attribute $A_j$, then the result is 1. After that, the map function multiplies the resultant (0 or 1) by all the other attribute values of the $i^{\mathit{th}}$ tuple. Thus, the map function creates a (virtual) relation of $n$ tuples and $m$ attributes, where all the tuples contain 0 across all the attributes except the tuple that contains $p$ in the attribute $A_j$. When the map function finishes on all the $n$ tuples, it adds all the secret-shares of each attribute, as: $S_1||S_2||\ldots||S_m$, and sends to the user, where $S_j$ is the sum of all the secret-shares value of the $j^{\mathit{th}}$ ($1\leq j\leq m$) attribute. (If there are multiple mappers are executed on different splits, then a single reducer adds all the values of the mappers, which provides the final answer to the selection query.)

\smallskip
\noindent\textit{User: result reconstruction.} At the user-side, the interpolation module, on receiving shares from servers, performs Lagrange interpolation that provides the desired tuple containing $p$ in the attribute $A_j$.

\B
\subsubsection{\textbf{Multiple Values with Multiple Tuples}}
\label{subsubsec:Multiple Occurrences of a Pattern}
When multiple tuples contain a selection predicate, say $p$, the user cannot fetch all those tuples obliviously without knowing their addresses. Therefore, we first need to design an algorithm to obliviously obtain the addresses of all the tuples containing the predicate $p$, and then, obliviously fetch the tuples. Throughout this section, we consider that $\ell$ tuples contain $p$. This section provides two search algorithms that have 2-phases, as:
\parskip 0pt
\setlength{\parindent}{15pt}

\noindent \textsc{Phase} 0: Count the occurrence of $p$.

\noindent \textsc{Phase} 1: Finding the addresses of the desired $\ell$ tuples.

\noindent \textsc{Phase} 2: Fetching all the $\ell$ tuples (according to their addresses).

The algorithms differ only in \textsc{Phase} 1 to know the addresses of the desired tuples, as follows: The first algorithm, called \textbf{one-round algorithm} requires only one-round of communication between the user and the server, while the second algorithm, called \textbf{tree-based algorithm} requires multiple rounds of communication, but has lower communication cost. Before going into details of algorithms, we first explore a tradeoff.

\medskip\noindent\textbf{Tradeoff.} When fetching multiple tuples containing $p$, there is a tradeoff between the number of communication rounds and the computational cost at the user, and this tradeoff will be clear after the description of one-round and tree-based algorithms. In particular, the user interpolates $n$ values to know the addresses of all the tuples containing $p$ in one-round algorithm. On the other hand, obtaining the addresses of tuples containing $p$ in multiple rounds requires executing a count query multiple times by the servers, while the user performs an interpolation (on less than $n$ values) to know the answer of count queries.

\medskip\noindent\textbf{One-round algorithm.} The one-round algorithm requires only one communication round between the user and the server for each of the two phases of the selection query algorithm. Assume that a selection query is: \texttt{SELECT * FROM R WHERE} $A_j=p$.

\smallskip
\noindent\textit{User: secret-shared query generation.} This step is identical to the step carried previously in case of one tuple per value (\S\ref{subsec:Unary Occurrence of a Pattern}).

\smallskip
\noindent\textit{Servers: finding addresses.} In \textsc{Phase} 1, the server executes a map function that performs the string-matching algorithm on each secret-share value of the attribute $A_j$, (as we did to count the occurrences of a predicate in \S\ref{subsec:Count query execution}). However, now, the mapper \emph{does not} accumulate string-matching results, and hence, sends $n$ string-matching resultant values corresponding to each tuple to the user.

\smallskip
\noindent\textit{User: result reconstruction and a vector creation for fetching tuples.} The interpolation module at the user executes Lagrange interpolation on each value, resulting in a vector of length $n$, where the $i^{\mathit{th}}$ entity has either 0 or 1 depending on the occurrence of $p$ in the $i^{\mathit{th}}$ tuple of the attribute $A_j$ of the relation $R$. Thus, the user knows addresses ($\mathit{RID}$) of all the desired tuples in a single round, but the user interpolates $n$ secret-shared values.\footnote{Experiments (\S\ref{sec:Experimental Evaluation}) show that interpolating $n$ values does not incur significant workload at the user.} Knowing the addresses of the desired tuples results in some information-leakage, which will be discussed in \S\ref{subsec:Information Leakage to the User}.

Suppose that $\ell$ tuples contain the predicate $p$ in the attribute $A_j$. Now, the secret-sharing module at the user creates a vector, say $\mathit{vec}$, of length $\ell$. Note that the secret-sharing module can create a vector of larger length too to hide the number of tuples satisfying the predicate $p$; we will discuss leakage issue in detail in \S\ref{subsubsec:leakage_Multiple Values with Multiple Tuples}. The vector $\mathit{vec}$ contains all the row-ids, $\mathit{RID}$,\footnote{We will use the words `row-id' and `tuple-id' interchangeably.} of tuples that contain the predicate $p$ in the attribute $A_j$. The secret-sharing module creates secret-shares of each value of the vector $\mathit{vec}$, by following the approach as suggested in \S\ref{subsec:Data Model Creation and Distribution of Secret-Shares of a Relation}, and sends them to the servers.

\smallskip
\noindent\textit{Servers.} In \textsc{Phase} 2, the map function at servers executes the same operation on each value of the attribute $\mathit{RID}$ and each value of the received secret-shared vector, as the map function executed in the case of one tuple per value. Finally, the map function provides $\ell$ tuples. Particularly, the map function executes string-matching operation between $i^{\mathit{th}}$ ($1\leq i\leq n$) value of the attribute $\mathit{RID}$ and $j^{\mathit{th}}$ ($1\leq j\leq \ell$) value of the secret-shared vector, which results in a value $o_{ij}$ (either 0 or 1 of secret-shared form). The string-matching resultant $o_{ij}$ is multiplied by all the remaining attribute values of the $i^{\mathit{th}}$ tuple. Finally, a sum operation is carried out on each attribute value. Note that the secret-shared vector contains secret-shared row-ids that each of them will surely match with only one of the $n$ values of the attribute $\mathit{RID}$; hence, after multiplying secret-shared string-matching resultants and adding all the values of each attribute, the mapper will produce secret-shared $\ell$ desired tuples, without knowing which tuples have been satisfied with the selection query.

\smallskip
\noindent\textit{User: result reconstruction.} The interpolation module executes Lagrange interpolation over $\ell$ secret-shared tuples and provides the final answer to the user.

\noindent\textbf{Example.} Suppose the selection query is: \texttt{SELECT
EId, FirstName, LastName, DateofBirth, Salary FROM Employee WHERE FirstName = `John'}. Here, we show how do the servers and user execute this query. Note that for explanation purpose, we use cleartext values; however, servers execute operations on secret-shares. In this example, the server, first, performs string-matching operations on \texttt{FirstName} attribute and sends a vector having $\langle 0,1,0,1\rangle$ to the user. Table~\ref{tab:example_multi_tuple_fetch} shows string-matching operation results.

\begin{table}[!h]
\B
\scriptsize
  \centering
  \begin{tabular}{|l|l|l|l|l|l|l|}
    \hline
    FirstName & String-matching resultant \\ \hline
    Adam & 0 \\\hline
    John & 1 \\\hline
    Eve  & 0 \\\hline
    John & 1 \\\hline
  \end{tabular}
  \B
  \caption{String-matching results, when searching \texttt{FirstName = `John'}.}
  \label{tab:example_multi_tuple_fetch}
\BBB
\end{table}
Based on the received vector $\langle 0,1,0,1\rangle$, the user knows that it needs to fetch second and fourth tuples. Thus, the secret-sharing module at the user creates a vector, $\mathit{vec}=\langle 2,4\rangle$, creates secret-shares of the vector $\mathit{vec}$, and sends them to the server. The server executes string-matching operations on each value of $\mathit{RID}$ attribute of the relation and secret-shared vector, multiplies the resultant to the tuple values, and adds all the values of each attribute. For example, Table~\ref{tab:Tuple retrieval process, shown only for the second tuple} shows the execution at the server for only row-id 2 in cleartext.

\begin{table}[!h]
\scriptsize
  \centering
  \begin{tabular}{|p{0.25cm}|p{0.32cm}|p{0.99cm}|p{0.95cm}|p{1.2cm}|p{1.45cm}|p{0.95cm}|l|}
    \hline
    RID & SMR& EId & FirstName & LastName & DateofBirth & Salary \\ \hline\hline
    1& 0& $0\times$(E101) & $0\times$(Adam) & $0\times$(Smith) & $0\times$(12/07/1975) & $0\times$(1000)  \\\hline
    2& 1& $1\times$(E102) & $1\times$(John) & $1\times$(Taylor)& $1\times$(10/30/1985) & $1\times$(2000) \\\hline
    3& 0&$0\times$(E103) & $0\times$(Eve)  & $0\times$(Smith)      & $0\times$(05/07/1985) & $0\times$(500)   \\ \hline
    4& 0&$0\times$(E104) & $0\times$(John) & $0\times$(Williams)   & $0\times$(04/04/1990) & $0\times$(5000)   \\ \hline\hline
    2& ~& E102 & John & Taylor& 10/30/1985 & 2000 \\\hline
  \end{tabular}
  \BB
  \caption{Tuple retrieval process, shown only for the second tuple. SMR: string-matching resultant.}
  \label{tab:Tuple retrieval process, shown only for the second tuple}
\BBB
\end{table}

\medskip\noindent\textbf{Tree-based algorithm.} In order to decrease the computational load at the user, unlike one-round algorithm, for knowing the desired tuple addresses, we propose a tree-based search (Algorithm~\ref{alg:Multi-tuple search Algorithm} in Appendix~\ref{app_sec:Pseudocodes}) that finds the addresses of the desired $\ell$ number of tuples in multiple rounds, and then, fetches all the desired $\ell$ tuples (by following the same method like one-round algorithm).

Taking inspiration form Algorithm~\ref{alg:single word fetch} for one tuple per value (given in \S\ref{subsec:Unary Occurrence of a Pattern}), we can also obtain the row-id in an oblivious manner, while a single tuple contains the selection predicate $p$. Thus, for finding addresses of $\ell$ tuples containing $p$, the user requests the servers to partition the whole relation into certain blocks ($\geq \ell$, depending on the privacy requirements) such that each block belongs to one of the following cases:
\begin{enumerate}[noitemsep,nolistsep,leftmargin=.4cm]
\item A block contains no occurrence of $p$, and hence, no fetch operation is needed.
\item A block contains one/multiple tuples but only a single tuple contains $p$.
\item A block contains $h$ tuples, and all the $h$ tuples contain $p$.
\item A block contains multiple tuples but fewer tuples contain $p$.
\end{enumerate}

\textit{Finding addresses}. We follow an idea of partitioning the database and counting the occurrences of $p$ in the blocks, until each block satisfies one of the above-mentioned cases. Specifically, the user initiates a sequence of Query \& Answer (Q\&A) rounds. In the first Q\&A round, the user asks the servers to count the occurrences of $p$ in the whole database (or in an assigned input split to a mapper) and then partitions the database into at least $\ell$ blocks, since we assumed that $\ell$ tuples contain $p$. In the second Q\&A round, the user again asks the servers to count the occurrences of $p$ in each block and focuses on the blocks satisfying Case 4 (due to performance; we will discuss leakages vs performance involved in this method in \S\ref{subsubsec:leakage_Multiple Values with Multiple Tuples}). There is no need to consider the blocks satisfying Case 2 or 3, since the user can apply Algorithm~\ref{alg:single word fetch} (one tuple per value) in both cases. However, if the multiple tuples of a block in the second round contain $p$, \textit{i}.\textit{e}., Case 4, the user again asks the servers to partition such a block until it satisfies either Case 1, 2 or 3. After that, the user can obtain the addresses of the related tuples using the method similar to Algorithm~\ref{alg:single word fetch}. Note that in this case, the user does not need to interpolate $n$ values corresponding to each row.

\textit{Fetching tuples}. After knowing the row-ids, the user applies the approach described in the one-round algorithm for fetching multiple tuples.

\begin{figure}[h]
\B
\centering
\includegraphics[scale=0.30]{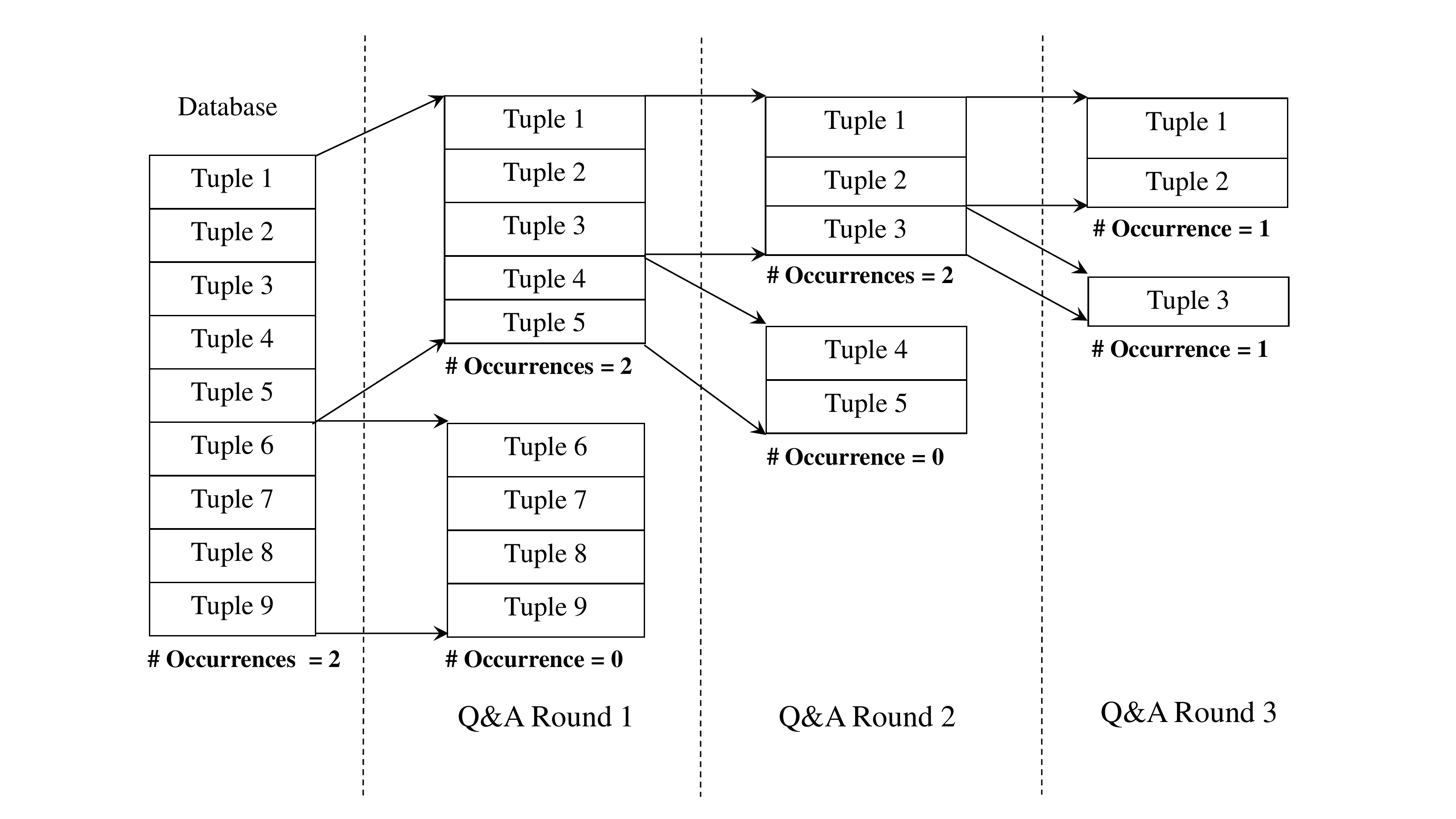}
\B
\caption{Example of Q\&A rounds.}
\BBB
\label{fig:Q/A}
\end{figure}

\medskip\noindent\textbf{\textit{Example}}. Figure~\ref{fig:Q/A} provides an example to illustrate the tree-based approach. Assume that an input split consists of 9 tuples, and the number of occurrences of $p$ is two. When the user knows the number of occurrences, the user starts Q\&A rounds. In each Q\&A round, a mapper partitions specific parts of the input split into two blocks, performs counting operation in each block, and sends results (which are occurrences of $p$ in each block) of the form of secret-shares, back to the user.

In this example, the user initiates the first Q\&A round, and a mapper divides the input split into two blocks. In each block, it counts the occurrences of $p$ and sends the results to the user. The interpolation module at the user executes Lagrange interpolation and provides that the first and second blocks contain two and zero tuples having $p$, respectively. In the second Q\&A round, the user requests to the servers to partition the first block into two blocks. The mapper performs an identical operation as it does in the first Q\&A round, and after three Q\&A rounds, the user knows the addresses of all the tuples containing $p$.

\BB
\subsection{Join Queries}
\label{subsec:Join Query Execution}
\B

Consider two relations $X(A,B)$ and $Y(B,C)$, where $B$ is the joining attribute. A trivial solution for an oblivious join is, as follows: (\textit{i}) fetch all the secret-shared values of the joining attribute from servers and interpolate them, (\textit{ii}) find tuples of both relations that have an identical joining value and fetch all those tuples, (\textit{iii}) interpolate all the secret-shared tuples, and (\textit{iv}) perform a MapReduce job for joining the tuples at the user. However, note that in this solution, the user performs most of the computation. To decrease the computational load at the user, the following sections present oblivious join algorithms for primary to foreign key (PK/FK) join and oblivious non-PK/FK equijoin at the servers.

\subsubsection{\textbf{PK/FK-based Oblivious Join}}
\label{subsubsec:A unique occurrence of the joining value}
Consider two relations: a parent relation $X(A,B)$ having the attribute $B$ as a primary key and $n_x$ tuples, and a child relation $Y(B,C)$ of $n_y$ tuples. We use string-matching operations (a variant of one tuple per value; Algorithm~\ref{alg:single word fetch}) on secret-shares for performing PK/FK join on $X(A,B)$ and $Y(B,C)$ relations based on the $B$ attribute. The following steps are executed:

\noindent 1. At the server:
\begin{enumerate}[label=\textit{\alph{*}.},noitemsep,nolistsep]
\item A mapper reads $i^{th}$ tuple $\langle \ast, b_i\rangle$ of the relation $X$ and provides $n_y$ pairs of $\langle\mathit{key,value}\rangle$, where the $\mathit{key}$ is an identity from 1 to $n_y$ and the $\mathit{value}$ is the $i^{\mathit{th}}$ secret-shared tuple $\langle \ast,b_i \rangle$. A ``$\ast$'' denotes any value.

\item A mapper reads $j^{th}$ tuple $\langle b_j, \ast\rangle$ of the relation $Y$ and provides one pair of $\langle\mathit{key,value}\rangle$, where the $\mathit{key}$ is the tuple id and the $\mathit{value}$ is $j^{\mathit{th}}$ secret-shared tuple $\langle b_j, \ast\rangle$.

\item All the tuples of the relation $X(A,B)$ and the $i^{\mathit{th}}$ tuple of the relation $Y(B,C)$ are assigned to the $i^{th}$ reducer. The reducer performs string-matching operations on the $B$ attribute, resulting in 0 or 1 of secret-shared form. Specifically, the reducer matches $\mathit{b}_i\in Y$ with each $b_j\in X$, and the resultant of the string-matching operation (of $b_i$ and $b_j$) is multiplied by the tuple $\langle \ast, b_j\rangle$ of the relation $X(A,B)$. After performing the string-matching operation on all the $B$ values of the $X(A,B)$ relation, the reducer adds all the secret-shares of the $A$ and $B$ attributes of the relation $X$. This will result in a tuple, say $\langle \ast^{\prime}, b^{\prime}\rangle$. The $C$ value of the tuple $\langle b, \ast\rangle$ of the relation $Y$ is appended to $\langle \ast^{\prime}, b^{\prime}\rangle$, and it results in the final joined tuple $\langle \ast^{\prime}, b^{\prime}, \ast\rangle$.
\end{enumerate}
\noindent 2. The user fetches all the outputs of reducers from the servers and interpolates them to obtains the outputs of the join.

\begin{table}[!ht]
\scriptsize

\centering
\begin{tabular}{|c|c|}\hline
$A$ & $B$ \\ \hline
 $a_1$ & $b_1$ \\ \hline
 $a_2$ & $b_2$ \\ \hline
 $a_3$ & $b_3$ \\ \hline
 \end{tabular}
    \qquad
\begin{tabular}{|c|c|}\hline
$B$ & $C$ \\ \hline
 $b_1$ & $c_1$  \\ \hline
 $b_2$ & $c_2$  \\ \hline
 $b_2$ & $c_3$  \\ \hline
 $b_2$ & $c_4$  \\ \hline
    \end{tabular}
    \qquad
\begin{tabular}{|c|c|c|}\hline
$A$   & $B$   & $C$ \\ \hline
$a_1$ & $b_1$ & $c_1$  \\ \hline
$a_2$ & $b_2$ & $c_2$  \\ \hline
$a_2$ & $b_2$ & $c_3$  \\ \hline
$a_2$ & $b_2$ & $c_4$  \\ \hline
    \end{tabular}
    \BB
\caption{Two relations $X(A,B)$ and $Y(B,C)$, and their joined outputs.}
\label{fig:two_relation}
\BBB
\end{table}

\medskip\noindent\textit{\textbf{Example.}} Table~\ref{fig:two_relation} shows two relations $X$ and $Y$, and their joined outputs. For explanation purpose, we show cleartext values. The mapper reads the tuple $\langle a_1,b_1\rangle$ and provides $\langle 1, [a_1,b_1]\rangle$, $\langle 2, [a_1,b_1]\rangle$, $\langle 3, [a_1,b_1]\rangle$, and $\langle 4, [a_1,b_1]\rangle$. A similar operation is also carried out on the tuples $\langle a_2,b_2\rangle$ and $\langle a_3,b_3\rangle$. Mappers read the tuples $\langle b_1, c_1\rangle$, $\langle b_2,c_2\rangle$, $\langle b_2, c_3\rangle$, and $\langle b_2, c_4\rangle$ and provide $\langle 1, [b_1,c_1]\rangle$, $\langle 2, [b_2,c_2]\rangle$, $\langle 3, [b_2,c_3]\rangle$, and $\langle 4, [b_2,c_4]\rangle$, respectively.

A reducer corresponding to key 1 matches $b_1$ of $X$ with $b_1$ of $Y$ that results in 1, $b_2$ of $X$ with $b_1$ of $Y$ that results in 0, and $b_3$ of $X$ with $b_1$ of $Y$ that results in 0. Remember 0 and 1 are of the form of secret-shares. Now, the reducer multiplies the three values (1,0,0) of the form of secret-shares by the tuples $\langle a_1,b_1\rangle$, $\langle a_2,b_2\rangle$, and $\langle a_3,b_3\rangle$, respectively. After that, the reducer adds all the $A$-values and the $B$-values. Note that we will obtain now only the desired tuple, \textit{i}.\textit{e}., $\langle a_1,b_1\rangle$. The reducer appends the $C$ value of the tuple $\langle b_1,c_1\rangle$ to the tuple $\langle a_1,b_1\rangle$. The same operation is carried out by other reducers. When the user performs the interpolation on the outputs, only the desired joined output tuples are obtained.

\noindent\textbf{Aside.} We assume that all the $A$, $B$, and $C$ values of the relations do not contain zero.

\subsubsection{\textbf{Non-PK/FK-based oblivious equijoin}}
\label{subsubsec:Multiple occurrences of the joining value}
This section presents an oblivious non-PK/FK equijoin algorithm. Consider two relations $X(A,B)$ and $Y(B,C)$ having $n$ tuples in each and $B$ as a joining attribute that can have multiple occurrences of a value in both relations.

Note that PK/FK-based join approach, \S\ref{subsubsec:A unique occurrence of the joining value}, will not work here. The reason is as follows: when the $B$ value of a tuple, say $\langle a_1, b_1\rangle$, of the relation $X$ is compared with all tuples of $Y$ that contains multiple tuples with the value $b_1$, say $\langle b_1,c_1\rangle$, $\langle b_1,c_2\rangle \ldots$, it results in multiple ones that will produce all the desired tuples of the relation $Y$ containing $b_1$. Here, the sum of all the values of an attribute cannot distinguish two or multiple occurrences of the $b_1$ value, and hence, the result will be incorrect.

In the proposed oblivious non-PK/FK join approach, the user performs a little bit more computation, unlike PK/FK join. In particular, the user
interpolates all secret-shared $B$-values of both relations, finds common $B$ values of both relations, and then, requests the servers to execute the join operation. Note that here the user will also know which $B$-values of the relations do not join and the addresses of the joining tuples in both relations. We will discuss information-leakage and their prevention in \S\ref{subsec:Information Leakage to the User}.

\begin{figure}[!t]
\centering
\includegraphics[scale=0.34]{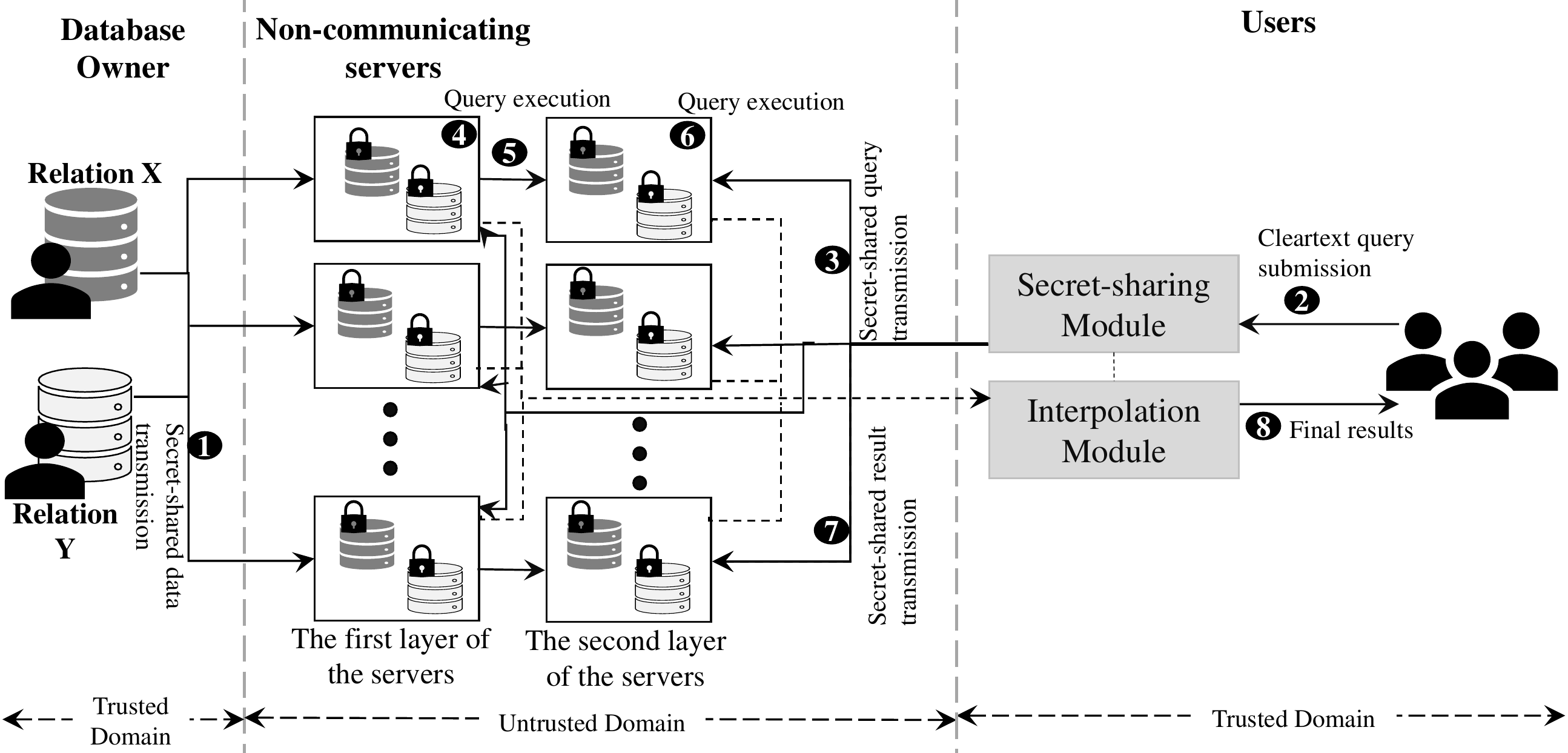}
\caption{The two layers of the servers for an oblivious equijoin. The first layer executes selection queries using the one-round algorithm and the second layer executes an equijoin.}
\BBB
\label{fig:new_system_settings}
\end{figure}

In order to perform oblivious non-PK/FK join, we need $c$ more non-communicating servers.\footnote{The use of $c$ additional servers makes the explanation easier. However, at the end of the algorithm, we show how to overcome these additional $c$ servers.} Here, we call a set of the first $c$ servers as \emph{the first layer}, and a set of the remaining $c$ servers as \emph{the second layer}. Note that the servers within a layer are non-communicating; however, the $i^{\mathit{th}}$ server of the first layer can communicate with only the $i^{\mathit{th}}$ server of the second layer. We assume that the first layer holds the relations, and the second layer provides outputs to an equijoin; see Figure~\ref{fig:new_system_settings}.

\medskip\noindent\textit{Approach.} The approach consists of the following three steps, where the second step that performs an equijoin is executed at the servers:

\noindent 1. The user fetches all the secret-shared $B$-values of the relations $X$ and $Y$ and performs an interpolation. Consequently, the user knows which $B$-values are identical in both relations and which tuples contain identical joining values.

\noindent 2. For each $B$-value (say, $b_i$) that is in both the relations:
\begin{enumerate}[label=\textit{\alph{*}.},noitemsep,nolistsep]
  \item The user executes \textsc{Phase 2} of the one-round algorithm (given in \S\ref{subsubsec:Multiple Occurrences of a Pattern}) for the value $b_i$ on the first layer servers that obliviously send all the tuples containing $b_i$ of the relations $X$ and $Y$ to the second layer servers. Figure~\ref{fig:new_system_settings} illustrates the algorithm's execution. (Note that after interpolating $B$-values of both the relations, the user also knows the $\mathit{RID}$ values of common $B$-values. Hence, here is no need to execute \textsc{Phase 1} of the one-round algorithm.)

  \item On receiving tuples containing the joining value $b_i$ from the first layer, the second layer servers create two new relations, say $X^{\prime}(A,B)$ and $Y^{\prime}(B,C)$, corresponding to the tuples of the relations $X$ and $Y$, respectively. Then, the second layer servers execute an oblivious equijoin MapReduce job on $X^{\prime}$ and $Y^{\prime}$, and provide the output of an equijoin.

      In particular, a mapper reads a tuple, say $\langle \ast, b_i\rangle$, of the relation $X^{\prime}$ and produces $k$ pairs of $\langle \mathit{key},\mathit{value}\rangle$, where $k$ is the number of tuples in the relation $Y^{\prime}$, the key is 1 to $k$, and the value is the secret-shared tuple $\langle \ast, b_i\rangle$. In addition, a mapper reads a tuple, say $\langle b_i,\ast\rangle$, of the relation $Y^{\prime}$ and produces one pair of $\langle \mathit{key},\mathit{value}\rangle$, where the key is the tuple id, and the value is the secret-shared tuple $\langle b_i,\ast\rangle$. A reducer $i$ is assigned all the tuples of the $X^{\prime}(A,B)$ relation and the $i^{\mathit{th}}$ tuple of the $Y^{\prime}(B,C)$ relation. The reducer concatenates each tuple of the $X^{\prime}(A,B)$ relation with the $i^{\mathit{th}}$ tuple of the $Y^{\prime}(B,C)$. Since the two new relations, $X^{\prime}(A,B)$ and $Y^{\prime}(B,C)$, have only one identical $B$-value, the concatenation provides the correct answer.
\end{enumerate}

\noindent 3. The user fetches the output tuples from the second layer servers and performs the interpolation.

\noindent\textbf{Note.} We used two layers of servers to better explain the approach. However, one can avoid the second layer servers trivially by placing the work of the second layer servers at the first layer servers. Thus, after executing Step 2a (\textit{i}.\textit{e}., fetching tuples having common $b_i$-values in both relations), the $c$ servers can execute the join operations (Step 2b).

\subsection{Range Query}
\label{subsec:Range Query Execution}
\B
A range query finds, for example, all the employees whose salaries are between \$1000 and \$2000. We propose an approach for performing privacy-preserving range queries based on 2's complement subtraction. A number, say $x$, belongs in a range, say $[a, b]$, if $\mathit{sign}(x-a)=0$ and $\mathit{sign}(x-b)=0$, where $\mathit{sign}(x-a)$ and $\mathit{sign}(b-x)$ denote the sign bits of $x-a$ and $x-b$, respectively, after 2's complement-based subtraction.

\parskip 0pt
\setlength{\parindent}{15pt}

Recall that in \S\ref{subsec:Data Model Creation and Distribution of Secret-Shares of a Relation}, we proposed an approach for creating secret-shares of a number, say $x$, using a unary representation that provides a vector, where all the values are 0 except only 1 according to the position of the number. The approach works well to count the occurrences of $x$ and fetch all the tuples having $x$. However, on unary vectors, we cannot perform a subtraction operation using 2's complement. Hence, to execute range queries, we present a number using a binary-representation, which results in a vector of length, say $l$. After that, we use SSS to make secret-shares of every bit in the vector by selecting $l$ different polynomials of an identical degree for each bit position.

\noindent\textbf{Approach.} The idea of finding whether a number, $x$, belongs to the range, $[a,b]$, is based on 2's complement subtraction. In~\cite{DBLP:conf/algocloud/DolevL15}, the authors provided an algorithm for subtracting secret-shares using 2's complement. However, we will provide a simple 2's complement-based subtraction algorithm for secret-shares. A mapper checks the sign bits after subtraction for deciding the number whether it is in the range or not, as follows:
\begin{equation}\label{eq:sig}
\scriptsize
\begin{array}{ll}
\text{If~}x\in[a,b], & \mathit{op}_1=sign(x-a)=0, \mathit{op}_2=sign(b-x)=0\\
\text{If~} x<a, & \mathit{op}_1=sign(x-a)=1, \mathit{op}_2=sign(b-x)=0\\
\text{If~} x>b, & \mathit{op}_1=sign(x-a)=0, \mathit{op}_2=sign(b-x)=1
\end{array}
\end{equation}
The mappers execute the above equation on each value of the desired attribute and add $\mathit{op}_1$ and $\mathit{op}_2$.

\noindent\textbf{Range-based count query.} For a range-based count query (pseudocode is given Algorithm~\ref{alg:range count} in Appendix~\ref{app_sec:Pseudocodes}), after executing the above Equation~\ref{eq:sig} and adding $\mathit{op}_1$ and $\mathit{op}_2$, the mapper has $n$ values, either 0 or 1 of secret-shared form, where $n$ is the number of tuples in a relation. Finally, the mapper adds all the $n$ values and provides the result to the answer.

On receiving the secret-shared answer, the user interpolates them. The answer shows how many tuples have not satisfied the range query, and subtracting the answer from $n$ provides the final answer to the range-based count query. (Recall that $n$ is assumed to be a publicly known \S\ref{subsec:Adversarial Settings}.)

\noindent\textbf{Range-based selection query.} The servers execute equation (1) and add $\mathit{op}_1$ and $\mathit{op}_2$, which results in $n$ values, either 0 or 1 of secret-shared form, where $n$ is the number of tuples in a relation. Finally, the server provides $n$ values to the user.

The user interpolates all the received $n$ values from servers. If the number $x$ in the $i^{th}$ tuples belongs in the range $[a,b]$, then the $i^{th}$ position in the array is zero. Otherwise, the $i^{th}$ position in the array is one. Finally, the user fetches all the tuples having value 0 in the array using the one-round algorithm for fetching multiple tuples; see \S\ref{subsubsec:Multiple Occurrences of a Pattern}.

\noindent\textbf{Note.} After obtaining secret-shares of sign bits of $\mathit{op}_1=x-a$ and $\mathit{op}_1=b-x$, instead of adding $\mathit{op}_1$ and $\mathit{op}_2$, the mapper can also perform the following:
\begin{equation}\label{eq:eta}
1-\big(sign(x-a)+sign(b-x)\big).
\end{equation}
According to Equation~\ref{eq:sig}, if $x\in [a,b]$, the result of Equation~\ref{eq:eta} is secret-share of 1; otherwise, the result is secret-share of 0. Also, note that after doing this operation, when mappers will send an answer to the count query, after interpolation at the user, it will produce the correct answer to the count query, unlike the above-mentioned method, where the user subtracts from $n$. Similarly, in the case of selection queries, the user creates a vector of 1 or 0 to find the qualified tuples, however, the user will fetch tuples according to the position of 1 in the vector.

\medskip\noindent{\bf 2's complement-based secret-shares subtraction.} Algorithm~\ref{algo:SS-SUB} (pseudocode is given in Appendix~\ref{app_sec:Pseudocodes}) provides a way to perform 2's complement-based subtraction on secret-shares. We follow the definition of 2's complement subtraction to convert $B-A$ into $B+\bar{A}+1$, where $\bar{A}+1$ is 2's complement representation of $-A$. We start at the least significant bit (LSB), invert $a_0$, calculate $\bar{a}_0+b_0+1$ and its carry bit (lines~\ref{ln:1}-~\ref{ln:3} of Algorithm~\ref{algo:SS-SUB}). Then, we go through the rest of the bits, calculate the carry and the result for each bit (line~\ref{ln:4} of Algorithm~\ref{algo:SS-SUB}). After finishing all the computations, the most significant bit (MSB) or the sign bit is returned (line~\ref{ln:5} of Algorithm~\ref{algo:SS-SUB}). This method is similar to the method presented in \cite{DBLP:conf/algocloud/DolevL15}, but simpler, as we only need the sign bit of the result.

\LinesNotNumbered

\BB
\section{Experimental Evaluation}
\label{sec:Experimental Evaluation}
\B
This section shows the experimental results of our proposed algorithms. We used servers having 64GB RAM, 3.0GHz Intel Xeon CPU with 36 cores in AWS EMR clusters, while a 16GB RAM machine was used as the DB owner, as well as, a user that communicates with AWS servers.

\noindent\textbf{Secret-share (SS) dataset generation.} We used three tables of TPC-H benchmark, namely \texttt{Nation} with columns
NationKey (NK), Name (NE), and RegionKey (RK), \texttt{Customer}
with columns CustKey (CK), CustName (CN), NationKey (NK), and MarketSegment (MS), and \texttt{Supplier} with columns SuppKey (SK), SuppName (SN), and NationKey (NK). We placed 1M and 9M rows in \texttt{Customer} table and 70K and 670K rows in \texttt{Supplier} table, while \texttt{Nation} table has 25 rows.

\smallskip
\noindent\textit{SS-\texttt{Customer} table generation}. We explain the method followed to generate SS data for 1M \texttt{Customer} rows. A similar method was used to generate SS data for 9M rows. The four columns of \texttt{Customer} table contain numbers: CK: 1 to 1,000,000 (9,000,000 in 9M dataset), CN: 1 to 1,000,000 (9,000,000 in 9M dataset), NK: 1 to 25, and MS: 1 to 5. The following steps are required to generate SS of the four columns in 1M rows:
\begin{enumerate}[noitemsep,nolistsep,leftmargin=0.1in]
  \item \textit{Step 1: Padding}. The first step is to pad each number of each column with zeros. Hence, all numbers in a column contain identical digits to prevent an adversary to know the distribution of values. For example, after padding the value 1 of CK column is represented as 0,000,001, since the maximum length of a value in CK column was seven. Similarly, values of CN and MS were padded. We did not pad NK values, since they took only one digit.

  \item \textit{Step 2: Unary representation}. The second step is representing each digit into a set of ten numbers, as mentioned in \S\ref{subsec:Data Model Creation and Distribution of Secret-Shares of a Relation}, having only 0s or 1s. For example, 0,000,001 (one value of CK) was converted into 70 numbers, having all zeros except the number at 1$^\mathit{st}$, 11$^\mathit{th}$, 21$^\mathit{st}$, 31$^\mathit{st}$, 41$^\mathit{st}$, 51$^\mathit{st}$, and 62$^\mathit{nd}$ positions. Here, a group of the first ten numbers shows the first digit, \textit{i}.\textit{e}., 0, a group of 11th to 20th number shows the second digit, \textit{i}.\textit{e}., 0, and so on. Similarly, each value of CN, NK, and MS  was converted. We also added $\mathit{RID}$ column.

  \item \textit{Step 3: Secret-share generation}. The third step is used to creating SS of the numbers, generated in step 2. We selected a polynomial $f(x)= \mathit{secret\_value}+a_1x$, where $a_1$ was selected randomly between 1 to 9M for each number, the modulus is chosen as 15,000,017, and $x$ was varied from one to sixteen to obtain sixteen shares of each value. Thus, we obtained $[R]^i$, $1\leq i\leq 16$. Recall that for matching a string of length $l$ in one round, we need at least $2l+1$ shares, while using polynomials of degree one, as mentioned in \S\ref{sec:Preliminarily Example}. Here, the largest string length is 7, thus, we need at least 15 shares for count queries. However, for selection queries based on $RID$ , we need at least 16 shares. To minimize query processing time for selection and join queries, we add four more attributes corresponding to each of the four attributes in \texttt{Customer} table. A value of each of the additional four attributes has only one secret-shared value, created using SSS (not after padding). But, one can also implement the same query on secret-shared values obtained after step 2.

  \item \textit{Step 4: Outsourcing the shares}. We placed $i^{\mathit{th}}$ share of the relation $[R]^i$ to $i^{\mathit{th}}$ AWS EMR cluster.
\end{enumerate}

\smallskip
\noindent\textit{SS-\texttt{Supplier} and SS-\texttt{Nation} tables generation}. The three columns of \texttt{Supplier} table contain numbers: SK and SN: 1 to 70,000 (670,000 in 9M), NK: 1 to 25. The three columns of \texttt{Nation} table contain numbers: NK and NE: 25 and RK: 1 to 5. We followed the same steps as followed on \texttt{Customer} table, except the difference of padding the values with 0, that has changed in according to the maximum length of a value in a column.

\noindent\textbf{Experiment 1: SS data generation time and size.} Table~\ref{tab:Data generation} shows the average size of generated secret-shared \texttt{Customer}, \texttt{Supplier}, and \texttt{Nation} tables, at the DB owner machine using one-threaded implementation. Note that the dataset size is increased as expected, due to unary representation. For example, in 1M cleartext \texttt{Customer} table, CK column has 1M numbers, which each took 32 bits, while a single value of CK column in SS data took $32\times70=2240$ bits.

\begin{table}
  \centering
  \scriptsize
        \begin{tabular}{|l|l|l|}
    \hline
    Tuples          &  Size (in GB) \\ \hline

\texttt{Customer}--1M &  2.3  \\ \hline
\texttt{Customer}--9M &  21  \\ \hline

\texttt{Supplier}--70K &   0.1275  \\ \hline
\texttt{Supplier}--670K&   1.2  \\ \hline

\texttt{Customer}--25 &  0.0000165  \\ \hline
    \end{tabular}
    \BB
    \caption{Experiment. 1. Average size for shared data generation by the DB owner (single threaded implementation).}
    \label{tab:Data generation}
    \BBB
\end{table}

\begin{figure}[!t]
\centering
\includegraphics[scale=0.45]{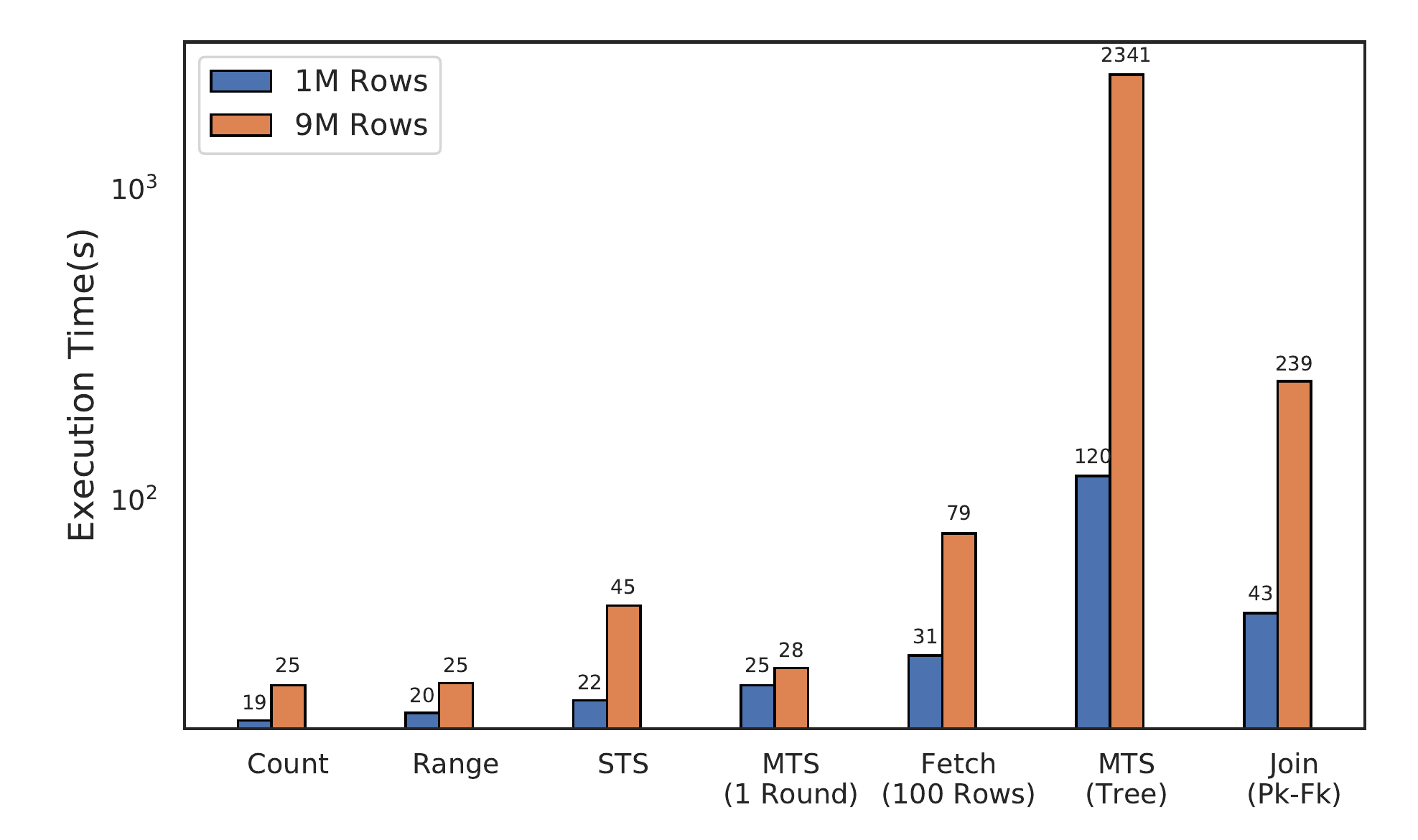}
\BB
\caption{Experiment 2. Performance test on 1M and 9M rows.}
\label{fig:fig_scalability}
\BBB
\end{figure}

\noindent\textbf{Experiment 2: Performance test.} We executed count, selection, join, and range queries using the proposed algorithms. Figure~\ref{fig:fig_scalability} shows server processing time for each query on 1M and 9M rows of \texttt{Customer} table and 25 rows of \texttt{Nation} table using a cluster of size four. We will discuss query processing below.

\noindent\textit{\textbf{Count queries.}} Figure~\ref{fig:fig_scalability} shows the time taken by a count query on NK column of \texttt{Customer} table on 1M and 9M rows. Interestingly for count queries, the increase in time is not proportional to the increase in dataset size, since parallel processing using multiple mappers on a cluster of size four reduced the computation time significantly, resulting in the total execution time to reach close to the time spent in the sequential part of the code, \textit{i}.\textit{e}., job setup time and disk I/O time.

\noindent\textit{\textbf{Selection queries.}} We executed algorithms for one tuple per value, as well as, one-round and tree-based algorithms for fetching multiple tuples. For one-tuple per value or single tuple selection (STS), see Figure~\ref{fig:fig_scalability}, we executed a selection query on CK column of \texttt{Customer} table. The computation time with 9M rows is two times more than the computation time on 1M rows. Furthermore, single tuple selection involves almost 4 times more disk I/O which cannot be made parallel by an increasing number of mappers.
For fetching multiple tuples, we used NK column of \texttt{Customer} table, while limiting the number of retrieved tuples to be 100. Figure~\ref{fig:fig_scalability} shows the time spent at servers for determining the tuple address using one-round algorithm (MTS-1round) and tree-based algorithm (MTS-Tree). Note that the entire query execution time of MTS-1round is less than MTS-Tree, since multiple rounds in MTS-Tree consume more time at the servers due to scanning and computing on the database many times. For MTS-Tree on 1M rows, there were 7930, 12161 and 1211 blocks in Q\&A rounds, while block sizes were 128, 32 and 8, respectively. For MTS-Tree on 9M rows, there were 75874, 115637 and 11325 blocks in the Q\&A rounds with block sizes of 128, 32 and 8 rows respectively. After knowing tuple addresses using either MTS-1round or MTS-Tree, the user retrieved 100 tuples, as shown in Figure~\ref{fig:fig_scalability} denoted by `Fetch (100 rows).' Note that the total processing time for fetching 100 tuples using MTS-1round is the sum of time taken by MTS-1round and Fetch (100 rows). Similarly, the total processing time for fetching 100 tuples using MTS-Tree is the sum of time taken by MTS-Tree and Fetch (100 rows).

\noindent\textit{\textbf{Join queries.}} We executed PK/FK joins over \texttt{Nation} and \texttt{Customer} tables on NK columns. Figure~\ref{fig:fig_scalability} shows the execution time for PK/FK based join query. For non-PK/FK join, we considered NK column of \texttt{Supplier} and \texttt{Customer} tables. The join is preceded by a selection, NK $=10$, on \texttt{Customer} and \texttt{Supplier} tables. We implemented this join query as a 3-step process: (\textit{i}) MTS-1round algorithm for finding the tuples matching the selection condition (NK $=10$) on both \texttt{Customer} and \texttt{Supplier} tables. (\textit{ii}) Fetching all the matching tuples using the tuple fetch algorithm from both the tables. (\textit{iii}) Execute a Cartesian product on the tuples fetched from each table. However, for executing non-PK/FK join, we used \texttt{Supplier} table of size 1000, 2000, and 3000, while \texttt{Customer} table was of size 10,000, 20,000, and 30,000; see Table~\ref{tab:nonpk-fk}. The reason for not executing non-PK/FK join over 1M or 9M rows was that we need to first fetch 40,000 or 360K tuples from 1M or 9M rows of \texttt{Customer} table having NK $=10$. Similarly, we need to first fetch 2800 or 26,800 tuples from 70K or 670K rows of \texttt{Supplier} table having NK $=10$. Finally, we need to execute Cartesian product over $40K\times2800$ and $360K\times26.8K$, which was taking more than 1 hour, which seems infeasible. Note that there is no work on secret-shared join processing that can handle a non-PK/FK join over $40K\times2800$ or $360K\times26.8K$ rows. Thus, designing an efficient algorithm for non-PK/FK join will be an interesting idea in the future.

\begin{figure*}[t!]
		\BB
		\begin{center}
		\begin{minipage}{\linewidth}
				\centering
				\includegraphics[scale=0.33]{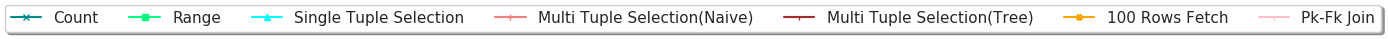}
			\end{minipage}
			\vspace{-1mm}
			\begin{minipage}{.24\linewidth}
				\centering
				\includegraphics[scale=0.29]{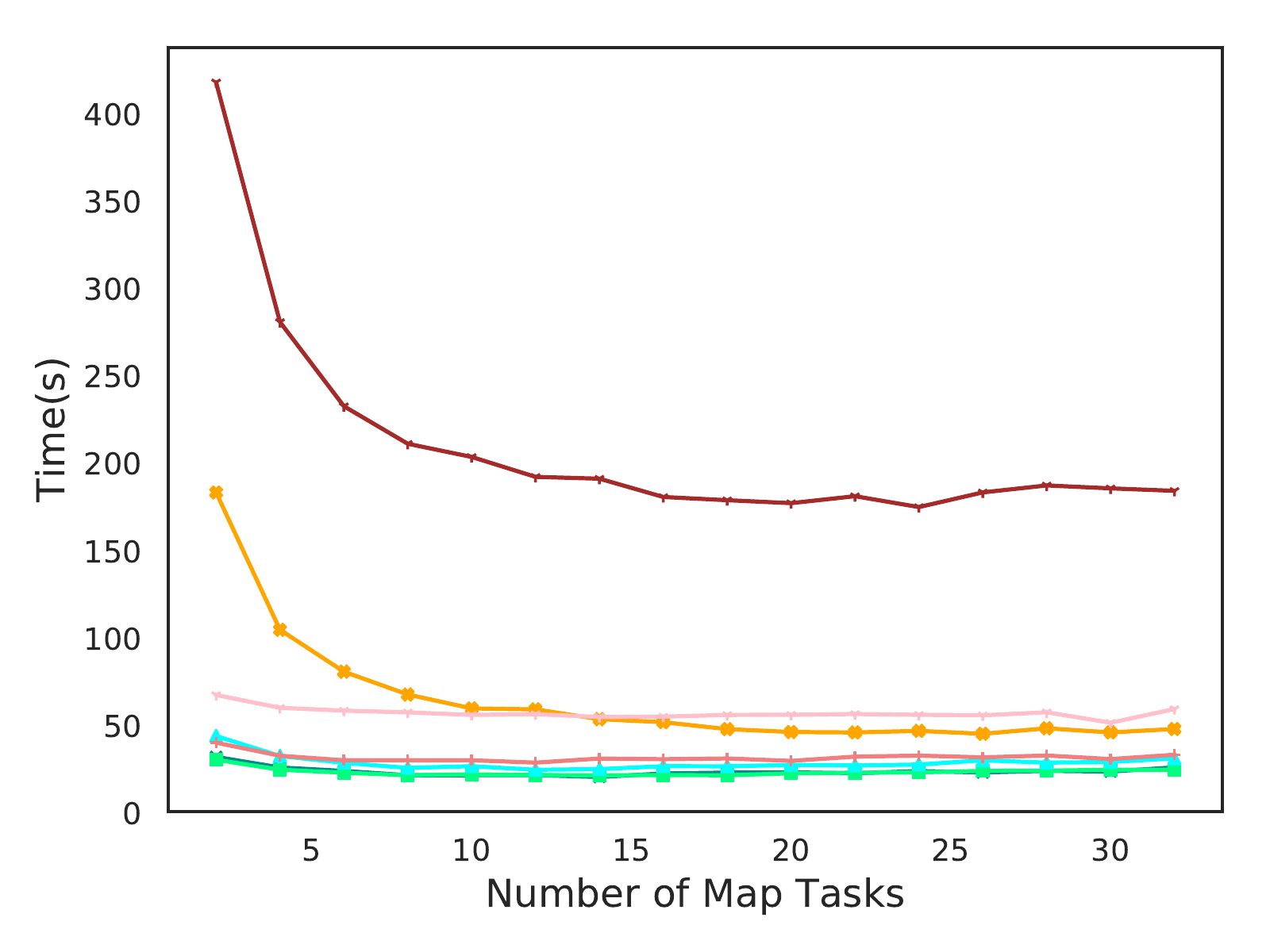}\BB
				\subcaption{Cluster size 1.}
				\label{fig:Cluster size 1}
			\end{minipage}
			\begin{minipage}{.24\linewidth}
				\centering
				\includegraphics[scale=0.29]{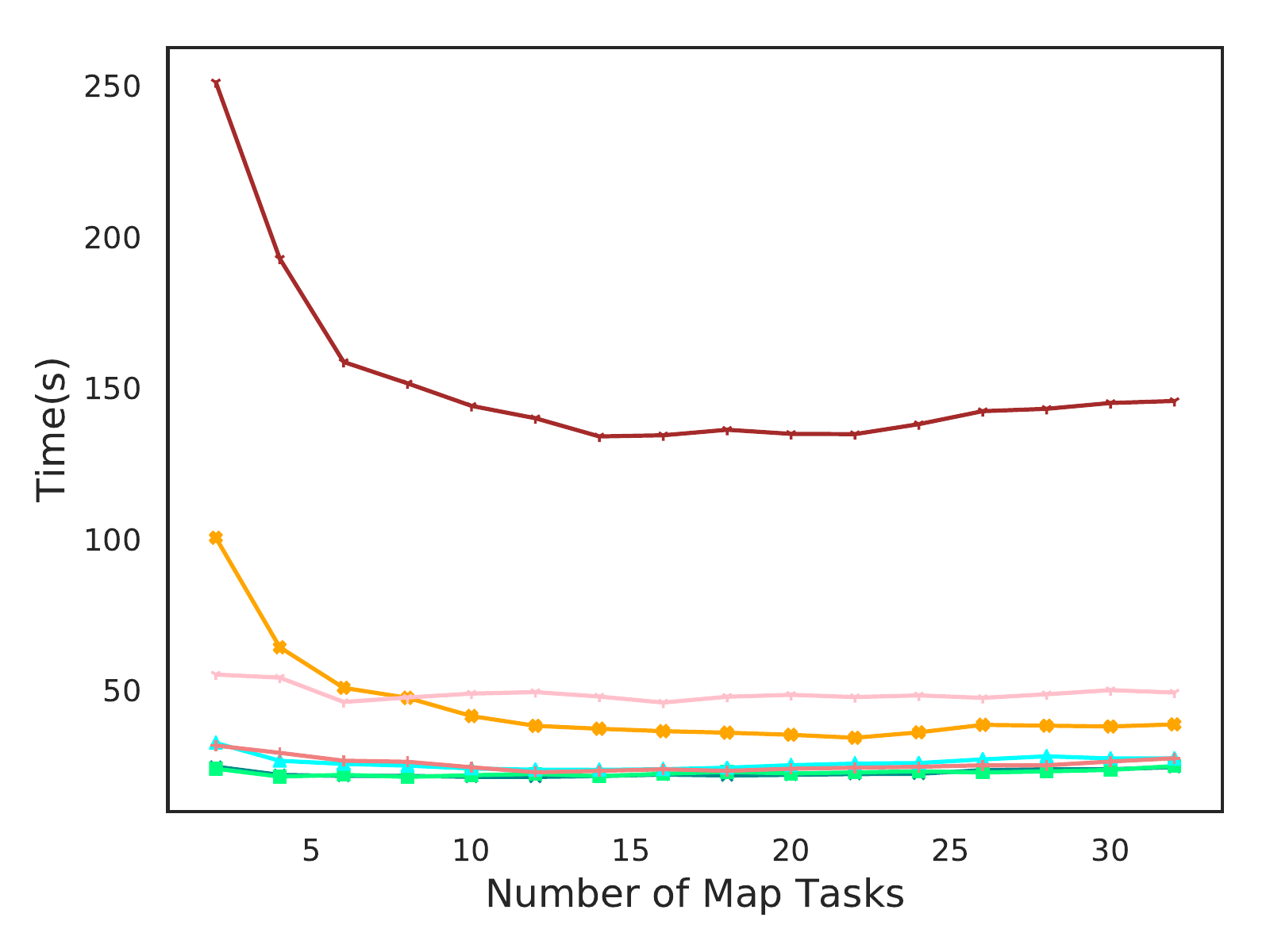}\BB
				\subcaption{Cluster size 2.}
				\label{fig:Cluster size 2}
			\end{minipage}			
			\begin{minipage}{.24\linewidth}
				\centering
				\includegraphics[scale=0.29]{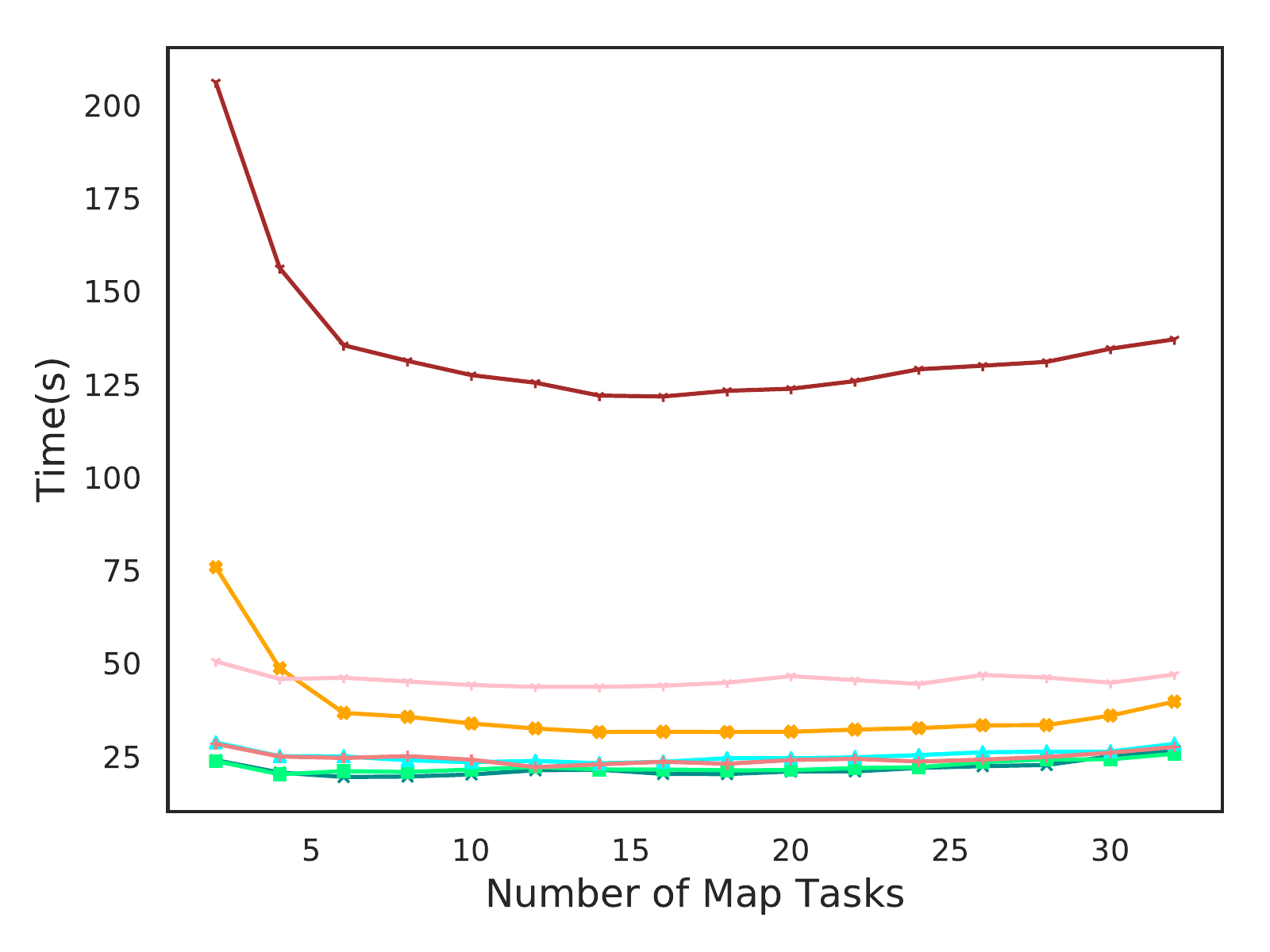}\BB
				\subcaption{Cluster size 3.}
				\label{fig:Cluster size 3}
			\end{minipage}
			\begin{minipage}{.24\linewidth}
				\centering
				\includegraphics[scale=0.3]{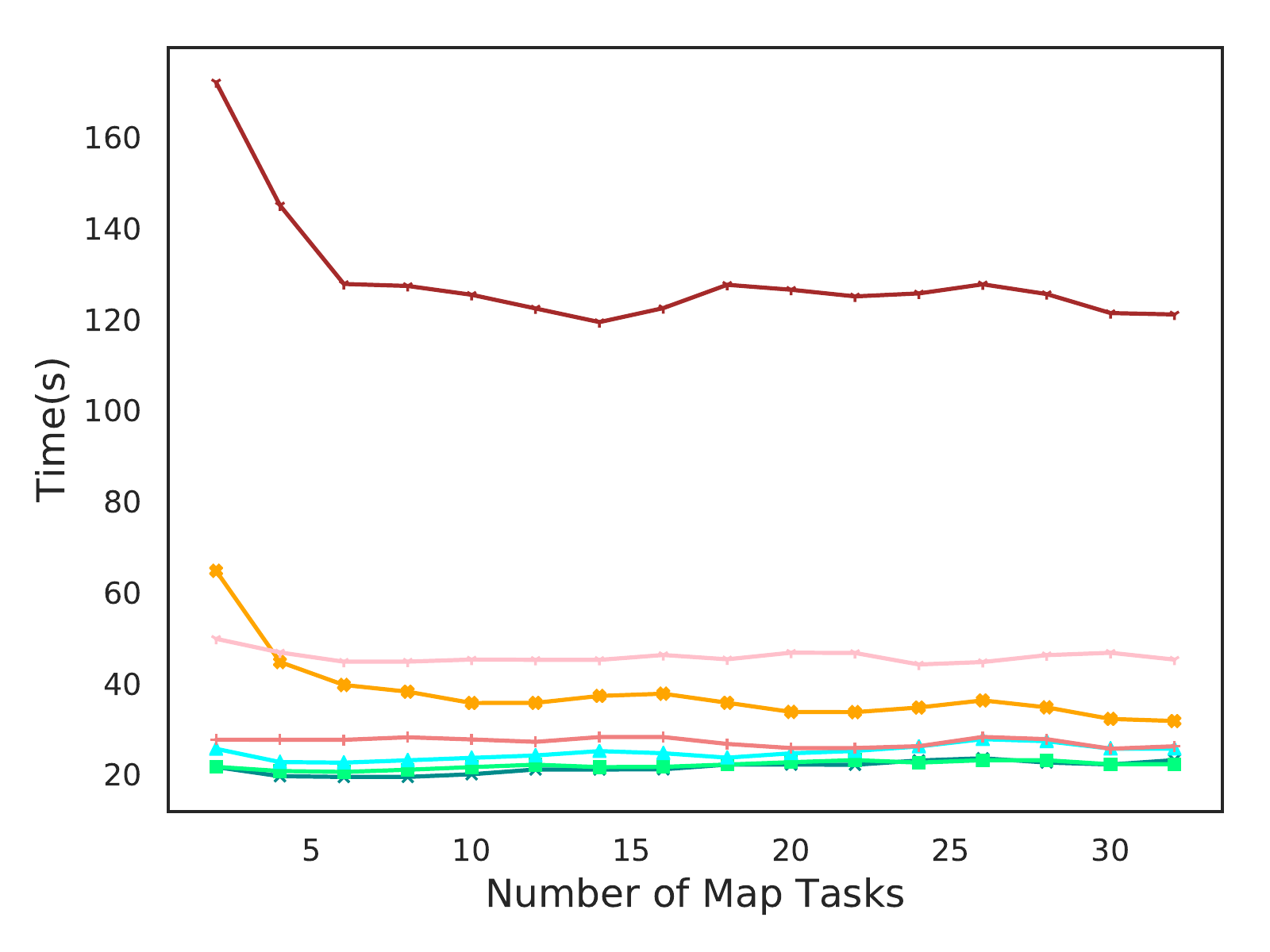}\BB
				\subcaption{Cluster size 4.}
				\label{fig:Cluster size 4}
			\end{minipage}
		
		\end{center}
	    \vspace{-3mm}
		\caption{Experiment 3. Impact of the number of mappers on 1M rows.}
		\label{fig:Impact of the number of mappers on 1m}
	\end{figure*}
	
\begin{figure*}[t]
		\BB
		\begin{center}
		    \begin{minipage}{\linewidth}
				\centering
				\includegraphics[scale=0.33]{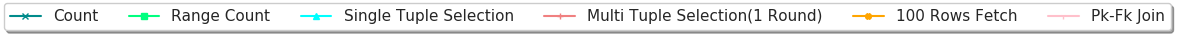}
			\end{minipage}
			\vspace{-1mm}
			\begin{minipage}{.24\linewidth}
				\centering
				\includegraphics[scale=0.29]{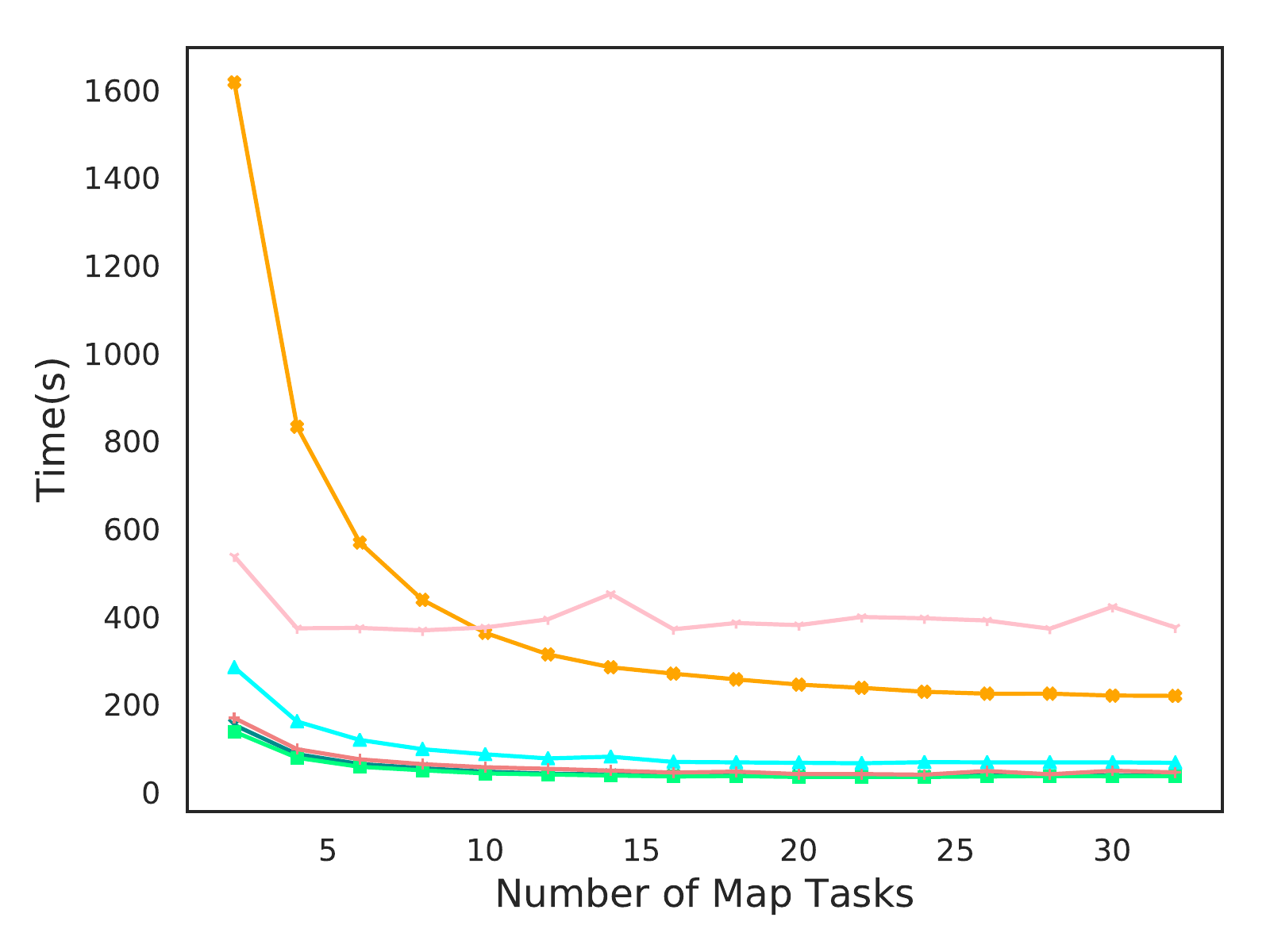}\BB
				\subcaption{Cluster size 1.}
				\label{fig:Cluster size 1}
			\end{minipage}
			\begin{minipage}{.24\linewidth}
				\centering
				\includegraphics[scale=0.29]{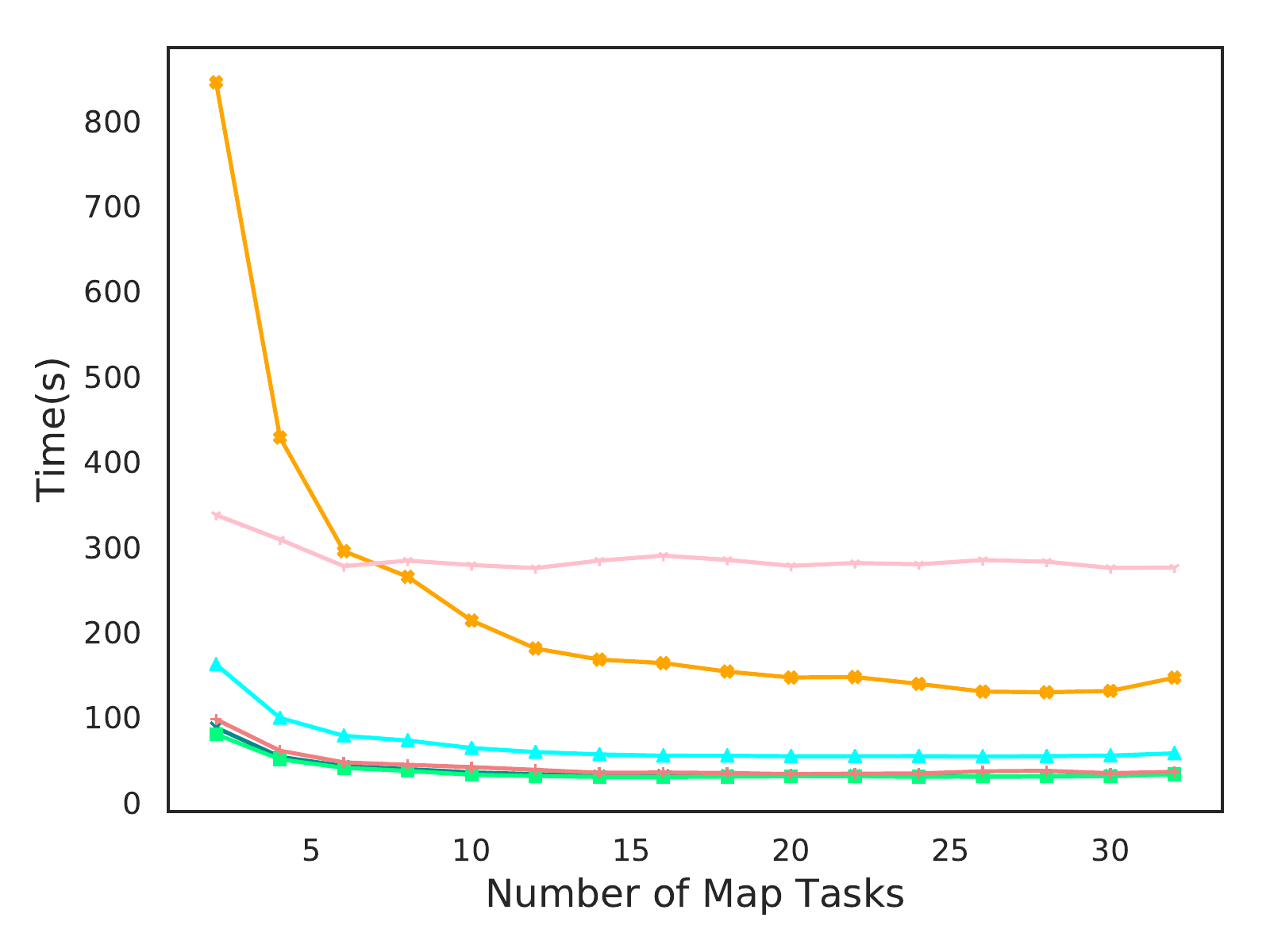}\BB
				\subcaption{Cluster size 2.}
				\label{fig:Cluster size 2}
			\end{minipage}			
			\begin{minipage}{.24\linewidth}
				\centering
				\includegraphics[scale=0.29]{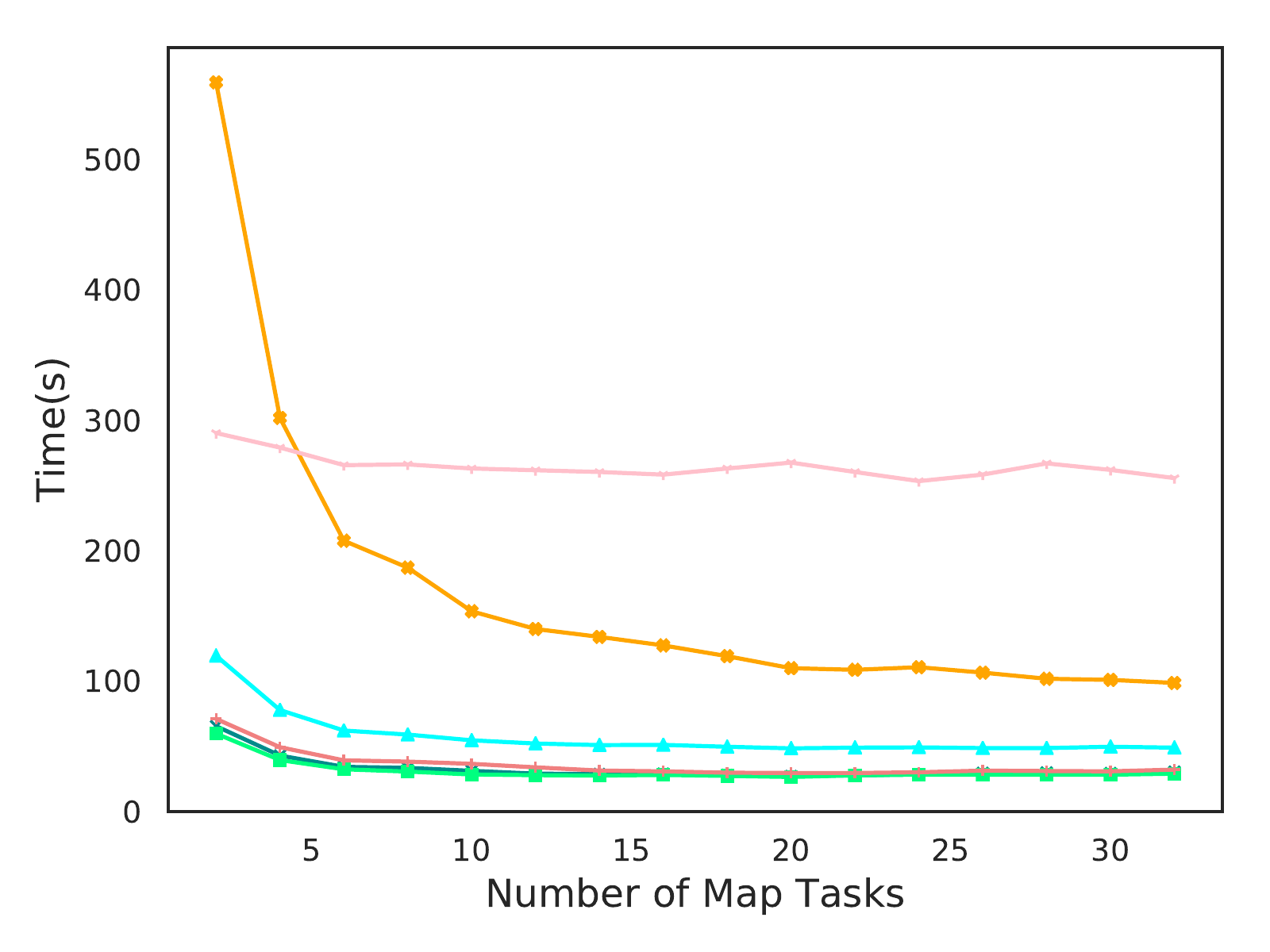}\BB
				\subcaption{Cluster size 3.}
				\label{fig:Cluster size 3}
			\end{minipage}
			\begin{minipage}{.24\linewidth}
				\centering
				\includegraphics[scale=0.3]{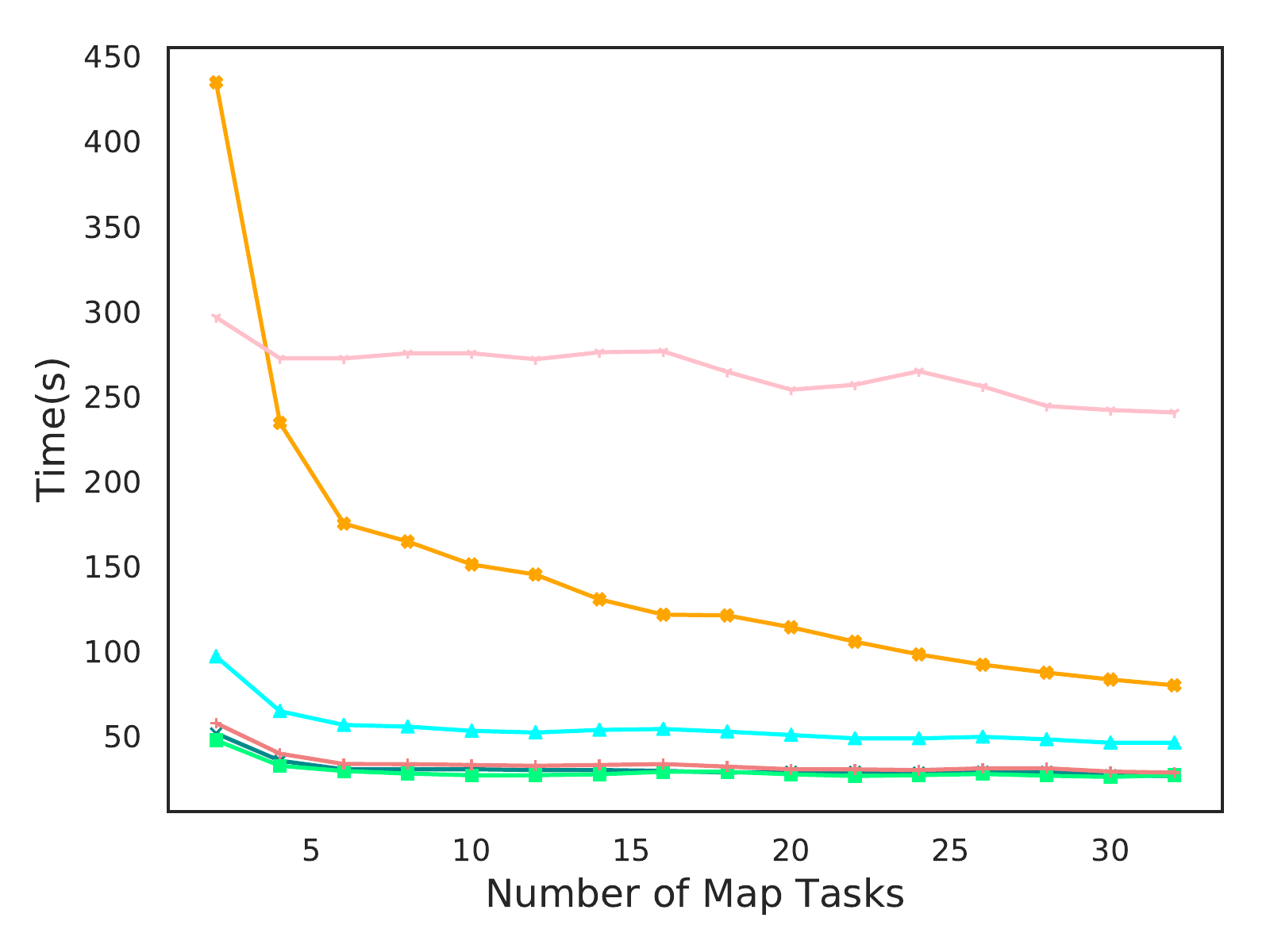}\BB
				\subcaption{Cluster size 4.}
				\label{fig:Cluster size 4}
			\end{minipage}
				\end{center}
			    \vspace{-3mm}
		\caption{Experiment 3. Impact of the number of mappers on 9M rows.}
		\label{fig:Impact of the number of mappers on 9m}
		\BBB\BB
	\end{figure*}

\begin{table}[h]
\BB
  \centering
  \scriptsize
      \begin{tabular}{|l|l|l|}
    \hline
   Customer Rows & Supplier Rows & Time \\ \hline

    10000 & 1000 & 147s  \\ \hline
    20000 & 2000 & 200s  \\ \hline
    30000 & 3000 & 242s  \\ \hline

    \end{tabular}
    \BB
    \caption{Experiment 2: Non-PK/FK join.}
    \label{tab:nonpk-fk}
    \BB
\end{table}

\noindent\textit{\textbf{Range queries.}} We executed range-based count and range-based selection queries on NK column of \texttt{Customer} table. For this query, we outsource NK column of \texttt{Customer} table using binary-representation, as mentioned in \S\ref{subsec:Range Query Execution}. Figure~\ref{fig:fig_scalability} shows the execution time of the range-based count query. Further, the range-based selection query took 51s and 81s for 1M and 9M rows respectively.

\noindent\textbf{Experiment 3: Impact of mappers.} The processing time at the server can be reduced by an increasing number of mappers. Figure~\ref{fig:Impact of the number of mappers on 1m} and Figure~\ref{fig:Impact of the number of mappers on 9m} show the execution time for each query with an increasing number of mappers per node for 1M and 9M rows, respectively. The execution time decreases for all queries (except PK/FK join query) as the number of mappers increases. PK/FK join is a reduce heavy computation, and hence, we will see in the next experiment that as the number of reducers increases, the computation time in the case of join query also decreases. Note that in Figure~\ref{fig:Impact of the number of mappers on 1m} and Figure~\ref{fig:Impact of the number of mappers on 9m}, the execution time does not decrease much after increasing the numbers of mappers per node beyond 20 (also, for the smaller dataset, the execution time starts to increase). This happens because the total execution time reaches close to the time spent in the sequential part of the code and in the disk I/O.

\begin{figure}[!h]
		\B
		\begin{center}
			\begin{minipage}{.48\linewidth}
				\centering
				\includegraphics[scale=.28]{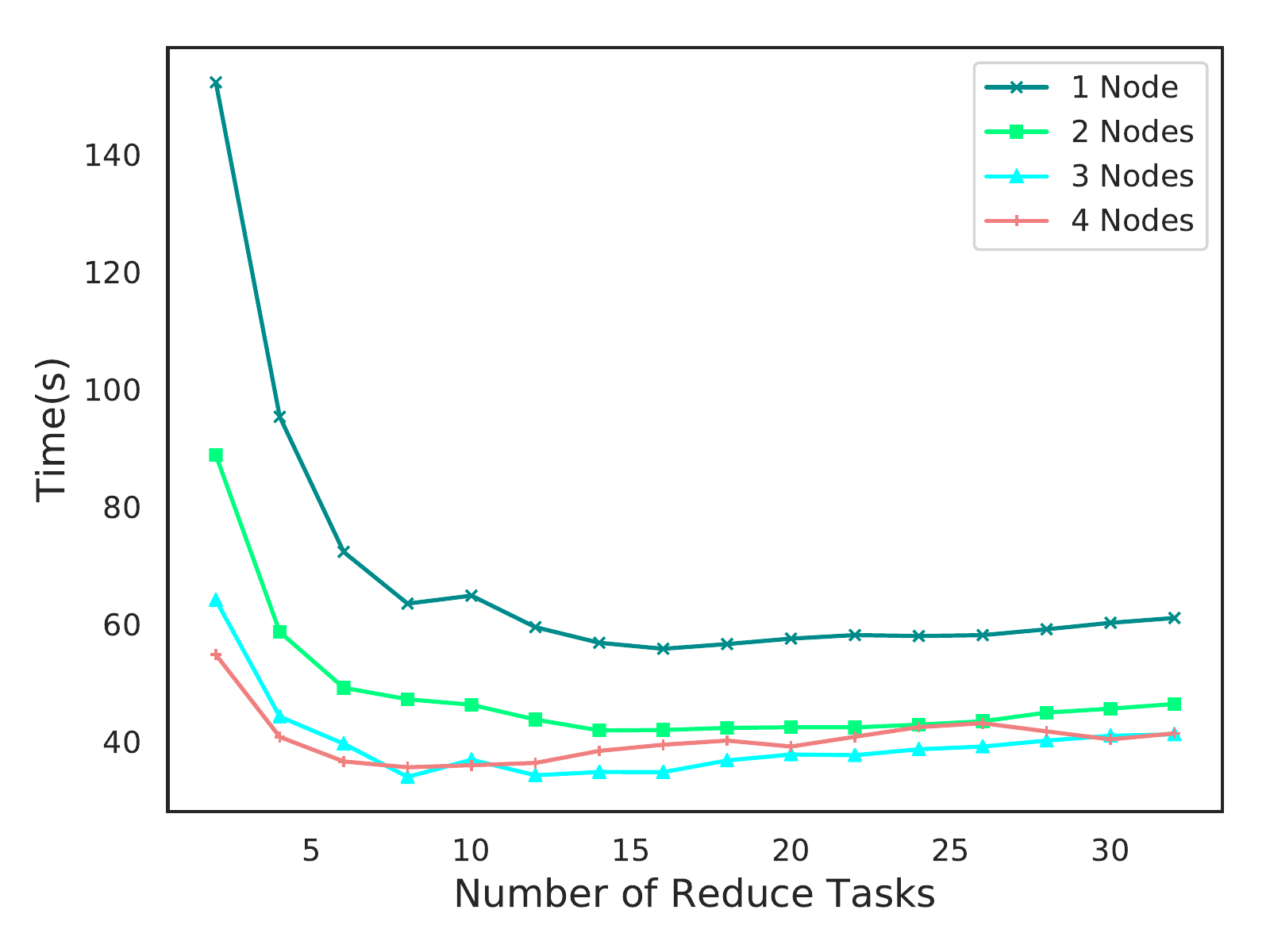}
				\BBB
				\subcaption{1M rows.}
				\label{fig:1M rows reducer size impact}
			\end{minipage}
			\begin{minipage}{.48\linewidth}
				\centering
				\includegraphics[scale=0.29]{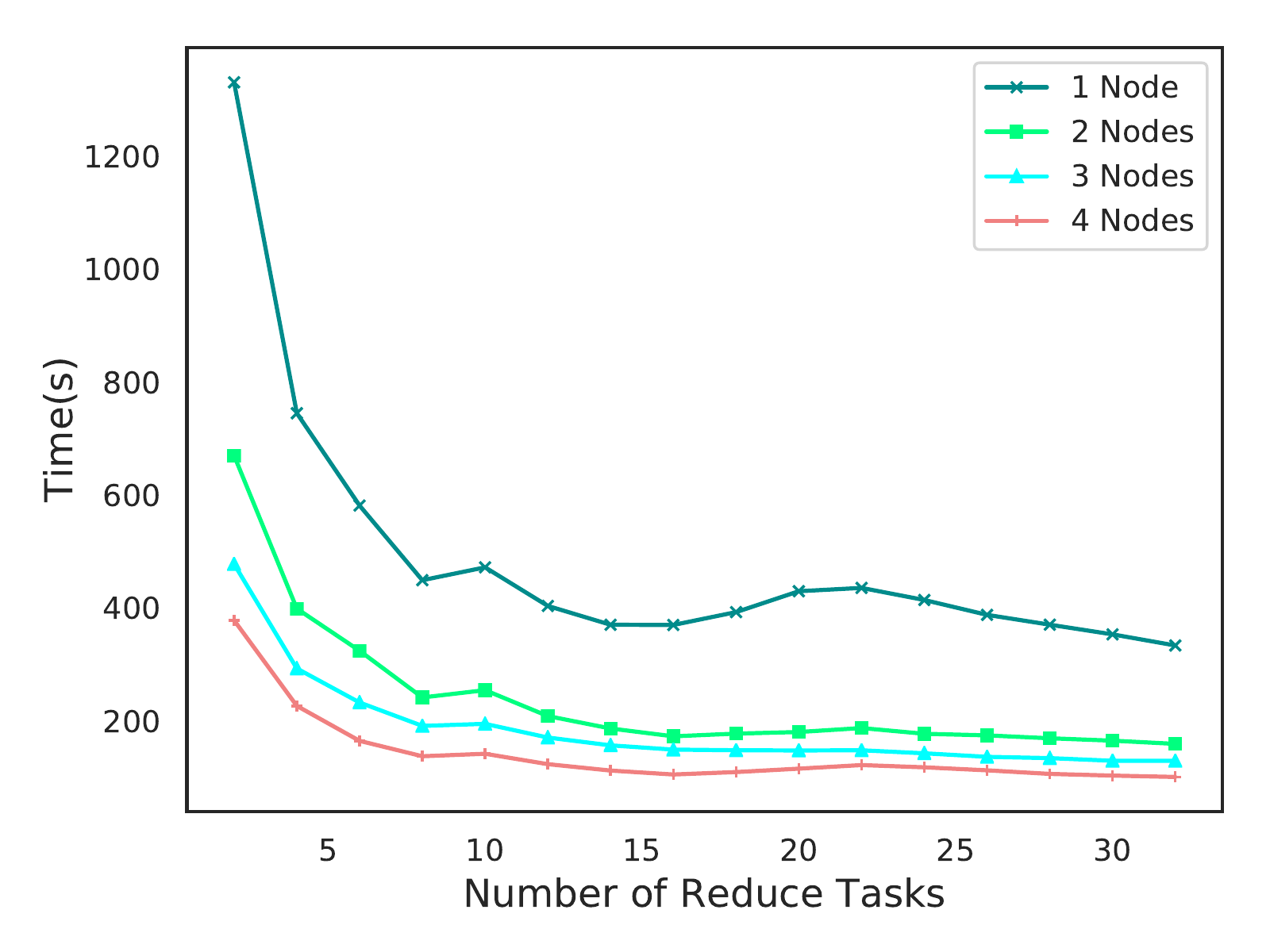}
				\BBB
				\subcaption{9M rows.}
				\label{fig:9M rows reducer size impact}
			\end{minipage}			
			  \end{center}
		\BBB\B
		\caption{Exp. 5. Impact of the number of reducers on PK/FK join.}
		\label{fig:Impact of number of reducers on PK/FK Join}
		\BB
	\end{figure}

\noindent\textbf{Experiment 4: Impact of reducers.} The processing time at the server can be reduced by an increasing number of reducers as well.  Note that for queries that generate a single tuple as an output, \textit{e}.\textit{g}., count, range-based count, and single tuple selection queries, there is only one unique key after the map phase. Thus, all the outputs of the map phase are assigned to one reducer only. However, in the case of a join query, reducers execute multiplication and addition operations over all the tuples of both the relations to compute the join outputs. Therefore, the execution time decreases as the number of reducers increases; see Figure~\ref{fig:Impact of number of reducers on PK/FK Join}. However, in the case of 1M rows, the execution time starts to increase after increasing the number of reducers more than 16, due to the increased overhead for maintaining multiple reducers per node.

\begin{figure}[!h]
\BB
			\begin{center}
		\begin{minipage}{\linewidth}
				\centering
				\includegraphics[scale=0.3]{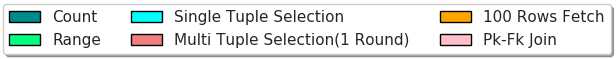}
			\end{minipage}
			\vspace{-1mm}
			\begin{minipage}{.48\linewidth}
				\centering
				\includegraphics[scale=.28]{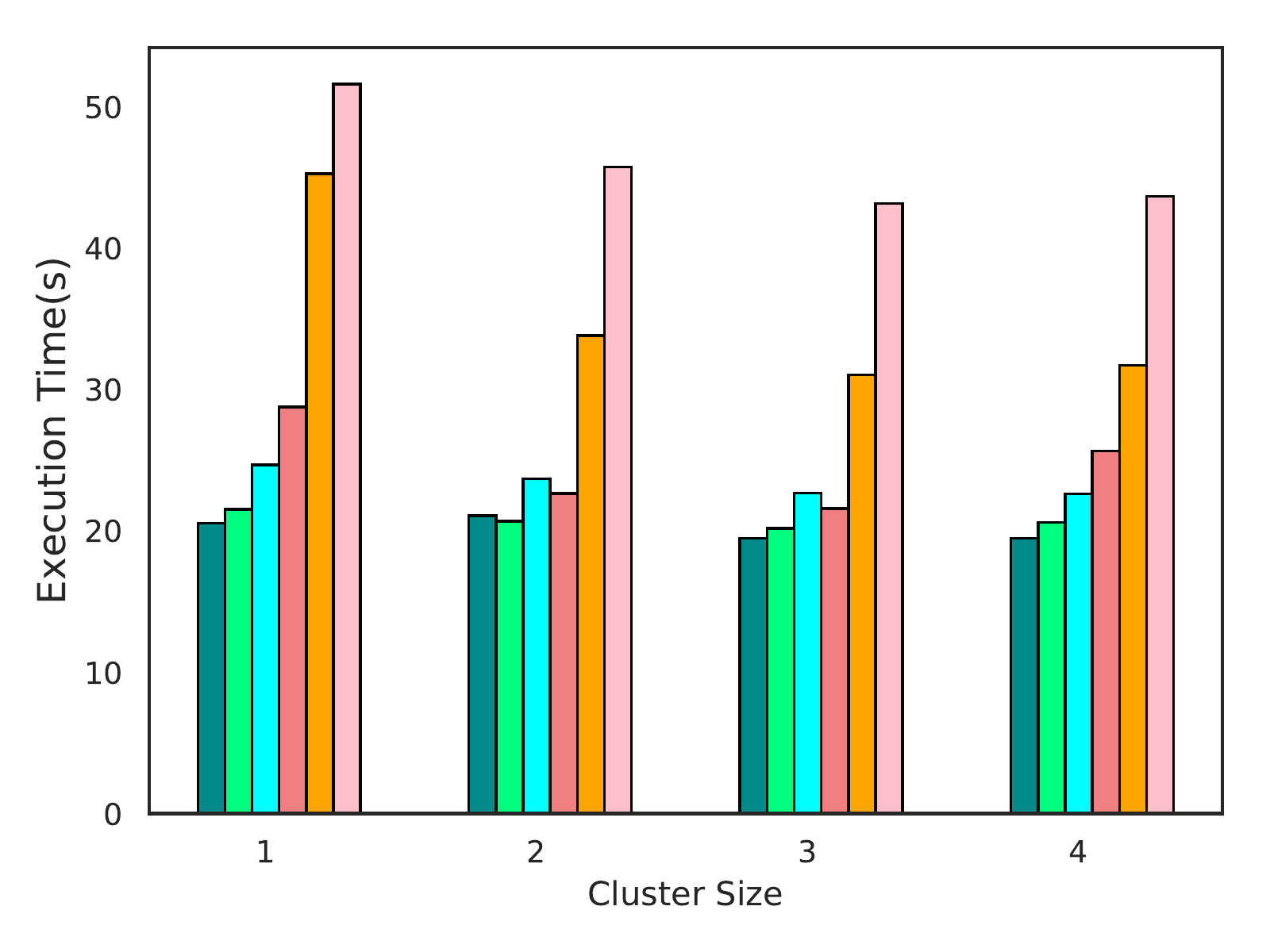}
				\BBB
				\subcaption{1M rows.}
				\label{fig:1M rows cluster size impact}
			\end{minipage}
			\begin{minipage}{.48\linewidth}
				\centering
				\includegraphics[scale=0.29]{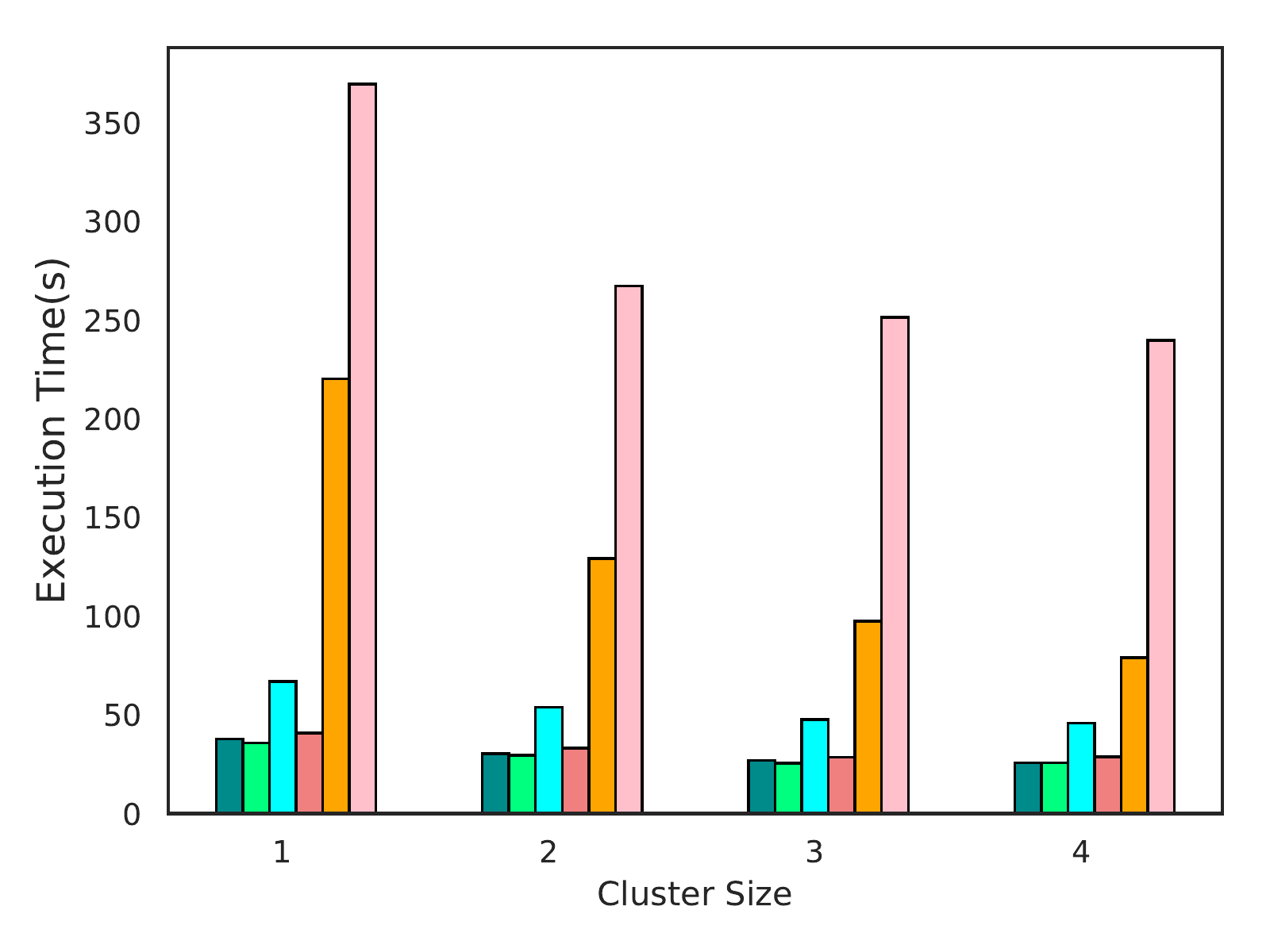}
				\BBB
				\subcaption{9M rows.}
				\label{fig:9M rows cluster size impact}
			\end{minipage}			
			  \end{center}
		\BBB
		\caption{Experiment 5. Impact of the cluster size.}
		\label{fig:Impact of the cluster size}
		\BB
	\end{figure}

\noindent\textbf{Experiment 5: Impact of cluster size.} The processing time at the server can be reduced by an increasing the size of a cluster. Figure~\ref{fig:Impact of the cluster size} shows the server processing time of various queries with increasing cluster size. An increase in the cluster size allows us to increase the total number of mappers (or reducers), thereby achieving  more parallelism. Also, an increase in the cluster size reduces I/O time, due to more available disks that support parallel read and write. Observe that the execution time consistently decreases as the cluster size increases; however, there is not much reduction in the execution time between a cluster of size three and a cluster of size four on 1M rows of \texttt{Customer}, since the increase in parallelism is not able to reduce the execution time further as the total execution time reaches the time spent in sequential part of the program.

\begin{figure}[!h]
\B
\centering
	\includegraphics[scale=0.35]{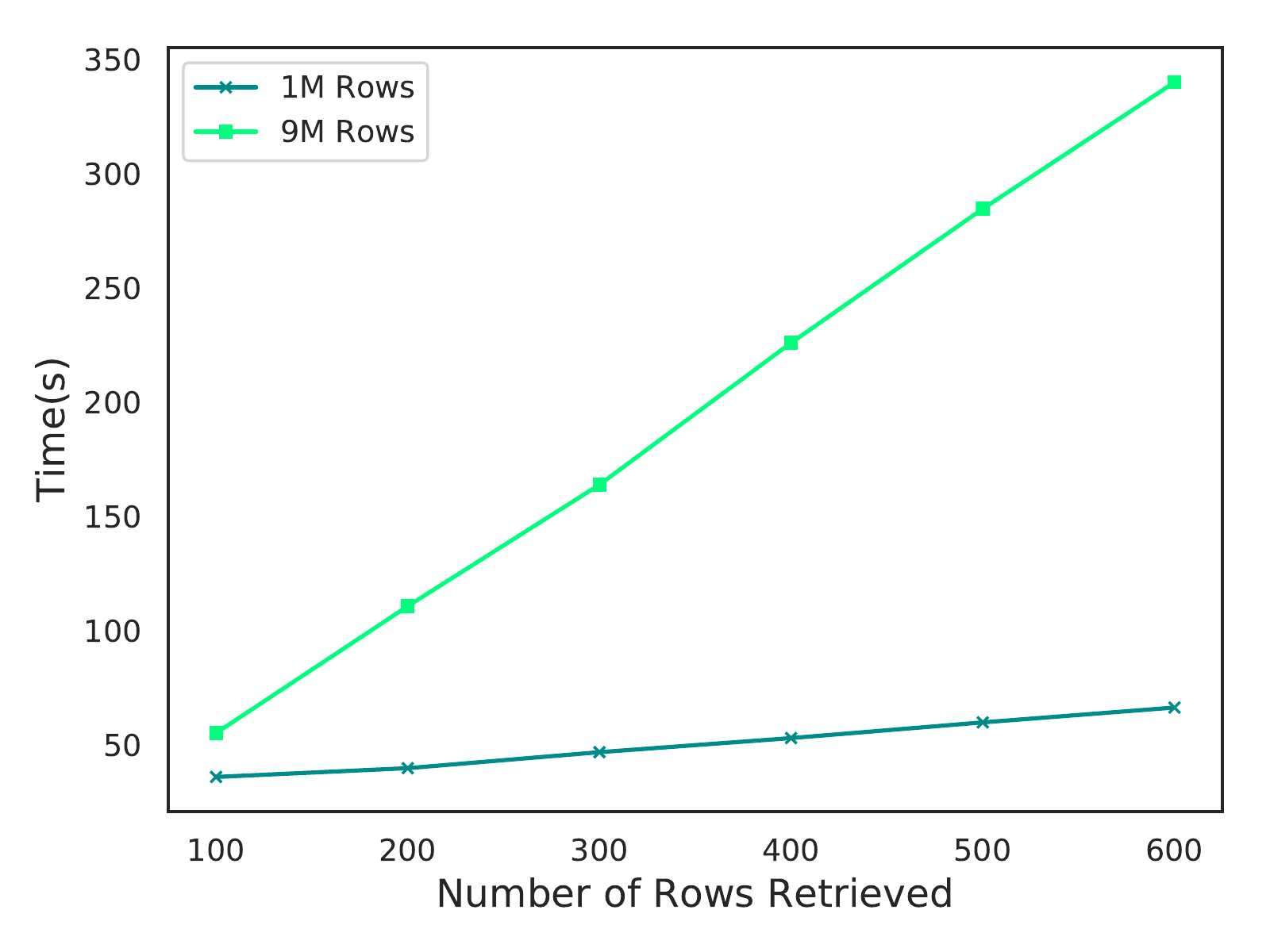}
	\BB
\caption{Exp. 6. Time for retrieving a different number of tuples.}
\label{fig:multiple_tuple_fetch}
\BBB
\end{figure}

\noindent\textbf{Experiment 6: Multi-tuple fetch.} We also measured the computation time at servers to fetch multiple tuples; see Figure~\ref{fig:multiple_tuple_fetch}. Note that as we increase the number of tuples to be retrieved, the computation time also increases, due to an increasing number of multiplication and addition operations involved in string-matching operations.

\noindent\textbf{Experiment 7: Reducing the size of secret-shared data using hashing.} As mentioned in \S\ref{subsec:Data Model Creation and Distribution of Secret-Shares of a Relation}, as the data contains long non-numerals, it results in an increased size of secret-shared dataset. For comparing secret-shared data of non-numerals with and without using a hash function, we used Big-Data Benchmark~\cite{bbd} \texttt{Ranking} table having 360K tuples and three attributes, namely URL, PageRank, and Duration. URL attribute can be very long with an average length of 55 characters. The use of unary representation for creating secret-shares of long URLs increases the dataset size significantly; see Table~\ref{tab:hashing}. Thus, to reduce the size of secret-shared data, we used hashing before creating secret-shares of URL attribute. We selected SHA1~\cite{eastlake2001us} cryptographic hash function to hash URL strings. Since \texttt{Ranking} table has 360K unique URLs, a smaller length hashed value is sufficient to prevent a collision. Thus, we used only the last eight digits of the hashed URL. As a result, the size of the secret-shared data reduced from 4GB to 695MB after implementing the above-mentioned hashing technique. Furthermore, we executed a count query on the hashed data that shows less time as compared to executing the same query on non-hash data, since executing operations on non-hashed data require more number of multiplications and additions as compared to the hashed data.

\begin{table}[h]
\BB
\centering
\scriptsize
\begin{tabular}{|l|l|l|}\hline
~    & Hashing & Without Hashing \\ \hline
Secret-shared dataset size & 685MB & 4GB  \\ \hline
Count query (4 mappers) &  34s &  110s  \\ \hline
Count query (16 mappers) & 18s & 25s  \\ \hline
    \end{tabular}
    \BB
    \caption{Experiment 7. Impact of hashing.}
    \label{tab:hashing}
    \BB
\end{table}

\noindent\textbf{Experiment 8: Impact of communication between user and servers.} We measured the communication impact on multi tuple selection queries. For the count query, the output is just one number, so the performance is not impacted by network communication. In the case of a selection query based on one-round algorithm, the user fetches the entire NK column of \texttt{Customer} table from three servers. The size of NK column was 7.9MB in 1M rows and 71MB in 9M rows of \texttt{Customer} table. When using slow (100MB/s), medium (500MB/s), and fast (1GB/s) speed of data transmission, the data transmission time in case of 1M and 9M rows was negligible on all transmission speeds.

\noindent\textbf{Experiment 9: User computation time.} In the proposed algorithms, the interpolation time in the case of count, range-based count, and single tuple selection queries was less than 1s. However, in the case of a selection query using one-round algorithm that requires interpolating all the values of the entire NK column, the user-side interpolation time was 1s on 1M rows and 7s on 9M rows. Recall that we are fetching at most 600 rows (Experiment 6 Figure~\ref{fig:multiple_tuple_fetch}), and therefore, the user needs to generate shares corresponding to at most 600 $\mathit{RID}$ values, which took less than 1s. Further, as a result of the selection query, the user interpolated 2400 values (600 rows, where each row with four attribute values), and it took again less than 1s. Thus, we can see that the user computation time during the entire selection query was less than 2s on 1M and 8s on 9M rows. In the case of the tree-based algorithm, the user needs to interpolate 8200, 31,720, 48,644 and 4846 numbers, for rounds with block sizes 128, 32, 8 and 2 respectively when the data contains 1M rows. With 9M rows the user needs to interpolate 78,516, 303,496, 462,548 and 45,302 numbers for rounds with sizes 128, 32, 8 and 2, respectively. The user spends at most 0.2s for 1M rows and 1.6s for 9M rows on interpolation in any of the rounds. Observe that from this experiment it is clear that the user processing time is significantly less than the server processing time (given in Figure~\ref{fig:fig_scalability}).

\noindent\textbf{Experiment 10: Comparing against downloading scheme.} We also evaluated the queries for the case when a user downloads the entire data, decrypts it, and then, executes a query. Decrypting the completely non-deterministically encrypted \texttt{Customer} table took 24s and 238s for 1M and 9M rows, respectively, while loading into MySQL database took 4s and 35s, respectively. Similarly, decrypting the entirely non-deterministically encrypted \texttt{Nation} table of 25 rows and loading it into MySQL took less than 1s. On these datasets, executing any query took at most 2s and 12s for 1M rows and 9M rows, respectively (PK/FK join query took the highest execution time). Note that it indicates that our proposed algorithms at public servers are significantly better than a trivial way of executing a query after downloading and decrypting the data. Observe that the maximum computation time by our proposed count, MTC-1-Round selection, and PK/FK-based join algorithms was 56s on 1M and 239s on 9M rows; see Figure~\ref{fig:fig_scalability}.

\BB
\section{Leakage Analysis}
\label{sec:Leakage Analysis}
\B
\noindent This section provides a discussion on side-information leakages at the servers due to query execution. We provide the adversarial objective with respect to different queries, and then, discuss leakages due to either query execution access-patterns (\textit{i}.\textit{e}., the way tuples are scanned) or the output-size.

\BB
\subsection{Count Query}
\label{subsec:leakage_count}
\B
In the count query, the adversarial objective is to know the frequency-count of each value, \textit{i}.\textit{e}., the number of tuples having the same value in an attribute of the relation, by observing the query execution or output-size.

\noindent\textit{Query execution based leakage.} The secret-shared count query predicate is different from the secret-shared values of the desired attribute on which the count query will be executed. Hence, the adversary cannot deduce by just looking at the secret-shared query and attribute values that which attribute value is identical to the query predicate. In addition, the adversary cannot distinguish two count queries, due to different polynomials used to create secret-shares of two count queries. Also, the string-matching operation is executed on all the values of the desired attribute, hence, access-patterns are hidden.

\noindent\textit{Output-size-based leakage.} The output of any count query, which is of secret-shared form, is identical in terms of the number of bits in our setting; hence, the adversary cannot know the exact count and differentiate two count queries based on the output-sizes.

\BB
\subsection{Selection Query}
\label{subsec:leakage_Selection Query}
\B
This section discusses leakages due to a selection query. Note that, likewise a count query, in a selection query, the adversary cannot deduce which attribute value is identical to the query predicate by just observing query or attribute values. Also, the adversary cannot distinguish two selection query predicates, due to different polynomials, used to create secret-shares. As we proposed different algorithms for selection queries, below we discuss information-leakage involved in each algorithm.

\subsubsection{One Value One Tuple}
\label{subsubsec:leakage_One Value One Tuple}
In the case of one tuple per value based selection query, the adversarial objective is to learn which tuple satisfies the query.

\noindent\textit{Query execution based leakage.} Servers perform identical operations (\textit{i}.\textit{e}., string-matching on secret-shared values against the query predicate and multiplication of the resultant with all  values) on each tuple of the relation, and finally, add all secret-shares of each attribute. Hence, access-patterns are hidden and so obscuring the tuple satisfying the query predicate from the adversary.

\noindent\textit{Output-size-based leakage.} The output of the selection query, which is of the form of secret-shares, is identical in terms of the number of bits in secret-sharing, like the count query; hence, the adversary cannot learn the desired tuple based on the final secret-shared tuple sent to the user.

\subsubsection{Multiple Values with Multiple Tuples}
\label{subsubsec:leakage_Multiple Values with Multiple Tuples}
In the case of multiple tuples satisfying the selection query predicate, the adversarial objective is to deduce which tuples satisfy the query based on the access-patterns and the number of tuples corresponding to the query.

\noindent\textbf{Leakage discussion of one-round algorithm.}

\noindent\textit{Query execution based leakage.} The server performs identical operations on all the tuples of the relation (\textit{i}.\textit{e}., (\textit{i}) string-matching operations on secret-shared attribute values and the selection query predicate in the first round to know the desired tuple ids, and (\textit{ii}) string-matching operations on secret-shared tuple ids, $\mathit{RID}$, in the second round to transmit the desired secret-shared tuples). Hence, access-patterns are hidden, and so, the adversary cannot know which tuples satisfy the selection query predicate.

Note that for fetching multiple tuples, the user sends a secret-shared vector of some length, say $\ell$. However, by just observing the vector, the adversary cannot learn that which $\ell$ tuples are satisfying the query, since an $i^{\mathit{th}}$ element of the vector does not indicate that the $i^{\mathit{th}}$ tuple is required by the user.

\noindent\textit{Output-size-based leakage.} The number of output tuples (or the output-size), however, can reveal the number of tuples satisfying the selection query, which was hidden from the adversary before the query execution. Here, if the adversary is aware of selection predicates based on background knowledge, then the adversary can execute frequency-count analysis, due to the number of returned tuples. However, knowing a secret-shared selection predicate is not a trivial task. A trivial way to overcome such an attack is to download the entire database and execute the query at the trusted user. However, such a solution incurs communication and computation overheads. To the best of our knowledge, there is no solution to prevent output-size attacks except sending fake tuples to the user. Such solutions work for users who are willing to pay a little bit higher communication cost, while preserving a desired level of privacy. For example, Opaque~\cite{opaque} and ObliDB~\cite{DBLP:journals/corr/abs-1710-00458} that send fake data to prevent output-size attacks.

We also adopt the same solution and request servers to send more tuples than the desired tuples (in an oblivious manner). Recall that the user sends a vector, having secret-secret row-ids, to the server (\S\ref{subsec:Unary Occurrence of a Pattern}). To prevent the output-size-based attack, the user sends a vector of an identical length having some fake row-ids, regardless of the number of tuples satisfying the selection query predicate. Thus, the user retrieves an identical number of tuples, regardless of the selection query predicate. Hence, the adversary cannot learn which tuples satisfy the query, due to preventing output-size attacks. In particular, due to hiding access-patterns, an $i^{\mathit{th}}$ element of the vector does not reveal that either the $i^{\mathit{th}}$ tuple is required by the user or the $i^{\mathit{th}}$ element of the vector is a fake value (row-id).

\smallskip\noindent\textbf{Leakage discussion of tree-based algorithm.}

\noindent\textit{Query execution based leakage.}
The server performs identical operations, \textit{i}.\textit{e}., counting the tuples satisfying the selection predicate in \textsc{Phase} 0, partitioning the relation into certain blocks in \textsc{Phase} 1, and transmission of tuples in \textsc{Phase} 2 (using the algorithm for one tuple per value \S\ref{subsec:Unary Occurrence of a Pattern}), on all the tuples of the relation. Hence, access-patterns are hidden, obscuring the selection predicate and the tuple containing the predicate from the adversary.

In the tree-based search algorithm, the user requests servers to partition the blocks into sub-blocks; see \S\ref{subsubsec:Multiple Occurrences of a Pattern}. However, to improve efficiency, the user may request the server not to partition the blocks satisfying either Case 1 (a block containing no tuple satisfying the selection query predicate, $p$), Case 2 (a block containing one/multiple tuples but only a single tuple contains $p$), or Case 3 (a block containing $h$ tuples, and all the $h$ tuples contain $p$); see \S\ref{subsubsec:Multiple Occurrences of a Pattern}. Thus, the user requests the server to partitions those blocks that contain some $h$ tuples, while all the $h$ tuples do not have $p$. Consequently, the adversary can deduce that which blocks contain the desired tuples and which blocks do not contain the desired tuples, without knowing the actual tuples. Thus, we suggest partitioning all the blocks until we know the row-ids of all the desired tuples or each block satisfies either Case 2 or Case 3. Of course, it incurs the computational cost, as studied in \S\ref{sec:Experimental Evaluation}.

\noindent\textit{Output-size-based leakage.} This discussion is identical to the discussion presented above for one-round algorithm.

\BB
\subsection{Join Queries}
\label{subsubsec:leakage_Join Queries}
\B
\S\ref{subsec:Join Query Execution} discussed PK/FK and non-PK/FK join algorithms. Here we discuss information-leakages due to both algorithms.

\noindent\textbf{Leakage discussion of PK/FK join.}

\noindent\textit{Query execution based leakage.}
Each mapper performs identical operations, \textit{i}.\textit{e}., generating key-value pairs of each tuple. Each reducer also performs identical operations, \textit{i}.\textit{e}., string-matching on secret-shared values of the joining attribute and multiplication of the resultant by each tuple, and finally adds all the secret-shares of each attribute. In addition, the dataflow between mappers and reducers is identical, \textit{i}.\textit{e}., the number of key-value pairs produced by a mapper and the number of key-value pairs allocated to a reducer are same. Therefore, access-patterns are hidden from the adversary, preventing the adversary from knowing which tuples of the parent relation join with which tuples of the child relation. However, the adversary can only know the identity of joining attribute. For example, if $R(A,B)$ and $S(B,C)$ are two joining relations, where $B$ is the joining attribute, then the adversary will only know that the second attribute of $R$ and the first attribute of $S$ are the joining attributes. Note that the adversary will not learn any value of the joining attribute, since they are secret-shared.

In most of the cases, the selection operation is pushed before the join operation to faster the query processing may reveal some information based on output-sizes of the selected tuples. To prevent leakages from selection query, we first select real tuples with some fake tuples, as mentioned in~\S\ref{subsubsec:leakage_Multiple Values with Multiple Tuples}, and then, execute the join algorithm on both real and fake tuples.

\noindent\textit{Output-size-based leakage.} The outputs of the reducers,
which is of the form of secret-shares, is identical in terms
of the number of bits; hence, the adversary cannot deduce any relationship between any two tuples of the two relations.

\noindent\noindent\textbf{Leakage discussion of non-PK/FK joins.}

\noindent\textit{Query execution based leakage.} In the oblivious non-PK/FK equijoin approach, the servers may know the number of joining tuples per value in both relations and the number of joining values by observing data transmission. Our approach can prevent the adversary from knowing the number of joining tuples per value and the number of joining values, by executing the join algorithm on some extra fake tuples, which actually do not produce any real output while preventing the information leakage. Particularly, for each joining value, the first layer servers obliviously send an identical number of tuples to the second layer servers. Thus, the adversary cannot know an exact number of joining tuples per value. Further, the same operation can be carried out for any fake joining value to prevent revealing the number of joining values. However, both the steps incur computational and communication overheads.

\noindent\textit{Output-size-based leakage.} Naive execution of non-PK/FK algorithm, \textit{i}.\textit{e}., without executing algorithm on fake tuples per joining value, provide a rough estimate to the adversary to know the number of tuples having common joining values in both the relation, due to different-sized outputs. However, as mentioned above, executing non-PK/FK for fake tuples can completely overcome output-sized-based attacks.

\BB
\subsection{Range Query}
\label{subsec:leakage_Range Query}
\B
Existing encryption-based~\cite{DBLP:conf/sigmod/AgrawalKSX04} or order-preserving secret-shared~\cite{DBLP:journals/isci/EmekciMAA14} approaches for a range search reveal the order of the values. Since we are outsourcing each value of the form of secret-shares created by using different polynomials, our approach does not reveal any order of the value, unlike~\cite{DBLP:conf/sigmod/AgrawalKSX04,DBLP:journals/isci/EmekciMAA14}.

\noindent\textit{Query execution based leakage.} The server executes an identical operation, regardless of range values. Hence, the adversary cannot learn the exact range values and the tuples satisfying the range query by observing string-matching operations, which hide access-patterns.

\noindent\textit{Output-size-based leakage.} In range-based count queries, the adversary cannot estimate the number of qualified tuples to range query, since the output for each range-based count query is identical. However, in the case of range-based selection queries, the adversary may estimate the query range based on the output-size and background knowledge. For example, consider an employee relation and two queries: (\textit{i}) find details of employees having age between 18-21, and (\textit{ii}) find details of employees having age between 22-75. The output-sizes of the two queries are significantly different. Hence, the adversary can trivially distinguish such two range queries. However, to the best of our knowledge, there is no technique to prevent such an attack without executing range-based selection query for fake values, and then, filter non-desired tuples at the user, as we suggested in a simple selection query (\S\ref{subsubsec:leakage_Multiple Values with Multiple Tuples}).

\BB
\subsection{Information Leakage to the User}
\label{subsec:Information Leakage to the User}
\B
In our selection queries for multiple tuples and non-PK/FK join algorithms, the user knows the addresses of the qualifying tuples to the query. Since users are trusted (as per our assumption), this knowledge does not violate data privacy.

There are two models that can support untrusted users, at different leakages/assumptions. In the first model, all the queries are submitted to the DB owner, who executes the query at the servers on behalf of the user, and then, returns the exact answers to the user. Here, the user will not learn anything except the desired output; however, the DB owner learns the user query. Unlike this model, in our model, the DB owner will not learn the query.

In the second model, a trusted proxy is deployed at a different server than the servers hosting secret-shares. The proxy replaces the role of the DB owner for answering queries (unlike the first model) and provides the exact answers to untrusted users.\footnote{The communication channels between the proxy and the majority of the servers follow the same restriction like the communication channels between the user and the majority of the server, as mentioned in \S\ref{subsec:Adversarial Settings}.} \textit{Note that having a trusted proxy does not change any of our proposed algorithms.} Further, note that the assumption of having a trusted proxy at the public servers is also considered in the previous systems such as Arx~\cite{DBLP:journals/iacr/PoddarBP16} and PPGJ~\cite{DBLP:journals/scn/MaYZ14}. Observe that the second model, where the trusted proxy learns the user query,  is similar to the first model from the data privacy perspective. To the best of our knowledge, there is no work on secret-sharing that assumes untrusted users that can execute their queries at the server without involving either the DB owner or a trusted proxy.

To completely prevent any type of information leakage to the user, we may use an approach, where the database owner outsources unique random \texttt{RID} values (in a range 1 to $x$, where $x\geq n$) of secret-shared form (instead of sequential \texttt{RID} values as shown in Figure~\ref{fig:database1}), along with a permutation function. After matching the query predicate over each $i^{\mathit{th}}$ ($1\leq i\leq n$) value of the desired attribute, the server multiplies the $i^{\mathit{th}}$ string-matching resultant with $i^{\mathit{th}}$ \texttt{RID} value, and it results in a vector of length $n$. Then, the server permutes all the $n$ values of the vector and sends them to the user. After interpolation, the user obtains a vector of $n$ values that contains the desired \texttt{RID} values with some zeros. Now, the user creates secret-shares of the \texttt{RID} values with some fake \texttt{RID} values that are larger than $x$ to retrieve the desired tuples. Thus, in this case, the user will also not learn anything about the position of the desired tuples.

\BB
\section{Complexity Analysis}
\label{sec:Complexity Analysis}
\B
This section presents the complexity analysis of different algorithms. For all the following theorems, we use the following notations: Let $n$ and $m$ be the number of tuples and attributes in a relation, respectively. Let $w$ be the maximum bit length (according to our unary data representation as mentioned in \S\ref{subsec:Data Model Creation and Distribution of Secret-Shares of a Relation}).

\BB
\begin{theorem}
{\textnormal{\textbf{(Count Query)}}} For the count operation, the communication cost is $\BigO(1)$, the computational cost at the server is $\BigO(\mathit{nw})$, and the computational cost at the user is $\BigO(1)$.
\end{theorem}
\BB
\begin{proof}
Since a user sends a predicate of bit length $w$ and receives $c$ values from the servers, the communication cost is constant that is $\BigO(1)$. The server works on a specific attribute containing $n$ values, each of bit length $w$; hence, the computational cost at a server is at most $\BigO(\mathit{nw})$. The user only performs Lagrange interpolation on the received $c$ values from the servers; hence, the computational cost at the user is also constant, $\BigO(1)$.
\end{proof}

\BB
\begin{theorem}
{\textnormal{\textbf{(Selection Query: One Value One Tuple)}}} For selection operation where only a single tuple can have the searching predicate, the communication cost is at most $\BigO(mw)$, the computational cost at a server is at most $\BigO(nmw)$, and the computational cost at the user is at most $\BigO(mw)$.
\end{theorem}
\BB
\begin{proof}
The user sends a selection predicate of bit length $w$ and receives $c$ secret-shares and, eventually, a tuple containing $m$ attributes of size at most $mw$. Thus, the communication cost is at most $\BigO(mw)$ bits. The server, first, counts the occurrences of the predicate in a specific attribute containing $n$ values, and then, again, performs a similar operation on the $n$ tuples, multiplying the resultant by each $m$ values of bit length at most $w$. Hence, the computational cost at the server is $\BigO(nmw)$. The user performs the interpolation on $c$ values (the output of count query) to know the occurrences of the predicate, and then, again, performs the interpolation on $c$ tuples containing $m$ attributes. Thus, the computational cost at the user is $\BigO(mw)$.\end{proof}

\BB
\begin{theorem}
{\textnormal{\textbf{(Selection Query: One-round Algorithm)}}} Let $\ell >1$ be the number of tuples containing a searching predicate $p$. After obtaining the addresses of the desired tuples containing a predicate, $p$, the fetch operation of the desired tuples results in the following cost: the communication cost is $\BigO(n+\ell mw)$, the computational cost at a server is $\BigO(\ell nmw)$, and the computational cost at the user is $\BigO(n+\ell mw)$.
\end{theorem}
\BB
\begin{proof}
In the first round of the one-round algorithm, the user receives $n$ secret-shares of a particular attribute. In the second round, the user sends a vector having $\ell$ row-ids each of at most $w$ bits and receives $\ell$ tuples, each of size at most $mw$. Thus, the maximum number of bits flow is $\BigO(n+\ell mw)$. Mappers perform string-matching operations on $n$ secret-shares of a particular attribute in the first round, then, string-matching on \texttt{RID} attribute in the second round, and multiply the string-matching resultant to the remaining secret-shared values of the tuple. Hence, the computational cost at the server is $\BigO(\ell nmw)$. The computational cost at the user is $\BigO(n+\ell mw)$, since the user works on the $n$ secret-shares of a specific attribute, creates a vector of length $\ell$ in the first round and, then, works on $\ell$ tuples containing $m$ values, each of size at most $w$ bits.
\end{proof}

\BB
\begin{theorem}
{\textnormal{\textbf{(Selection Query: Tree-based Algorithm)}}} Let $\ell>1$ be the number of tuples containing a searching predicate $p$. The number of rounds for obtaining addresses of tuples containing a selection predicate, $p$, using tree-based search (Algorithm~\ref{alg:Multi-tuple search Algorithm}) is at most $\lfloor \log_{\ell}n\rfloor +\lfloor \log_2{\ell} \rfloor+1$, and the communication cost for obtaining such addresses is $\BigO\big((\log_{\ell}n+\log_2{\ell})\ell\big)$. The computational cost at a server is $\BigO\big((\log_{\ell}n+\log_2{\ell})\ell nw\big)$, and the computational cost at the user is $\BigO\big((\log_{\ell}n+\log_2{\ell})\ell\big)$.
\end{theorem}
\BB
\begin{proof}
In each round of Q\&A of tree-based algorithm, the user obtains the number of occurrences of $p$ in each block. If the block contains zero or one occurrence, then the user can know the tuple-id. However, if there are some blocks containing more than one occurrence of $p$, the user asks the servers to partition these blocks into $\ell$ sub-blocks. After $i^{th}$ round of Q\&A, each sub-block contains at most $\frac{n}{\ell^i}$ tuples. Therefore, it is clear that after $\lfloor\log_\ell n \rfloor$ rounds, the number of tuples contained in each block is less than $\frac{n}{\ell^{\lfloor\log_\ell n \rfloor}}<\ell$. At this time, there may be some blocks still contain more than one tuple having $p$. Similarly, we split these blocks until all the sub-blocks contain only one/zero occurrence of $p$. It requires at most $\lfloor\log_2\ell\rfloor$ rounds of Q\&A. Finally, it may need one more round for obtaining the tuple-ids of desired tuples. Thus, the Q\&A round is at most $\lfloor \log_\ell n \rfloor+\lfloor\log_2\ell\rfloor + 1$.

Note that for each round, there may be at most $\frac{\ell}{2}$ blocks containing more than two occurrences, which indicate that at most $\frac{\ell}{2}$ blocks need further partitioning. Thus, in every Q\&A round (except the first round), the server only needs to perform count operation for $\frac{\ell}{2}$ sub-blocks and sends the results to the user. When the server finishes partitioning all the sub-blocks, it may need to determine the tuple-ids, which require at most $\ell$ words transmission between the server and the user. Therefore, the communication cost is $\BigO((\log_{\ell}n+\log_2\ell)\cdot\ell)$.

A server performs count operation in each round; hence, the computational cost at the server is $\BigO((\log_{\ell}n+\log_2\ell)\ell nw)$. In each round, the user performs the interpolation
for obtaining the occurrences of $p$ in each block; hence, the computational cost at the user-side is at most $\BigO((\log_{\ell}n+\log_2\ell)\cdot\ell)$.
\end{proof}

\noindent\textit{\textbf{Example.}} In figure~\ref{fig:Q/A}, in order to fetch tuples containing $p$, the user needs 4 rounds (include the last fetch round), which is smaller than $\lfloor\log_2 9\rfloor + \lfloor\log_2 2\rfloor + 1 = 5$.
\BB
\begin{theorem}
{\textnormal{\textbf{(PK-FK Joins)}}} For performing PK/FK join of two relations $X$ and $Y$, where each relation is having $n$ tuples and each tuple of the parent relation joins with one tuple of the child relation, the communication cost is $\BigO(nmw)$, the computational cost at a server is $\BigO(n^2mw)$, and the computational cost at the user is $\BigO(nmw)$.
\end{theorem}
\BB
\begin{proof}
Since the user receives the whole relation of $n$ tuples and at most $2m-1$ attributes, the communication cost is $\BigO(nmw)$, and due to the interpolation on the $n$ tuples, the computational cost at the user is $\BigO(nmw)$. A reducer compares each value of the joining attribute of the relation $Y$ with all $n$ values of the joining attribute of the relation $X$, and it results in at most $n^2$ comparisons. Further, the output of the comparison is multiplied by $m-1$ attributes of the corresponding tuple of the relation $Y$. Hence, the computational cost at a server is $\BigO(n^2mw)$.
\end{proof}

\BB
\begin{theorem}
{\textnormal{\textbf{(Non-PK/FK Joins)}}} Let $X$ and $Y$ be two relations. Let $k$ be the number of identical values of the joining attribute in the relations, and let $\ell$ be the maximum number of occurrences of a joining value. For performing non-PK/FK equijoin of two relations $X$ and $Y$, the number of rounds is $\BigO(2k)$, the communication cost is $\BigO(2nwk+2k\ell^2mw)$, the computational cost at a server is $\BigO(\ell^2kmw)$, and the computational cost at the user is $\BigO(2nw+2k\ell^2mw)$.
\end{theorem}
\BB
\begin{proof}
Since there are at most $k$ identical values of the joining attribute in both relations and all the $k$ values can have a different number of occurrences in the relations, the user has to send at most $2k$ matrices (following an approach of the one-round algorithm for fetching multiple tuples~\S\ref{subsec:Unary Occurrence of a Pattern}) in $\BigO(2k)$ rounds.

The user sends at most $2k$ matrices, each of $n$ rows and of size at most $w$; hence, the user sends $\BigO(2knw)$ bits. Since at most $\ell$ tuples have an identical value of the joining attribute in one relation, equijoin provides at most $\ell^2$ tuples. The user receives at most $\ell^2$ tuples for each $k$ value having at most $2m-1$ attributes; hence, the user receives $\BigO(2k\ell^2mw)$ bits. Therefore, the communication cost is $\BigO(2nwk+2k\ell^2mw)$ bits.

The server of the first layer executes the one-round algorithm for fetching multiple tuples for all $k$ values of both relations having $2n$ tuples; hence the servers of the first layer performs $\BigO(2nkw)$ computation. In the second layer, a server performs an equijoin (or concatenation) of at most $\ell$ tuples for each $k$ value; thus, the computational cost at the server is $\BigO(\ell^2kmw)$.

The user first interpolates at most $2n$ values of bit length $w$ of the joining attribute, and then, interpolates $k\ell^2$ tuples containing at most $2m-1$ attributes of bit length $w$. Therefore, the computational cost at the user is $\BigO(2nw+2k\ell^2mw)$.
\end{proof}

\BB
\section{Conclusion}
\label{sec:Conclusion}
\B
The database outsourcing to public servers is a prominent solution to deal with a resource-constrained database owner and avoid overheads for maintaining and executing queries at the database owner. However, the public servers do not ensure security and privacy of data/computation. The modern data processing frameworks such as Hadoop or Spark, which are designed for a trusted environment, do not deal with security and privacy of data and computations. This paper presented information-theoretically secure data and computation outsourcing techniques, especially, algorithms for count, selection, projection, join, and range queries, while using MapReduce as an underlying programming model. The techniques are designed in a way that a heavy computation is executed at the servers instead of the user, without revealing the query to the server or the database owner. We experimentally evaluated our proposed algorithms on 1M and 9M rows of TPC-H benchmark using AWS servers. Experimental results show that our algorithms are better than a trivial way of executing queries after downloading and decrypting the entire data at the user.

\noindent\textbf{Future directions.} The algorithms proposed in this paper should be considered in different domains. For example, an interesting direction may focus on the data outsourcing model, where an untrusted user executes queries on the secret-shared data without involving either the database owner or a trusted-third-party, and the user receives only the desired output. Another interesting direction may focus on malicious adversaries, while not using any Byzantine fault-tolerant protocol. Such algorithms should also be information-theoretically secure and may allow an untrusted user to verify the result of their queries without involving a third-party (\textit{e}.\textit{g}., the database owner). Nevertheless, designing algorithms for these new models should also prevent access-patterns- and output-size-based attacks to achieve complete data privacy.


\newpage
\begin{IEEEbiography}[{\includegraphics[width=1in,height=1.25in,clip,keepaspectratio]{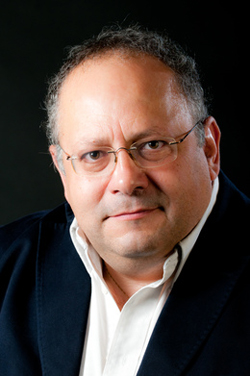}}]​{Shlomi Dolev} received his DSc in Computer Science in 1992 from the Technion. He is the founder and the first department head of the Computer Science Department at Ben Gurion University, established in 2000. Shlomi is the author of a book entitled Self Stabilization published by MIT Press in 2000. His publications include more than three hundreds publications in computer science, distributed computing, networks, cryptography, security, optical and quantum computing, nanotechnology, brain science and machine learning. He served in more than a hundred program committees, chairing several including the two leading conferences in distributed computing, DISC 2006, and PODC 2014. Prof. Dolev is the head of BGU Negev faculty High-Tech Accelerator and holds the Ben-Gurion University Rita Altura trust chair. From 2011 to 2014, Prof. Dolev served as the Dean of the Natural Sciences Faculty at Ben-Gurion University of the Negev. From 2010 to 2016, he has served as Head of the Inter University Computation Center of Israel. Shlomi currently serves as the steering committee head of the Computer Science discipline of the Israeli ministry of education. He is fellow of the European Alliance for Innovation (EAI from 2019) and serial entrepreneur, currently co-founder, board member and Chief Scientific Officer of Secret Double Octopus.
\end{IEEEbiography}
\BBB\BBB\BBB\BBB\BBB

\begin{IEEEbiography}[{\includegraphics[width=1in,height=1.25in,clip,keepaspectratio]{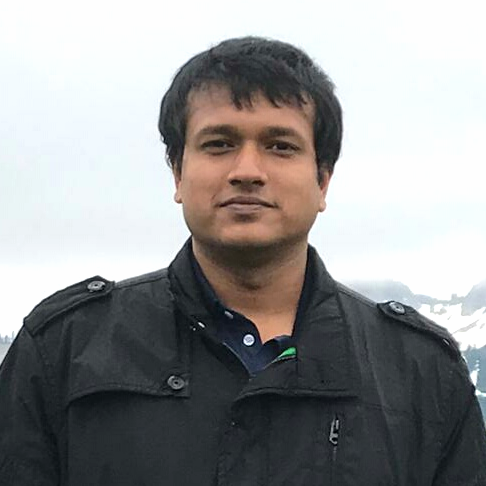}}]{Peeyush Gupta} is a Ph.D. student, advised by Prof. Sharad Mehrotra, at University of California, Irvine, USA. He obtained his Master of Technology  degree in Computer Science from Indian Institute of Technology, Bombay, India, in 2013. His research interests include IoT data management, time series database systems, and data security and privacy.
\end{IEEEbiography}
\BBB\BBB\BBB\BBB\BBB

\begin{IEEEbiography}[{\includegraphics[width=1in,height=1.25in,clip,keepaspectratio]{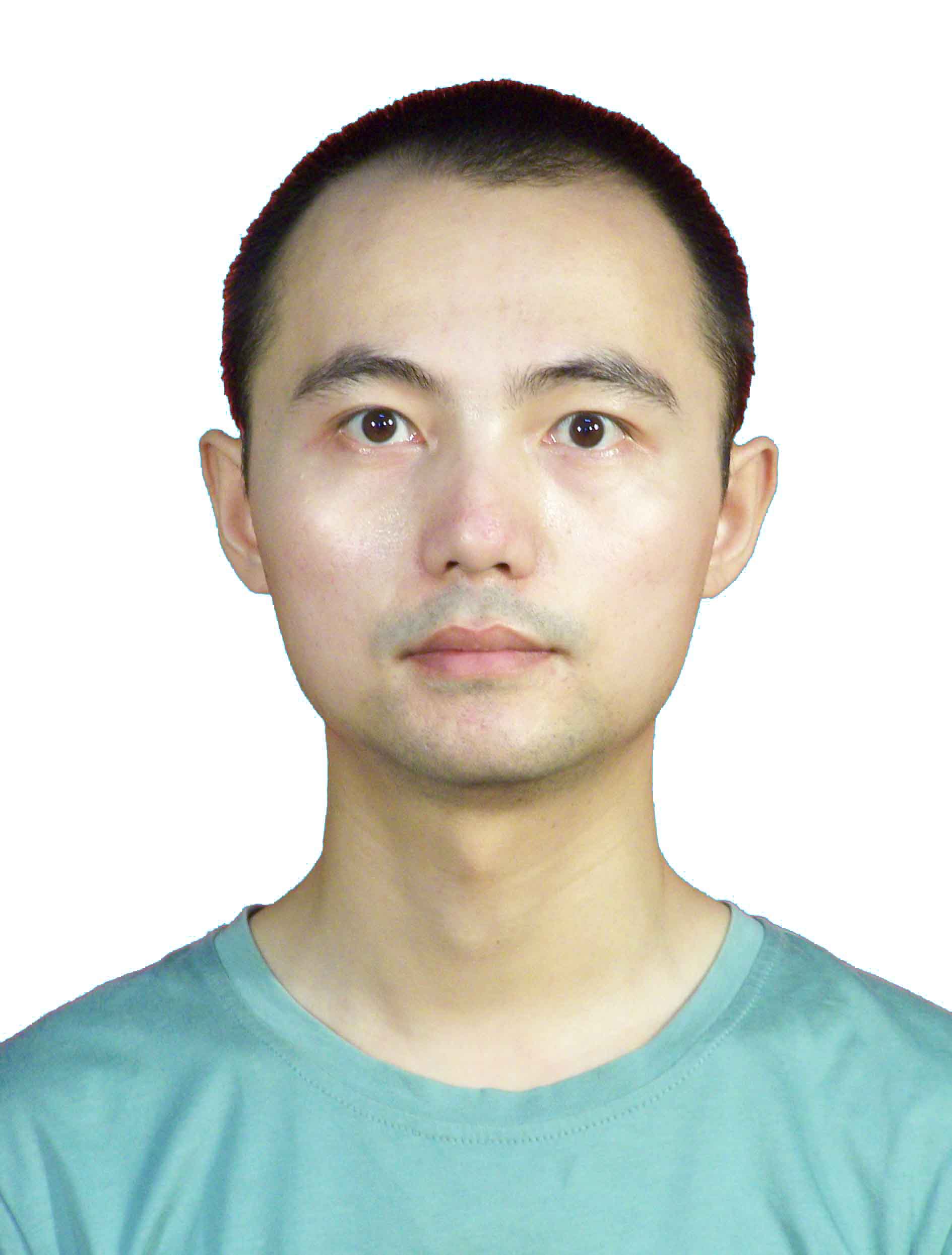}}]{Yin Li} received his BSc degree in Information Engineering, and MSc degree in Cryptography from Information Engineering University, Zhenzhou, in 2004 and 2007, and the PhD in Computer Science from Shanghai Jiaotong University (SJTU), Shanghai (2011). He was a postdoc in Department of Computer Science, Ben-Gurion University of the Negev, Israel, assisted by Prof. Shlomi Dolev. Now, he is an associated professor in Department of Computer Science and Technology, Xinyang Normal University, Henan, China. His current research interests include algorithm and architectures for computation in finite field, computer algebra, and secure cloud computing.
\end{IEEEbiography}
\BBB\BBB\BBB\BBB\BBB

\begin{IEEEbiography}[{\includegraphics[width=1in,height=1.25in,clip,keepaspectratio]{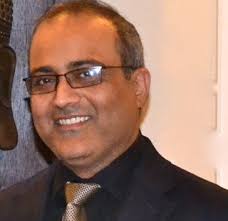}}]{Sharad Mehrotra} received the PhD degree in computer science from the University of Texas, Austin, in 1993. He is currently a professor in the Department of Computer Science, University of California, Irvine. Previously, he was a professor with the University of Illinois at Urbana Champaign. He has received numerous awards and honors, including the 2011 SIGMOD Best Paper Award, 2007 DASFAA Best Paper Award, SIGMOD test of time award, 2012, DASFAA ten year best paper awards for 2013 and 2014, 1998 CAREER Award from the US National Science Foundation (NSF), and ACM ICMR best paper award for 2013. His primary research interests include the area of database management, distributed systems, secure databases, and Internet of Things.
\end{IEEEbiography}
\BBB\BBB\BBB\BBB\BBB

\begin{IEEEbiography}[{\includegraphics[width=1in,height=1.25in,clip,keepaspectratio]{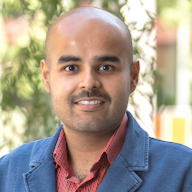}}]{Shantanu Sharma} received his Ph.D. in Computer Science in 2016 from Ben-Gurion University, Israel. During his Ph.D., he worked with Prof. Shlomi Dolev and Prof. Jeffrey Ullman. He obtained his Master of Technology (M.Tech.) degree in Computer Science from National Institute of Technology, Kurukshetra, India, in 2011. He was awarded a gold medal for the first position in his M.Tech. degree. Currently, he is pursuing his Post Doc at the University of California, Irvine, USA, assisted by Prof. Sharad Mehrotra. His research interests include data security and privacy, building secure and privacy-preserving systems on sensor data for smart buildings,
designing models for MapReduce computations, distributed algorithms, mobile computing, and wireless communication.

\end{IEEEbiography}
\BBB\BBB\BBB\BBB\BBB

\newpage

\appendix

\section{Algorithms' Pseudocodes}
\label{app_sec:Pseudocodes}
\LinesNotNumbered
\begin{algorithm}[!h]
\DontPrintSemicolon
\scriptsize
\textbf{Inputs:} $\mathcal{R}$: a relation having $n$ tuples and $m$ attributes, $c$: the number of non-communicating servers

{\bf Variables:} $\mathit{letter}$: represents a letter

\nl{\bf Function $\mathit{create}\_\mathit{secret}\textnormal{-}\mathit{shares}(\mathcal{R})$} \nllabel{ln:function_create_secret_shares}
\Begin{

\nl \For{$(i,j) \in (n,m)$}{

\nl \lForEach{$\mathit{letter}[i,j]$}{$\mathit{Make\_shares(letter[i,j])}$}
}}

\nl{\bf Function $\mathit{Make\_shares(letter[i,j])}$}
\Begin{
\nl $x \gets$ length of $\mathit{letter}[i,j]$

\nl Create $x$ unary-vectors, where the position of the letter has value 1 and all the other values are 0

\nl Use $x$ polynomials of an identical degree for creating secret-shares of 0 and 1

\nl Send secret-shares to $c$ servers \nllabel{ln:assign_each_pair_of_derived_input}
}

\caption{Algorithm for creating secret-shares}
\label{alg:Algorithm for creating secret-shares}
\end{algorithm}
\setlength{\textfloatsep}{0pt}


\LinesNotNumbered \begin{algorithm}[!h]
\DontPrintSemicolon
\scriptsize
\textbf{Inputs:} $R$: a relation of the form of secret-shares having $n$ tuples and $m$ attributes, $p$: a counting predicate, $c$: the number of servers, $v_i$: an $i^{\mathit{th}}$ value, and $\mathit{Len}(v_i)$: length of $v_i$. \\

\textbf{Output:} $\mathit{fcount}$: the number of occurrences of $p$\\

\textbf{Interfaces:} $\mathit{attribute(p)}$: which attribute of the relation has to be searched for $p$\\

\textbf{Variables:} $\mathit{int\_result_i}$: the output at $i^{\mathit{th}}$ server after executing the map function\\

\textbf{Notation:} $\odot$: string-matching operations\\

\smallskip
{\bf User:}\\
\nl Compute secret-shares of $p$: $p^{\prime}\leftarrow\mathit{Make}\_\mathit{shares(p)}$ {\tiny\tcp*{Algorithm~\ref{alg:Algorithm for creating secret-shares}}} \nllabel{ln:creat_shares}

\nl Send $\mathit{p^{\prime}, m^{\prime} \leftarrow attribute(p)}$ to $c$ servers \nllabel{ln:send_shares}

\smallskip
{\bf Server $i$:} \\
\nl $\mathit{int\_result_i}\leftarrow \mathit{MAP}\_\mathit{count}(p^{\prime},m^{\prime})$ \nllabel{ln:call_map}

\nl Send $\mathit{int\_result}_i$ back to the user \nllabel{ln:send_result}

\smallskip
{\bf User:}\\
\nl $\mathit{result}[i] \leftarrow \mathit{int\_result}_i$, $\forall i\in\{1,2,\ldots,c\}$ \nllabel{ln:store_cloud_outputs}

\nl Compute the final output: $\mathit{fcount} \leftarrow$ $\mathit{Interpolate(result[])}$ \nllabel{ln:call_reduce_count}

\medskip
\nl \textbf{Function $\mathit{MAP\_count(p^{\prime},m^{\prime})}$} \nllabel{ln:function_map}
\Begin{
\nl \For{$\mathrm{r} \in (1,n)$}{\nllabel{ln:map}

\textsc{Step} $1^{\mathrm{r}}$: $N_1^{\mathrm{r}}=m^{\prime}[[v_{\mathrm{r}}]_1] \odot [p^{\prime}]_1$\\
\textsc{Step} $2^{\mathrm{r}}$: $N_2^{\mathrm{r}}=m^{\prime}[[v_{\mathrm{r}}]_2] \odot [p^{\prime}]_2$\\

\indent \vdots

\textsc{Step} ${\mathit{Len(v_r)}}^{\mathrm{r}}$: $N_{\mathit{Len(v_r)}}^{\mathrm{r}}=m^{\prime}[[v_{\mathrm{r}}]_{\mathit{Len(v_r)}}] \odot [p^{\prime}]_{\mathit{Len(v_r)}}$\\

\textsc{Step} $(\mathit{Len(v_r)}+1)^{\mathrm{r}}$: $\mathit{count}= \mathit{count}+ N_1^{\mathrm{r}} \times N_2^{\mathrm{r}} \times \ldots \times N_{\mathit{Len(v_r)}}^{\mathrm{r}}$\\
}

\nl $\KwRet(count)$\nllabel{ln:output_of_a_cloud}}
\caption{Algorithm for privacy-preserving count queries}
\label{alg:count algorithm}
\end{algorithm}
\setlength{\textfloatsep}{0pt}

\noindent
\textbf{Count query --- Algorithm~\ref{alg:count algorithm}'s pseudocode description.}

\noindent 1. The user creates secret-shares of $p$ (see line~\ref{ln:creat_shares}), sends them along with the attribute identity ($m^{\prime}$, on which the servers will execute the count query) to $c$ servers (see line~\ref{ln:send_shares}).

\noindent 2. The server executes a map function (line~\ref{ln:map}) that reads each value of the form of secret-share of the $m^{\prime}$ attribute, executes string-matching, and produces the final secret-shared output (line~\ref{ln:output_of_a_cloud}).

\noindent 3. The user interpolates the received results from the servers and obtains an answer to the count query (line~\ref{ln:call_reduce_count}).


\LinesNotNumbered
\begin{algorithm}[!h]
\DontPrintSemicolon
\scriptsize
\textbf{Inputs:} $R$, $n$, $m$, $p$, and $c$ are defined in Algorithm~\ref{alg:count algorithm}\\

\textbf{Output:} A tuple $t$ containing $p$\\

\textbf{Variables:} $\ell$: the number of occurrences of $p$\\
$\mathit{int\_result\_search_i}$: the output at a server $i$ after fetching a single tuple in a privacy-preserving manner\\
$\mathit{result\_search}[]$: an array to store outputs of all the servers\\

\textbf{Interfaces:} $\mathit{attribute(p)}$: which attribute of the relation has to be searched for $p$\\

\smallskip
{\bf User:}\\
\nl Compute secret-shares of $p$: $p^{\prime}\leftarrow\mathit{Make}\_\mathit{shares(p)}$ \\
Send $p^{\prime}$ and $m^{\prime}\leftarrow \mathit{attribute}(p)$ to $c$ servers to execute Algorithm~\ref{alg:count algorithm} for obtaining the number of occurrences ($\ell$) of $p$ \nllabel{ln:count_single}

\nl \lIf{$\ell > 1$}{Execute Algorithm~\ref{alg:Multi-tuple search Algorithm}}

\nl \lElse{Send $p^{\prime}$ and $m^{\prime}\leftarrow attribute(p)$ to $c$ servers \nllabel{ln:resend_shares_for_single_fetch_single}}

\smallskip
{\bf Server $i$:} \\
\nl $\mathit{int\_result\_search}_i\leftarrow \mathit{MAP\_single\_tuple\_fetch}(\mathit{p^{\prime},m^{\prime}})$

\nl Send $\mathit{int\_result\_search}_i$ back to the user

\smallskip
{\bf User:}\\
\nl $\mathit{result\_search}[i] \leftarrow \mathit{int\_result\_search}[i]$, $\forall i\in\{1,2,\ldots,c\}$

\nl Obtain the tuple $\mathit{t}\leftarrow\mathit{Interpolate(result\_search[])}$\nllabel{ln:call_reduce_single}\\

\medskip
\nl \textbf{Function $\mathit{MAP\_single\_tuple\_fetch(p^{\prime},m^{\prime})}$} \nllabel{ln:map_function_single_fetch}
\Begin{
\nl \For{$i\in (1,n)$}{$\mathit{temp} \leftarrow \mathit{string\_matching}(m^{\prime}[[v_i]] \odot [p^{\prime}])$ {\scriptsize\tcp*{Likewise Algorithm~\ref{alg:count algorithm}}}\nllabel{ln:aa_single_search}

\nl \lFor{$j \in (1,m)$}{$\mathit{temp} \times R^i[\ast,j]$} \nllabel{ln:map_multiply_single}
}

\nl \lFor{$(j,i) \in (m,n)$}{$S_j \leftarrow$ add all the shares of $j^{th}$ attribute\nllabel{ln:sum_shares_single}}

\nl $\KwRet(S_1||S_2||\ldots||S_m)$ \nllabel{ln:return_multiply_sum_result_single}

}
\caption{Algorithm for privacy-preserving selection queries in the case of one tuple per value}
\label{alg:single word fetch}
\end{algorithm}
\setlength{\textfloatsep}{0pt}

\medskip
\noindent\textbf{Selection query for one tuple per value --- Algorithm~\ref{alg:single word fetch}'s pseudocode description.}

\noindent 1. The user executes Algorithm~\ref{alg:count algorithm} for counting occurrences, say $\ell$, of $p$; line~\ref{ln:count_single}.

\noindent 2. If $\ell$ is one, the user sends secret-shares of $p$ and the attribute, say $m^{\prime}$, where $p$ occurs, to $c$ servers; line~\ref{ln:resend_shares_for_single_fetch_single}.

\noindent 3. Each server executes a map function that

\begin{enumerate}[label=\textit{\alph{*}.},noitemsep,nolistsep]
\item Executes string-matching on the $i^{th}$ value of the $m^{\prime}$ attribute (line~\ref{ln:aa_single_search}), and this provides a value, say $\mathit{temp}$, either 0 or 1 of the form of secret-shares. Then, it multiplies $\mathit{temp}$ by all the attribute values of the $i^{th}$ tuples; line~\ref{ln:map_multiply_single}.

\item When the above step is completed on all the $n$ secret-shared tuples, it adds all the secret-shares of each attribute; line~\ref{ln:sum_shares_single}.

\item Each server sends the sum of each attribute's secret-shares to the user; line~\ref{ln:return_multiply_sum_result_single}.

\end{enumerate}
\noindent 4. The user receives $c$ secret-shared tuples and interpolates them to obtain the desired tuple; line~\ref{ln:call_reduce_single}.

\begin{algorithm}[!h]
\DontPrintSemicolon
\scriptsize
\textbf{Inputs:} $R$, $n$, $m$, $p$, and $c$ are defined in Algorithm~\ref{alg:count algorithm}\\
\textbf{Outputs:} Tuples containing $p$\\
\textbf{Variables:} $\ell$: the number of occurrences of $p$\\
$\mathit{int\_result\_block\_count_i}[j]$: at $i^{th}$ server to store the number of occurrences of the form of secret-shares in $j^{th}$ block\\
$\mathit{result\_block\_count[]}$: at the user to store the count of occurrences of $p$ of the form of secret-shares in each block at each server\\
$\mathit{count[]}$: at the user to store the count of occurrences of $p$ in each block\\
$\mathit{Address}[]$: stores the addresses of the desired tuples

\smallskip
{\bf User:}\\
\nl Compute secret-shares of $p$: $p^{\prime}\leftarrow\mathit{Make}\_\mathit{shares(p)}$ and execute Algorithm~\ref{alg:count algorithm} for obtaining the number of occurrences ($\ell$) of $p$ \nllabel{ln:count}

\nl \lIf{$\ell = 1$}{Execute Steps~\ref{ln:resend_shares_for_single_fetch_single} to~\ref{ln:call_reduce_single} of Algorithm~\ref{alg:single word fetch}\nllabel{ln:single_tuple}}

\nl \lElse{Send $p^{\prime}$, $m^{\prime}\leftarrow attribute(p)$, $\ell$ to $c$ servers \nllabel{ln:multiple}}

\smallskip
{\bf Server $i$:} \\
\nl Partition $R$ into $\ell$ equal blocks, where each block contains $h=\frac{n}{\ell}$ tuples\nllabel{ln:partition}\\

\nl $\mathit{int\_result\_block\_count_i[j]}\leftarrow$ Execute $\mathit{MAP\_count}(p^{\prime},m^{\prime})$ $j^{th}$ block, $\forall j \in \{1, 2,\ldots, \ell\}$\nllabel{ln:call_map_multi}\\

\nl Send $\mathit{int\_result\_block\_count_i[j]}$ back to the user\nllabel{ln:send_block_count}

\smallskip
{\bf User:}\\

\nl $\mathit{result\_block\_count}[i,j] \leftarrow \mathit{int\_result\_block\_count_i}[j]$, $\forall i\in\{1,2,\ldots,c\}$, $\forall j\in\{1,2,\ldots,\ell\}$

\nl Compute $\mathit{count[j]}\leftarrow\mathit{Interpolate(result\_block\_count[i,j])}\nllabel{ln:count_in_block}$

\nl \If{$\mathit{count[j]}\notin\{0,1,h\}$}{Question the servers about $j^{th}$ block and send $\langle p^{\prime}, \mathit{count[j]}, m^{\prime}, j\rangle$ to servers}\nllabel{ln:call_recursive_partition}

\nl Fetch the tuples whose addresses are known using the one-round algorithm \nllabel{ln:multiple fetch}

\smallskip
{\bf Server $i$:}

\nl \lIf {Receive $\langle p^{\prime}, \mathit{count[j]}, m^{\prime}, j\rangle$} {Perform Steps~\ref{ln:partition} to~\ref{ln:send_block_count} to $j^{th}$ block recursively\nllabel{ln:recursive_division}}
\caption{Tree-based algorithm for privacy-preserving selection queries in the case of multiple values having tuples.}
\label{alg:Multi-tuple search Algorithm}
\end{algorithm}
\setlength{\textfloatsep}{0pt}

\medskip
\noindent\textbf{Tree-based selection query --- Algorithm~\ref{alg:Multi-tuple search Algorithm}'s pseudocode description.} A user creates secret-shares of $p$ and obtains the number of occurrences, $\ell$; see line~\ref{ln:count}. When the occurrences $\ell=1$, we can perform Algorithm~\ref{alg:single word fetch} for fetching the only tuple having $p$; see line~\ref{ln:single_tuple}. When the occurrences $\ell>1$, the user needs to know the addresses of all the $\ell$ tuples contain $p$. Thus, the user requests to partition the input split/relation to $\ell$ blocks, and hence, sends $\ell$ and $p$ of the form of secret-shares to the servers; see line~\ref{ln:multiple}.

The mappers partition the whole relation or input split into $\ell$ blocks, perform privacy-preserving count operations in each block, and send all the results back to the user; see lines~\ref{ln:partition} -~\ref{ln:send_block_count}. The user interpolates the results and knows the number of occurrences of $p$ in each block; see line~\ref{ln:count_in_block}. Based on the number of occurrences of $p$ in each block, the user decides which block needs further partitioning, and there are four cases, as follows:
 \begin{enumerate}[noitemsep,nolistsep,leftmargin=.4cm]
 \item The block contains no occurrence of $p$: it is not necessary to handle this block.

 \item The block contains only one tuple containing $p$: in this case, the user can execute Algorithm~\ref{alg:single word fetch}.

 \item The block contains $h$ tuples and each $h$ tuple contains $p$: directly know the addresses, \textit{i}.\textit{e}., all the $h$ tuples are required to fetch.

 \item The block contains $h$ tuples and more than one, but less than $h$ tuples contain $p$: we cannot know the addresses of these tuples. Hence, the user recursively requests to partition that block and continues the process until the sub-blocks satisfy the above-mentioned Case 2 or Case 3; see line~\ref{ln:call_recursive_partition}.
\end{enumerate}
When the user obtains the addresses of all the tuples containing $p$, she fetches all the tuples using a method described for the one-round algorithm; see line~\ref{ln:multiple fetch}.

\LinesNotNumbered
\begin{algorithm}[!h]
\DontPrintSemicolon
\scriptsize
\textbf{Inputs:} $R$, $n$, $m$, and $c$: defined in Algorithm~\ref{alg:count algorithm}, $[a,b]$: a searching range\\
\textbf{Output:} $\mathit{fcount}$: the number of occurrence in $[a,b]$ \\

\textbf{Variables:} $\mathit{int\_result_i}$: is initialized to 0 and the output at $i^{th}$ server after executing the $\mathit{MAP\_range\_count}$ function\\

\smallskip
{\bf User:}\\
\nl Compute secret-shares of $a, b$: $a^{\prime}\leftarrow\mathit{Make\_shares(a)}$, $b^{\prime}\leftarrow\mathit{Make\_shares(b)}$ \nllabel{ln:make_share_range} \\
\nl Send $\mathit{a^{\prime}, b^{\prime}, m^{\prime} \leftarrow attribute(a)}$ to $c$ servers \nllabel{ln:send_shares_range}

\smallskip
{\bf Server $i$:} \\

\nl \lFor{$i \in (1,n)$}{$\mathit{int\_result_i}\leftarrow \mathit{int\_result_i}+\mathit{MAP\_range\_count}(a^{\prime},b^{\prime},m^{\prime}[[v_i]])$ \nllabel{ln:call_map_range}}

\nl Send $\mathit{int\_result}_i$ back to the user \nllabel{ln:send_result_range}

\smallskip
{\bf User:}\\
\nl $\mathit{result}[i] \leftarrow \mathit{int\_result}_i$, $\forall i\in\{1,2,\ldots,c\}$ \nllabel{ln:store_cloud_outputs_rangle}

\nl $\mathit{fcount} \leftarrow$ $n-\mathit{Interpolate(result[])}$ \nllabel{ln:call_reduce_range}

\medskip
\nl \textbf{Function $\mathit{MAP\_range\_count(a^{\prime}, b^{\prime},m^{\prime}[[v_i]])}$} \nllabel{ln:function_range}
\Begin{

\nl $\mathit{op}_1 \leftarrow \mathit{SS}\textnormal{-}\mathit{SUB}(m^{\prime}[[v_i]],a^{\prime})$ {\scriptsize\tcp*{Algorithm~\ref{algo:SS-SUB}}}

\nl $\mathit{op}_2\leftarrow \mathit{SS}\textnormal{-}\mathit{SUB}(b^{\prime},m^{\prime}[[v_i]])$ {\scriptsize\tcp*{Algorithm~\ref{algo:SS-SUB}}}

\nl $\KwRet(\mathit{op}_1+\mathit{op}_2)$

}

\caption{Algorithm for privacy-preserving range-based count query}
\label{alg:range count}
\end{algorithm}
\setlength{\textfloatsep}{0pt}

\medskip \noindent{\bf Range-based Count Query --- Algorithm \ref{alg:range count}'s pseudocode description.} A user creates secret-shares of the range numbers $a, b$ and sends them to $c$ servers; see lines~\ref{ln:make_share_range} and~\ref{ln:send_shares_range}. The server executes a map function that checks each number in an input split by implementing Algorithm~\ref{algo:SS-SUB}; see lines~\ref{ln:call_map_range} and ~\ref{ln:function_range}. The map function; see line~\ref{ln:function_range}, provides 0 (of the form of secret-share) if $x\in[a,b]$; otherwise, 1 (of the form of secret-share). The server provides the number of occurrences (of the form of secret-shares) of tuples that do \emph{not} belong in the ranges to the user; see line~\ref{ln:send_result_range}. The user receives all the values from servers and interpolates them; see lines~\ref{ln:store_cloud_outputs_rangle} and~\ref{ln:call_reduce_range}. Since the obtained interpolated answer indicates the number of tuples that do not satisfy the range condition, subtracting it from $n$ provides the final answer to the range-based count query.

\begin{algorithm}[t]
\DontPrintSemicolon
\scriptsize
\textbf{Inputs:} $A=[a_{t-1}a_{t-2}\ldots a_1a_0], B=[b_{t-1}b_{t-2}\ldots b_1b_0]$ where $a_i, b_i$ are secret-shares of bits of 2's complement represented number, $t$: the length of $A$ and $B$ in the binary form\\
\textbf{Outputs:} $\mathit{rb}_{t-1}$: the sign bit of $B-A$\\
\textbf{Variable:} $\mathit{carry}[]$: to store the carry for each bit addition \\
$\mathit{rb}$: to store the result for each bit addition \\

\nl $a_0\leftarrow 1-a_0$ {\scriptsize\tcp*{Invert of the LSB of $A$}}\nllabel{ln:1}

\nl $\mathit{carry}[0]\leftarrow a_0+b_0-a_0\cdot b_0$ \nllabel{ln:2}

\nl $\mathit{rb}_0\leftarrow a_0+b_0-2\cdot \mathit{carry}[0]$ {\scriptsize\tcp*{$\bar{a}_0+b_0+1$}} \nllabel{ln:3}

\nl \For{$i\in (i,t-1)$}{\nllabel{ln:4} $a_i\leftarrow 1-a_i$ {\scriptsize\tcp*{invert each bit $A\rightarrow \bar{A}$}}

$\mathit{rb}_i\leftarrow a_i+b_i-2a_ib_i$

$\mathit{carry}[i]\leftarrow a_ib_i+ \mathit{carry}[i-1]\cdot \mathit{rb}_i$ {\scriptsize\tcp*{The carry bit}}

$\mathit{rb}_i+=\mathit{carry}[i-1]-2\cdot \mathit{carry}[i-1]\cdot \mathit{rb}_i$}

\nl $\KwRet(\mathit{rb}_{t-1})$ {\scriptsize\tcp*{The sign bit of $B-A$} } \nllabel{ln:5}
\caption{$\mathit{SS}\textnormal{-}\mathit{SUB}(A, B)$: 2's complement-based subtraction of secret-sharing}
\label{algo:SS-SUB}
\end{algorithm}
\setlength{\textfloatsep}{0pt}

\end{document}